\documentclass[twocolumn,tighten,astrosymb]{aastex631}

\usepackage{amsmath}
\usepackage{enumerate}
\usepackage{multirow}
\usepackage{CJKutf8}
\usepackage{todonotes}
\usepackage{soul}
\usepackage{hyperref}

\newcommand{\kms}{km s$^{-1}$}
\newcommand{\perpix}{pixel$^{-1}$}

\newcommand{\FeH}{\text{[Fe/H]}}

\newcommand{\pmratext}{$\mu_{\alpha*}$}

\newcommand{\pmdectext}{$\mu_{\delta}$}
\newcommand{\BooI}{Boo~{\sc I}}
\newcommand{\Sfive}{$S^{5}$}
\newcommand{\gaia}{\textit{Gaia}}
\defcitealias{koposov:2011}{K11}
\newcommand{\citetKoposov}{\citetalias{koposov:2011}}
\defcitealias{jenkins:2021}{J21}
\newcommand{\citetJenkins}{\citetalias{jenkins:2021}}
\defcitealias{longeard:2022}{L22}
\newcommand{\citetLongeard}{\citetalias{longeard:2022}}
\defcitealias{walker:2023}{W23}
\newcommand{\citetWalker}{\citetalias{walker:2023}}
\defcitealias{weinberg:2017}{WAF17}
\newcommand{\citetWAF}{\citetalias{weinberg:2017}}

\submitjournal{ApJ}

\shorttitle{Chemodynamics of Bo\"otes {\sc I} with $S^{5}$}
\shortauthors{Sandford et al.}
\graphicspath{{./}{figures/}}
\begin{document}
\begin{CJK*}{UTF8}{gbsn}
% \title{S$^{5}$: A Comprehensive Chemodynamic Analysis of UFD Bo\"otes {\sc I}}
\title{Chemodynamics of Bo\"otes {\sc I} with $S^{5}$:\\ Revised Velocity Gradient, Dark Matter Density, and Galactic Chemical Evolution Constraints}

\correspondingauthor{Nathan Sandford}
\email{nathan.sandford@utoronto.ca}

\author[0000-0002-7393-3595]{Nathan~R.~Sandford}
\affiliation{Department of Astronomy and Astrophysics, University of Toronto, 50 St. George Street, Toronto ON, M5S 3H4, Canada}

\author[0000-0002-9110-6163]{Ting~S.~Li}
\affiliation{Department of Astronomy and Astrophysics, University of Toronto, 50 St. George Street, Toronto ON, M5S 3H4, Canada}
\affiliation{Dunlap Institute for Astronomy \& Astrophysics, University of Toronto, 50 St George Street, Toronto, ON M5S 3H4, Canada}
\affiliation{Data Sciences Institute, University of Toronto, 17th Floor, Ontario Power Building, 700 University Ave, Toronto, ON M5G 1Z5, Canada}

\author[0000-0003-2644-135X]{Sergey~E.~Koposov}
\affiliation{Institute for Astronomy, University of Edinburgh, Royal Observatory, Blackford Hill, Edinburgh EH9 3HJ, UK}
\affiliation{Institute of Astronomy, University of Cambridge, Madingley Road, Cambridge CB3 0HA, UK}

\author[0000-0002-8758-8139]{Kohei~Hayashi}
\affiliation{National Institute of Technology, Sendai College, 4-16-1 Ayashi-Chuo, Sendai, Japan}
\affiliation{Astronomical Institute, Tohoku University, Aoba-ku, Sendai 980-8578, Japan}
\affiliation{ICRR, The University of Tokyo, Kashiwa, Chiba 277-8582, Japan}

\author[0000-0002-6021-8760]{Andrew~B.~Pace}
\affiliation{Department of Astronomy, University of Virginia, 530 McCormick Road, Charlottesville, VA 22904, USA}

\author[0000-0002-8448-5505]{Denis~Erkal}
\affiliation{Department of Physics, University of Surrey, Guildford GU2 7XH, UK}

\author[0000-0001-6855-442X]{Jo~Bovy}
\affiliation{Department of Astronomy and Astrophysics, University of Toronto, 50 St. George Street, Toronto ON, M5S 3H4, Canada}
\affiliation{Dunlap Institute for Astronomy \& Astrophysics, University of Toronto, 50 St George Street, Toronto, ON M5S 3H4, Canada}

\author[0000-0001-7019-649X]{Gary~S.~Da~Costa}
\affiliation{Research School of Astronomy and Astrophysics, Australian National University, Canberra, ACT 2611, Australia}

\author[0000-0001-8536-0547]{Lara~R.~Cullinane}
\affiliation{Leibniz-Institut f{\"u}r Astrophysik Potsdam (AIP), An der Sternwarte 16, D-14482 Potsdam, Germany}

\author[0000-0002-4863-8842]{Alexander~P.~Ji}
\affiliation{Department of Astronomy \& Astrophysics, University of Chicago, 5640 S Ellis Avenue, Chicago, IL 60637, USA}
\affiliation{Kavli Institute for Cosmological Physics, University of Chicago, Chicago, IL 60637, USA}
\affiliation{NSF-Simons AI Institute for the Sky (SkAI), 172 E. Chestnut St., Chicago, IL 60611, USA}

\author[0000-0003-0120-0808]{Kyler~Kuehn}
\affiliation{Lowell Observatory, 1400 W Mars Hill Rd, Flagstaff,  AZ 86001, USA}

\author[0000-0003-1124-8477]{Daniel~B.~Zucker}
\affiliation{School of Mathematical and Physical Sciences, Macquarie University, Sydney, NSW 2109, Australia}

\author[0000-0002-9269-8287]{Guilherme~Limberg}
\affiliation{Kavli Institute for Cosmological Physics, University of Chicago, Chicago, IL 60637, USA}

\author[0000-0003-0105-9576]{Gustavo~E.~Medina}
\affiliation{Department of Astronomy and Astrophysics, University of Toronto, 50 St. George Street, Toronto ON, M5S 3H4, Canada}
\affiliation{Dunlap Institute for Astronomy \& Astrophysics, University of Toronto, 50 St George Street, Toronto, ON M5S 3H4, Canada}

\author[0000-0002-4733-4994]{Joshua~D.~Simon}
\affiliation{Observatories of the Carnegie Institution for Science, 813 Santa Barbara St., Pasadena, CA 91101, USA}

\author[0000-0001-7609-1947]{Yong~Yang}
\affiliation{Sydney Institute for Astronomy, School of Physics, A28, The University of Sydney, NSW 2006, Australia}

\collaboration{20}{(S$^{5}$ Collaboration)}

%%% ABSTRACT %%%
\begin{abstract}  % (max 250 words)
We combine new spectroscopic observations of the ultra faint dwarf galaxy (UFD) Bo\"otes I (Boo~I) from the Southern Stellar Stream Spectroscopic Survey ($S^{5}$) with $\sim$15 years of archival spectroscopic data to create the largest sample of stellar kinematics and metallicities to date in any Milky Way UFD. 
Our combined sample includes 148 members extending out to $\sim7$~half-light radii ($r_h$), including 24 newly confirmed members, 18 binary candidates, 15 RR Lyrae stars, and 92 [Fe/H] measurements. 
Using this larger and more spatially extended sample, we provide updated constraints on Boo~I's systemic properties, including its radial population gradients. 
Properly accounting for perspective rotation effects in a UFD for the first time, we detect a $4\sigma$ line-of-sight velocity gradient of $1.2\pm0.3$~km~s$^{-1}$~$r_h^{-1}$ aligned along Boo~I's orbit and discuss its potential tidal origins.
We also infer a metallicity gradient of $-0.10\pm0.02$~dex~$r_h^{-1}$ in agreement with previous studies.
Using an axisymmetric Jeans model, we provide updated constraints on Boo~I's dark matter density profile, which weakly favor a cusped ($\gamma=1.0^{+0.5}_{-0.6}$) dark matter profile.
Lastly, we re-analyze Boo~I's metallicity distribution function with a one-zone galactic chemical evolution model and place new constraints on its rapid, inefficient star formation and strong galactic outflows.

%
%Our chemodynamical analysis supports the hypothesis that BooI is experiencing tidal disruption though it cannot rule out scenarios involving a past dry merger event.
\end{abstract}


%%% INTRODUCTION %%%
\section{Introduction}
\label{sec:introduction}
\end{CJK*}

Ultra faint dwarf galaxies (UFDs) are among the oldest ($\gtrsim$13 Gyr), most metal-poor ($\text{[Fe/H]} \lesssim -2.0$), lowest mass ($M_* \lesssim 10^5$ M$_\odot$), and dark matter-dominated ($M/L \gtrsim 100$) galaxies in the local universe \citep[e.g.,][and references therein]{simon:2019a}. As a result, their stellar populations provide exquisite testing sites for a broad range of astrophysical topics including star formation in the early Universe \citep[e.g.,][]{frebel:2015}, the nature of dark matter \citep[e.g.,][]{bullock:2017, battaglia:2022b}, the role of stellar feedback in low-mass galaxy evolution \citep[e.g.,][]{collins:2022a}, and hierarchical galaxy formation \citep[e.g.,][]{frebel:2010}.

In the last two decades, the number of known UFDs around the Milky Way (MW) has dramatically increased, in large part due to wide area imaging campaigns, including the Sloan Digital Sky Survey (SDSS; \citealt{york:2000}; e.g., \citealt{belokurov:2007}), the Dark Energy Survey (DES; \citealt{abbott:2018}; e.g., \citealt{bechtol:2015}), the Dark Energy Camera Legacy
Survey (DECaLS; \citealt{dey:2019}; e.g., \citealt{collins:2022b}), the DECam Local
Volume Exploration survey (DELVE; \citealt{drlica-wagner:2021}; e.g., \citealt{cerny:2023}), and the Ultraviolet Near Infrared Optical Northern Survey (UNIONS; \citealt{ibata:2017}; e.g., \citealt{smith:2023}). However, the majority of these systems are too distant and/or too faint for detailed spectroscopic studies of their stellar populations. UFDs with more than a dozen or so spectroscopic members are incredibly rare, which substantially inhibits analysis of the dynamic and chemical evolution.

As a relatively nearby ($d\sim66$~kpc; \citealt{dallora:2006}) and luminous ($M_V\sim-6$; \citealt{munoz:2018}) UFD, Bo\"otes I (\BooI) has been the subject of enormous observational investment since its early discovery in SDSS \citep{belokurov:2006}. Deep ground- and space-based imaging have revealed an ancient metal-poor stellar population with an extended and modestly eccentric morphology \citep[e.g.,][]{okamoto:2012, brown:2014, roderick:2016, munoz:2018, durbin:2025}.

Meanwhile, spectroscopic observations of \BooI\ \citep[e.g.,][]{munoz:2006b, martin:2007, norris:2008, norris:2010a, norris:2010b, lai:2011, koposov:2011, gilmore:2013, ishigaki:2014, frebel:2016, jenkins:2021, longeard:2022, waller:2023, walker:2023} have measured radial velocities and spectroscopic metallicities for $\sim90$ and $\sim70$ member stars respectively, which confirm \BooI\ as a dark matter-dominated ($\sigma_v=5.1$~\kms; $M/L\sim400$), metal-poor ($\text{[Fe/H]} = -2.3$; $\sigma_\text{[Fe/H]} = 0.3$) galaxy \citep{jenkins:2021}. Spectroscopic samples of this size are extremely rare in UFDs, making \BooI\ a particularly valuable laboratory to test theories of dark matter and galaxy evolution in low-mass galaxies and the early Universe.

\BooI's metallicity distribution function (MDF) as well as the individual elemental abundances of a handful of stars have been the subject of numerous galactic chemical evolution (GCE) studies, seeking to understand the baryonic processes at play during its short-lived episode of star formation \citep[e.g.,][]{lai:2011, gilmore:2013, vincenzo:2014, romano:2015, webster:2015, frebel:2016, romano:2019, lacchin:2020, jenkins:2021, rossi:2024}. Similarly, \BooI's stellar velocity dispersion has been used to infer its underlying dark matter content and constrain alternative dark matter models \citep[e.g.,][]{pace:2019, hayashi:2021b, hayashi:2021a, horigome:2023}.

Combined with \gaia\ proper motions \citep{gaiacollaboration:2022}, \BooI's systemic line-of-sight velocity place its time of infall into the MW as 7--10 Gyr ago \citep[e.g.,][]{fillingham:2019, miyoshi:2020, barmentloo:2023} with its most recent pericentric passage of 30--40 kpc occurring $\sim$300 Myr ago \citep[e.g.,][]{battaglia:2022a, pace:2022}. \BooI's orbit, in conjunction with its elongated morphology, extended stellar population, and internal dynamics suggest that \BooI\ is experiencing tidal disruption (e.g., \citealt{roderick:2016, munoz:2018, longeard:2022, pace:2022}; see also \citealt{munoz:2008}). However, others have argued that similar features could be indicative of a past dry merger event \citep[e.g.,][]{koposov:2011, frebel:2016}.

Searching for additional clues to \BooI's dynamical evolution in the galaxy's outskirts, an increasing number of studies have undertaken searches for member stars at large radii \citep[e.g.,][]{vivas:2020, filion:2020, filion:2022, longeard:2022, waller:2023, jensen:2024, tau:2024, pan:2025}. Despite these observational investments, spectroscopic data in \BooI\ beyond $\sim$3 half-light radii ($r_h$) remains sparse, and our understanding of the galaxy's chemodynamics is largely limited to its central region.

In this paper, we build upon the existing wealth of observational data in \BooI, combining $\sim$15 years of archival spectroscopic data with new data acquired with the Two-degree Field (2dF; \citealt{lewis:2002}) fiber-fed AAOmega spectrograph \citep{sharp:2006} on the Anglo-Australian Telescope (AAT) as a part of the Southern Stellar Stream Spectroscopic Survey (\Sfive; \citealt{li:2019}). This expanded dataset, which extends from the inner half-light radius out to beyond 5 half-light radii and includes over 140 member stars and 90 spectroscopic [Fe/H] measurements, provides valuable new insight into the chemodynamic evolution of this fossil galaxy.

This paper is organized as follows.
In Section \ref{sec:data}, we present the collection and analysis of new \Sfive\ spectroscopic observations, the re-reduction of archival AAT observations, and the additional archival datasets analyzed in this paper. In Section \ref{sec:binary}, we leverage the long time baseline of these observations to identify new binary star systems in \BooI. In Section \ref{sec:GMM}, we present our sample of new and recovered \BooI\ member stars and \BooI\ systemic properties inferred through Gaussian mixture modeling. Additional analysis and discussion of radial population trends in \BooI, including stellar velocity and metallicity gradients, are presented in Section \ref{sec:pop_gradients}. In Section \ref{sec:jeans_modeling}, we present updated constraints on \BooI's dark matter density profile using axisymmetric Jeans modeling, and in Section \ref{sec:GCE}, we provide new constraints on the baryonic evolution of \BooI\ using an analytic galactic chemical evolution model. We summarize and conclude our analysis in Section \ref{sec:summary}.

%%% DATA %%%
\section{Data}
\label{sec:data}

\subsection{$S^5$ AAT Observations}
\label{sec:data_S5}
Observations of \BooI\ were obtained as part of \Sfive\ (see \citealt{li:2019} for survey details), which uses the dual-arm AAOmega spectrograph \citep{sharp:2006} fed by the Two-degree Field \citep[2dF;][]{lewis:2002} fiber positioner facility on the 3.9 meter AAT. The standard \Sfive\ observational setup employs the 580V (3700--5800 \AA; $R\sim1300$) and 1700D (8420--8820 \AA; $R\sim10,000$) AAOmega gratings to combine the metallicity sensitivity of the absorption feature-rich blue optical region with higher resolution calcium triplet (CaT) observations in the red optical for radial velocities.
\BooI\ was observed using three 50-minute exposures of a single AAT pointing on both 2023 March 18 and 19.  %A second AAT pointing offset 2 degrees southwest along the direction of \BooI's orbital motion was attempted on March 22 but was interrupted due to poor observing conditions and is therefore excluded from this study.

When selecting targets in \BooI\, we first consider all stars from \citet{pace:2022} with membership probabilities $\texttt{mem\_fixed} > 0.1$ or $\texttt{mem\_gauss\_nospatial} > 0.1$ as high priority targets\footnote{\texttt{mem\_fixed} is the membership probability when the background model used by \citet{pace:2022} is fixed to the proper motion distribution of stars at large radii, while \texttt{mem\_gauss\_nospatial} is the membership probability when the background model uses a multivariate Gaussian distribution with free proper motion dispersion terms.}. Then additional \BooI\ targets are selected based on the photometry, proper motions, and parallax from \gaia\ Data Release 3 \citep[DR3;][]{gaiacollaboration:2022} following the approach described in \citet{li:2019}. The selection criteria are as follows:
We remove nearby foreground stars by only selecting sources with
\begin{equation}
    \varpi - 3\sigma_\varpi < 0,
\end{equation}
where $\varpi$ and $\sigma_\varpi$ are the parallax measurements and uncertainties from \gaia\ DR3.
A color-magnitude cut is applied by selecting stars within 0.15 magnitude of either a 10 gyr-old metal-rich ($\text{[Fe/H]} = -1.4$) or a 12.5 Gyr-old metal-poor ($\text{[Fe/H]} = -2.2$) Dotter isochrone \citep[][]{dotter:2016}.
Lastly, we require targets have proper motions (\pmratext\ and \pmdectext)\footnote{$\mu_{\alpha*} = \mu_{\alpha}\cos\delta$} satisfying
% Proper Motion Selection
\begin{align}
    |\mu_{\alpha*} - \mu_{\alpha*,\text{B}}| &< \text{max}(3\sigma_{\mu_{\alpha*}}, ~1.0)~\text{mas}~\text{yr}^{-1} \nonumber\\
    |\mu_{\delta} - \mu_{\delta,\text{B}}| &< \text{max}(3\sigma_{\mu_{\delta}}, ~1.0)~\text{mas}~\text{yr}^{-1} , \\
\end{align}
where $\mu_{\alpha*,\text{B}}$ and $\mu_{\delta,\text{B}}$ are the proper motions of \BooI\ as reported by \citet{pace:2022}, and $\sigma_{\mu_{\alpha*}}$ and $\sigma_{\mu_{\delta}}$ are the individual measurement uncertainties from \gaia. 

The above selection criteria yield a sample of $\sim$500 targets in the AAT field, which is used as input to the \texttt{configure} fiber allocation software \citep{miszalski:2006}. %Approximately 335 fibers are assigned to the \BooI\ targets along with 25 sky fibers and 8 guide star fibers.
In total, we observed 320 targets in \BooI, of which 270 are classified as stars with good measurements by our reduction pipeline (i.e., $\texttt{good\_star} = \text{True}$; \citealt{li:2019}). Of these, 139 have $\text{signal-to-noise (S/N)} > 3$ \perpix\ in the red arm and are used in our analysis.

\subsection{Archival AAT Observations}
\label{sec:data_AATArx}
Though the focus of this work is on new \Sfive\ observations, we also consider earlier AAT observations of \BooI\ that are publicly available on the AAT archive. As summarized in Table \ref{tab:observations}, this includes ten 15--40 minute exposures taken over 3 nights in May 2006, 25 30-minute exposures taken over 4 nights in April 2007, and six 45-minute exposures taken over 3 nights in June 2020. The archival observations collected in 2006 and 2007 were first analyzed in \citet{norris:2010b} and utilized the 1700B (3700--4500 \AA; $R\sim3500$) and 1700D gratings, while the 2020 observations were first analyzed in \citet[][hereafter \citetLongeard]{longeard:2022} and used the 580V and 1700D gratings (the same as \Sfive).

After applying the same selection cuts adopted for the \Sfive\ observations (see Section \ref{sec:data_S5}), we are left with archival AAT spectra for 233 unique targets, of which 202 are classified as stars by our reduction pipeline. Of these, 90 have $\text{S/N} > 3$ \perpix, including 70 that overlap with the \Sfive\ sample.

\begin{deluxetable*}{lcccl}
	\centerwidetable
	\caption{AAT/AAOmega-2dF Observations in \BooI}
	\label{tab:observations}
	\tablehead{
	    \colhead{Field} &  
            \colhead{Gratings} & 
	    \colhead{Date}  &         
            \colhead{Exposures} &         
            \colhead{Reference}
	}
	\startdata
	    BOO-AAT & 1700B, 1700D & 2006-05-23 & 1000s, $3\times1800$s & \citet{norris:2010b} \\
	    BOO-AAT & 1700B, 1700D & 2006-05-25 & $3\times2000$s & \citet{norris:2010b}  \\
	    BOO-AAT & 1700B, 1700D & 2006-05-29 & $3\times2500$s & \citet{norris:2010b}  \\
	    %BOO-AAT plate 0 & 1700B, 1700D & 2006-05-29 & 2500s & \citet{norris:2010}  \\
	    BOO-AAT-1 & 1700B, 1700D & 2007-04-18 & $7\times1800$s & \citet{norris:2010b}  \\
	    BOO-AAT-1 & 1700B, 1700D & 2007-04-19 & $6\times1800$s & \citet{norris:2010b}  \\
	    BOO-AAT-1 & 1700B, 1700D & 2007-04-20 & $6\times1800$s & \citet{norris:2010b}  \\
	    BOO-AAT-1 & 1700B, 1700D & 2007-04-22 & $6\times1800$s & \citet{norris:2010b}  \\
	    Field210.0214.5135 & 580V, 1700D & 2020-06-16 & $2\times2700$s & \citet{longeard:2022}  \\
	    Field210.0214.5135 & 580V, 1700D & 2020-06-17 & $2\times2700$s & \citet{longeard:2022}  \\
	    Field210.0214.5135 & 580V, 1700D & 2020-06-19 & $2\times2700$s & \citet{longeard:2022}  \\
	    BootesI-field-0 & 580V, 1700D & 2023-03-18 & $3\times3000$s & This work  \\
	    BootesI-field-0 & 580V, 1700D & 2023-03-19 & $3\times3000$s & This work  \\
	\enddata
\end{deluxetable*}

\subsection{AAT Data Reduction and Analysis}
\label{sec:data_rdx}
The \Sfive\ observations considered in this paper are reduced and analyzed as a part of the \Sfive\ internal data release iDR3.7, which is scheduled to be the second \Sfive\ public data release (DR2; T.S.\ Li et al.\ in prep.). As previously described in \citet{ji:2021} and \citet{li:2022}, this data release features several improvements to the spectral fitting procedure of the first public data release (DR1; \citealt{li:2019}). Here, we briefly summarize the most substantial improvements to the pipeline below and refer the reader to the upcoming DR2 paper for a more thorough description of the pipeline and its validation.

As in DR1, the data is analyzed using the \texttt{rvspecfit} pipeline \citep{koposov:2019}, which forward-models the stellar spectrum using the PHOENIX-2.0 stellar spectra library \citep{husser:2013} to determine the radial velocity and stellar parameters ($T_\text{eff}$, $\log g$, and [Fe/H]) of each star. The first notable improvement in this pipeline over DR1 is that we now simultaneously model both the red and blue spectral arms of AAOmega as well as observations taken across multiple nights (with appropriate consideration for the heliocentric corrections). This change substantially improves the recovery of radial velocity measurements for stars observed across multiple nights, especially those near the $\text{S/N} \sim 3$ pixel$^{-1}$ threshold. Another major improvement in this data release is the use of a neural-network interpolator to interpolate the PHOENIX model spectra grid. This substantially alleviates the ``gridding" issues of previous releases where the measured parameters tended to cluster around grid points in $T_\text{eff}$, $\log g$, and [Fe/H].
%\elaborate{Are there any other improvements or details worth mentioning?}

We reduce and analyze the archival AAT observations (see Section \ref{sec:data_AATArx}) similarly to the \Sfive\ observations. This includes applying the measurement calibrations previously determined in \citet{li:2019} of:
\begin{equation}
v_\text{cal} = v_\text{raw} - 1.11~\text{\kms},
\end{equation}
\begin{equation}
\sigma_{v,\text{cal}} = \left[(1.28\sigma_{v,\text{raw}})^{2} + 0.66^{2}\right]^{1/2},~\text{and}
\end{equation}
\begin{equation}
\sigma_{\text{[Fe/H],cal}} = 1.3\sigma_{\text{[Fe/H],raw}}, 
\end{equation}
where the ``raw" and ``cal" subscripts refer to the measurements and measurement uncertainties reported directly by \texttt{rvspecfit} and those calibrated by analysis of repeat observations, respectively. 

For the individual epoch measurements used to identify binary stars in Section \ref{sec:binary}, we perform a similar reduction except that we only simultaneously model the spectra that were acquired in the same AAT ``field" (see Table \ref{tab:observations}). Because observations in each field were all acquired within a few days, this approach should be suitable for long-period binaries but may lead to unexpected results for binaries with periods on the order of a week or less.

\subsection{Literature Datasets}
\label{sec:data_lit}
To supplement our observations further, we also consider published datasets from \citet[][hereafter \citetJenkins]{jenkins:2021} and \citet[][hereafter \citetWalker]{walker:2023} respectively. The dataset from \citetJenkins\ includes [Fe/H] and radial velocity measurements for $\sim$70 likely \BooI\ member stars, which was constructed from a re-analysis of multi-epoch FLAMES/GIRAFFE spectroscopy originally observed by \citet[][hereafter \citetKoposov]{koposov:2011} on the Very Large Telescope (VLT). Because of smaller field-of-view of the FLAMES fiber positioner, this sample is nearly entirely located within \BooI's inner half-light radius.
The dataset from \citetWalker\ includes [Fe/H] and radial velocity measurements for $\sim$100 red giant branch stars within $\sim$0.5 degrees ($\sim3r_h$) of \BooI\ from multi-epoch Hectochelle spectroscopy on the MMT. Because \citetWalker\ does not assign membership probabilities to the stars in their catalog, we identify likely members in this dataset following the same methodology that we adopt for the \Sfive\ and archival AAT datasets.% (see Appendix \ref{app:archival}).

\subsection{Creating a Combined Dataset}
\label{sec:data_comb}
Throughout this work, we focus our analysis primarily on two datasets: an \Sfive-only dataset, which contains only new \Sfive\ AAT observations taken in 2023, and a ``combined" dataset, which includes re-reduced archival AAT observations, archival MMT measurements from \citetWalker, and archival VLT measurements from \citetJenkins. We limit the combined dataset to high-probability \BooI\ member stars as determined using a Gaussian mixture model applied to \Sfive\ data (see Section \ref{sec:GMM}) and archival AAT and MMT data (see Appendix \ref{app:archival}). Member stars for the archival VLT data are taken directly from \citetJenkins. For stars appearing in multiple datasets, we combine velocity and metallicity measurements and uncertainties using a weighted mean after applying a zero-point correction to each dataset (see Section \ref{sec:binary}). %The composition of the combined dataset is summarized in Table \ref{tab:combined_data_sum}.

\subsection{Simulation Dataset}
\label{sec:data_sim}
To aid in the dynamical interpretation of our observational data described above, we also consider a mock sample of stars generated from an $N$-body simulation of \BooI. To generate this dataset, we use \textsc{gadget-3}, an improved version of the \textsc{gadget-2} $N$-body simulation code \citep{springel:2005}, to simulate the dynamical evolution of a \BooI-mass galaxy on a \BooI-like orbit through a Milky Way-like gravitational potential.

We initialize the phase-space coordinate of the simulated \BooI\ by rewinding a tracer particle 3.3~Gyr from \BooI's present-day location through the combined potential of the MW and the Large Magellanic Cloud (LMC). We choose this time since this corresponds to \BooI's apocenter and because it experiences two pericenters between this time and the present day, allowing for it to be tidally stripped. We model the MW using the \texttt{MWPotential2014} potential from \citet{bovy:2015}. For the LMC, we use a Hernquist profile \citep{hernquist:1990} with a mass of $1.5\times10^{11} \text{M}_\odot$ and a scale radius of 17.13 kpc so that the enclosed mass of the LMC within 8.7 kpc matches the observed value of $1.7\times10^{10}\text{M}_\odot$ \citep{vandermarel:2014}. The total LMC mass is based on the results of previous fits of the LMC mass with stellar streams \citep[][]{erkal:2019,shipp:2021,vasiliev:2021a,koposov:2023}. For the present-day phase-space coordinates of the LMC, we use measurements of its proper motion, distance, and radial velocity from \cite{kallivayalil:2013,pietrzynski:2019,vandermarel:2002}, respectively. We also include the effect of dynamical friction on the LMC from the MW using the prescription in \cite{jethwa:2016}.

We model \BooI\ using a Plummer sphere \citep{plummer:1911} embedded in an NFW profile using \textsc{agama} \citep{vasiliev:2019}. For the Plummer sphere, we use a mass of $4.81\times10^3~\text{M}_\odot$, a scale radius of 0.19~kpc, and $10^6$ particles, and a softening of 10.1~pc. For the NFW, we use a mass of $10^8 ~\text{M}_\odot$, a concentration of $c_{\rm 200} = 21.2$ from \cite{dutton:2014}, and a Hubble value of $H_0 = 67.9 \, \rm{km}\,\rm{s}^{-1}\rm{Mpc}^{-1}$ \citep[consistent with observations,][]{planck:2020} to set the present-day critical density of the Universe. To avoid artificially extended profiles, we truncate both profiles with an exponential cutoff in density, which is implemented in \texttt{agama} by setting \texttt{outerCutoffRadius} to ten times the scale radius (1.9 kpc) for the Plummer sphere and to the virial radius (9.77 kpc) for the NFW profile \citep{vasiliev:2019}. After evolving the system to the present day, the simulated galaxy is only slightly offset from the correct present-day location of \BooI\ by 0.02~kpc in heliocentric distance, 0.7$^{\circ}$ (0.81~kpc) in the plane of the sky, and 3.2~\kms\ in velocity.

In order to be able to easily compare to our observed data, we downsample the $N$-body simulation to approximately match the number and radial distribution of \BooI\ member stars identified in the \Sfive\ dataset (see Section \ref{sec:GMM_members}). We do this by splitting the sample of simulated stars into 4 elliptical annuli: $r/r_h < 1$, $1\leq r/r_h < 3$, $3\leq r/r_h < 5$, and $5\leq r/r_h < 8$. In each annulus we randomly draw the number of samples equal to the number of \Sfive\ member stars within that annulus. We repeat this sampling 10 times and perform our dynamical analysis on each. 

\begin{figure}[ht!] 
    \centering
	\includegraphics[width=0.5\textwidth]{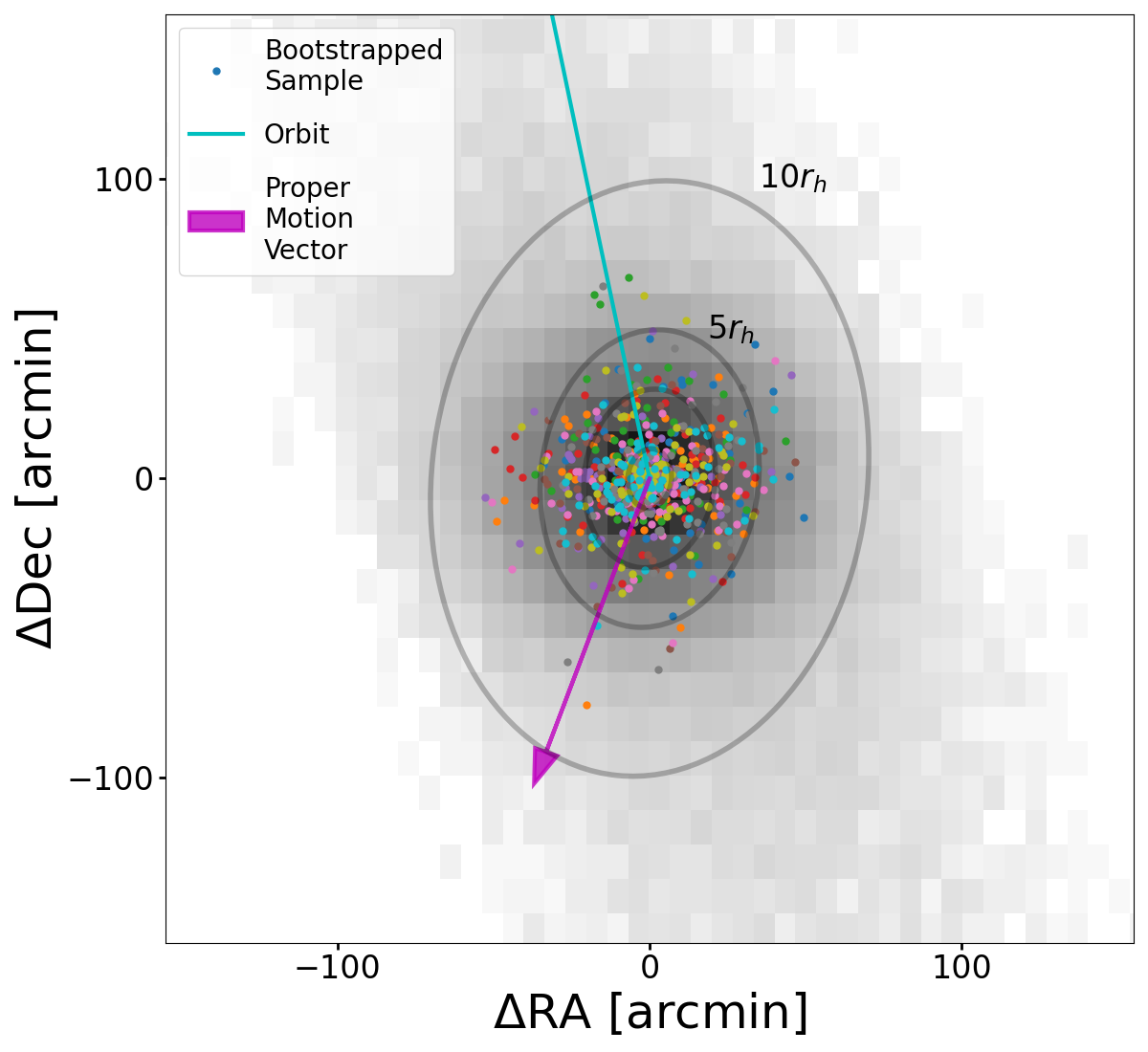}
    \caption{
        The log-number density of the \BooI\ $N$-body simulation. The simulated orbit and apparent proper motion (without Solar reflex correction) of the system are represented by the cyan line and magenta arrow respectively. Ten bootstrapped mock samples are included as colored points. 
        Gray ellipses indicate the observed 1, 3, 5, and 10 half-light radii of \BooI\ from \citet{munoz:2018}. %\textbf{Right.} Median line-of-sight velocity of $N$-body particles corrected for perspective rotation effects.
    } 
    \label{fig:nbody}
\end{figure}

The full results of the $N$-body simulation are illustrated in Figure \ref{fig:nbody}. Extended tidal features are clearly visible along the galaxy's orbit. The orientation of these tidal features is similar, though not identical, to the observed position angle of \BooI, which is just slightly misaligned eastward of the galaxy's orbit. The 10 bootstrapped samples are included as color points in Figure \ref{fig:nbody}.

\section{Binary Identification}
\label{sec:binary}
The presence of binary stars in a galaxy, if unaccounted for, can inflate the observed velocity dispersion through their additional orbital motion. This would make a galaxy appear more massive than it truly is and complicate other dynamical analyses. Previous studies by \citetKoposov\ and \citetJenkins\ identified several binaries in \BooI\ using multi-epoch VLT spectroscopy. However, the time baseline of these observations was only $\sim$1 month, limiting the sensitivity of the binary search to systems with $\lesssim$ 1 year periods.

We extend this binary search to longer periods by combining our new \Sfive\ observations with archival AAT observations (see Section \ref{sec:data_AATArx}) and previously published datasets from VLT and MMT observations (see Section \ref{sec:data_lit}). 
This combined dataset contains measurements for $\sim$600 stars over a 16-year time span. Roughly one third of these stars were observed more than once, and $\sim$5\% were observed between five and 9 times. Fifteen stars previously identified as RR Lyrae (RRL) stars in \citet{clementini:2023} are excluded from this analysis.

Similar to previous binary searches \citep[e.g.,][]{spencer:2017b, spencer:2018}, we perform null-hypothesis significance testing for each star in our sample with multiple velocity epochs. Here the null hypothesis is that each star has constant velocity and the reduced 
chi-squared statistic is given by
\begin{equation}
\chi^{2}_{\kappa} = \frac{1}{\kappa}\sum_i^n\left(\frac{v_i - \langle v\rangle}{\sigma_i}\right)^{2},
\end{equation}
where $\kappa=n-1$ is the number of degrees of freedom, $n$ is the number of observations, $v_i$ and $\sigma_i$ are the individual epoch velocity and velocity uncertainty respectively, and $\langle v\rangle$ is the velocity averaged across all epochs. If the probability, $P(\chi^2, \kappa)$, of exceeding this $\chi^2_\kappa$ for a star given the null hypothesis (i.e., the $p$-value) is very small ($\lesssim$0.01), then we can confidently reject the null hypothesis of constant velocity and declare the star a likely binary.

We proceed with binary identification iteratively, first finding all stars with at least weak evidence of variability, $P(\chi^2, \kappa) < 0.1$. These are then temporarily removed from the sample in order to determine the relative zero-point velocity offsets between the datasets, which are calculated using the weighted average velocity differences of shared non-variable stars. We then re-calculate the $p$-values using the complete sample with applied velocity offsets, and re-determine the zero-point offsets. This process is repeated until the number of candidate binaries and the velocity offsets converge. Metallicity offsets for each archival dataset relative to \Sfive\ are determined analogously from the [Fe/H] measurements of shared stars. The zero-points for each dataset and the number of stars used to determine each offset are summarized in Table \ref{tab:offsets}. The final $p$-value and binary classification for each star is included in Table \ref{tab:stars}.

\begin{deluxetable}{lccc}
	\centerwidetable
	\caption{Offsets Between Datasets}
	\label{tab:offsets}
	\tablehead{
	    \colhead{Dataset} &  
            \colhead{$\Delta v$ (km s$^{-1}$)} &   
            \colhead{$\Delta\text{[Fe/H]}$} &         
            \colhead{$N_*$}
	}
	\startdata
	    AAT (BOO-AAT)            & \phantom{$+$}$0.00$  & $-0.12$             & 10 \\
            AAT (BOO-AAT-1)          & $-0.09$              & \phantom{$+$}$0.02$ & 30 \\
            AAT (Field210.0214.5135) & \phantom{$+$}$0.00$  & $-0.19$             & 20 \\
            VLT                      & \phantom{$+$}$0.14$  & \phantom{$+$}$0.00$ & 10 \\
            MMT                      & $-0.26$              & $-0.28$             & 41 \\
	\enddata
	\tablecomments{
	    Velocity and metallicity offsets ($\Delta v$ and $\Delta\text{[Fe/H]}$) of each dataset relative to \Sfive\ as well as the number of overlapping stars used in the offset calculations, $N_*$. Potential binary stars and RR Lyrae were excluded when determining the offsets.
	}
\end{deluxetable}

\begin{figure*}[ht!] 
	\includegraphics[width=1.0\textwidth]{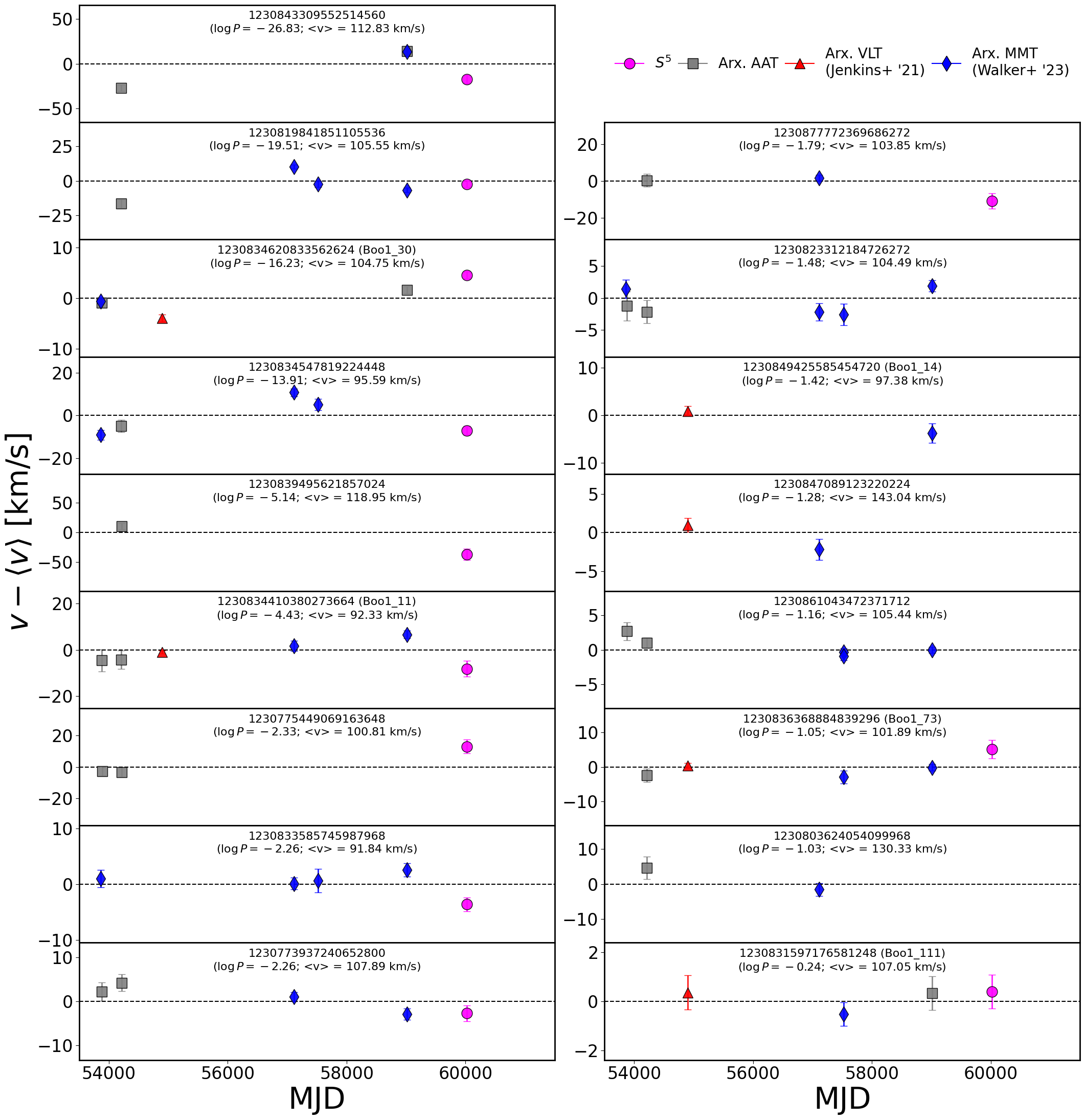}
    \caption{
        Difference between individual epoch velocity measurements and the mean velocity as a function of time for each of the potential \BooI\ binary stars in our combined sample. The stars are ordered by decreasing $p$-value. Star 1230831597176581248 (Boo1\_111) is included despite its low binary probability because there is weak evidence of binarity from individual epoch VLT \citep{koposov:2011, jenkins:2021} and MIKE observations (J.\ Simon, private communication).
    } 
    \label{fig:binaries}
\end{figure*}

In Figure \ref{fig:binaries}, we show the velocity curves for the 16 sources in our sample, with $P(\chi^2, \kappa) < 0.1$ and $\langle v\rangle$ within $\pm$50 km s$^{-1}$ of \BooI's systemic velocity. Nine of these sources show strong evidence of velocity variability, $P(\chi^2, \kappa) \leq 0.01$, while seven of these sources show only weak evidence, $0.01 < P(\chi^2, \kappa) < 0.1$.
Of the 9 sources with strong evidence for binarity, 8 are newly identified. The ninth, \gaia\ ID 1230834620833562624\footnote{Source Boo1\_30 in \citetKoposov\ and \citetJenkins.}, was first identified as a potential binary using individual epoch VLT data over a 4-month baseline by \citetKoposov\ but was later discounted as such in a reanalysis of the data by \citetJenkins. We believe that this source is a long-period binary, which is why its detection in VLT data is uncertain. This hypothesis is corroborated by preliminary analysis of recent MIKE observations (J.\ Simon, private communication). We attempt to fit the orbits of these likely binaries using \texttt{the Joker}\footnote{\url{https://thejoker.readthedocs.io/en/latest/index.html}} \citep{price-whelan:2017} but are unable to place meaningful constraints on any orbital parameters due to the sparse sampling of the velocity curves.

The seven sources with weak evidence for velocity variability have, on average, fewer and/or less precise measurements. For the three stars with VLT observations, \gaia\ sources 1230849425585454720, 1230847089123220224, and 1230836368884839296\footnote{Sources Boo1\_14, Boo1\_20, and Boo1\_73 respectively in \citetKoposov\ and \citetJenkins.}, both \citetKoposov\ and \citetJenkins\ previously found no evidence of variability. It is possible that these sources are also in long-period binaries, but additional observations are required to confirm or reject this. We also note that 1230847089123220224 has a mean velocity of $\sim$143 km s$^{-1}$, which compared with \BooI's systemic velocity of 103 km~$s^{-1}$ \citep{jenkins:2021,longeard:2022} makes its association with \BooI\ somewhat tenuous.

In Figure \ref{fig:binaries}, we also include the velocity curve of \gaia\ source 1230831597176581248\footnote{Source Boo1\_111 in \citetKoposov\ and \citetJenkins.}, which was identified as a weak binary candidate based on VLT data in \citetKoposov\ and \citetJenkins. Preliminary analysis of recent MIKE observations also indicates weak evidence for binarity (J.\ Simon, private communication). However, we do not find any evidence for velocity variability based on our combined dataset. Additional epochs of high-precision velocity measurements will prove useful in addressing this disagreement. Erring on the side of caution, we treat this star as a binary in our dynamical analysis of \BooI.

Not shown in Figure \ref{fig:binaries} are three binary candidates identified by \citetJenkins: \gaia\ IDs 1230853376955406848, 1230833860623898496, and 1230783626686364672\footnote{Sources Boo1\_26, Boo1\_61, and Boo1\_114 respectively in \citetKoposov\ and \citetJenkins.}. While these three sources were observed with AAT, the acquired spectra do not meet our $\text{S/N} > 3$ \perpix\ quality cut, and thus do not appear in this analysis.

% In Figure \ref{fig:binaries_odd}, we show the velocity curves for two stars of interest in our sample: \gaia\ sources 1230834410380273664\footnote{Source Boo1\_11 in \citetKoposov\ and \citetJenkins.} and 1230833585745987968. While neither of these stars pass the threshold of $P(\chi^2, \kappa) < 0.1$ used to identify radial velocity variability, both exhibit a noticeably lower heliocentric velocity in the \Sfive\ data compared to the archival epochs---particularly those of \citetWalker. It is possible that these two stars are in binary systems that escaped detection because of the sparse cadence of observations, but it is also possible that this difference in measured velocities is an artifact of either the archival or \Sfive\ data. Without additional spectroscopic follow-up, it is difficult to say which of these scenarios is more likely. Thus, in our analysis of \BooI\ that follows, we consider both the default case that these stars are not binaries as well as the alternative case where they are binaries. The impact of their inclusion and exclusion from our kinematic and dynamical analysis is discussed further in Sections \ref{sec:GMM_global}, \ref{sec:vel_disp_prof}, and \ref{sec:jeans_modeling}. 

% \begin{figure}[ht!] 
% 	\includegraphics[width=0.45\textwidth]{figures/binaries_odd.png}
%     \caption{
%         Same as Figure \ref{fig:binaries} except for stars 1230834410380273664 and 1230833585745987968. While these two stars do not pass our binary candidate threshold ($P<0.1$), their \Sfive\ radial velocity measurements are $\sim$8 and 3 km s $^{-1}$ smaller than archival measurements. We designate these two stars as ``ambiguous binaries" and repeat our analysis both with and without these stars treated as binaries.
%     } 
%     \label{fig:binaries_odd}
% \end{figure}

\section{Membership Modeling and Systemic Properties}
\label{sec:GMM}
In this section, we present the identification of \BooI\ member stars in our \Sfive\ sample and the inference of \BooI's systemic chemodynamic properties. We describe the final quality cuts made to our \Sfive\ dataset in Section \ref{sec:GMM_data} and introduce the adopted Gaussian mixture model (GMM) and sampling techniques in Section \ref{sec:GMM_like}. We present the results of our analysis, including likely members and systemic properties in Sections \ref{sec:GMM_members} and \ref{sec:GMM_global}.
In Appendix \ref{app:archival}, we apply the same fitting techniques to the archival AAT and MMT datasets individually and compare those results to the fiducial \Sfive-only and combined results.

\subsection{Data and Quality Cuts}
\label{sec:GMM_data}
For the purposes of this analysis, our dataset includes spectroscopic measurements of stellar metallicities, heliocentric line-of-sight (l.o.s.) velocities, and their respective uncertainties ([Fe/H], $\sigma_\text{[Fe/H]}$, $v$, $\sigma_{v}$) as well as proper motions, their uncertainties, and their correlation coefficients from \gaia\ DR3 ($\mu_{\alpha*}$, $\mu_{\delta}$, $\sigma_{\mu_{\alpha*}}$, $\sigma_{\mu_{\delta}}$, $C_{\mu_{\alpha*},\mu_{\delta}}$). We also consider the stars' distance from the center of \BooI, $r$, and the stars' positions along and perpendicular to the axis of \BooI's systemic proper motion, $\phi_{1*}$ and $\phi_{2*}$ (not to be confused with their positions along the major and minor axes, $\phi_{1}$ and $\phi_{2}$.) The structural parameters adopted for these coordinate transformations are summarized together with the inferred parameters from the combined dataset in Table \ref{tab:BooI_par}.

\begin{deluxetable*}{lcll}
	\centerwidetable
	\caption{\BooI\ Systemic Properties}
	\label{tab:BooI_par}
	\tablehead{
	    \colhead{Parameter} &  
            \colhead{Value} & 
            \colhead{Description} & 
	    \colhead{Reference}
	}
	\startdata
        \multicolumn{4}{c}{Literature Properties} \\
        \hline
	    $\alpha$ [deg] & 210.02 & \BooI\ center RA & \citet{munoz:2018} \\
	    $\delta$ [deg] & 14.51 & \BooI\ center Dec & \citet{munoz:2018} \\
	    $\mu$ [mag] & $19.11 \pm 0.08$ & Distance modulus & \citet{dallora:2006} \\
	      $d$ [kpc] & $66.4^{+2.5}_{-2.4}$ & Distance & \citet{dallora:2006} \\
	    $r_h$ [arcmin] & $9.97 \pm 0.27$ & Circularized half-light Radius & \citet{munoz:2018} \\
        $b_*$ [pc] & $191 \pm 8$ & Physical Half-light Radius & \citet{munoz:2018} \\
	      $e=(1- b/a)$ & $0.30 \pm 0.03$ & Ellipticity & \citet{munoz:2018}\\
	    $\theta_\text{PA}$ [deg] & $6 \pm 3$ & Position Angle (East of North) & \citet{munoz:2018} \\
        \hline
        % \multicolumn{4}{c}{\Sfive\ GMM Inferred Properties} \\
        \multicolumn{4}{c}{Inferred Properties} \\
        \hline
		$v_{0,\mathrm{B}}$ [km s$^{-1}$] & $103.0^{+0.4}_{-0.4}$ & \BooI\ central heliocentric l.o.s.\ velocity & Section \ref{sec:GMM_global} \\
		$\Delta v_{1,\mathrm{B}}$ [km s$^{-1}$ arcmin$^{-1}$] & $0.02^{+0.03}_{-0.03}$ & Heliocentric l.o.s.\ velocity gradient along $\phi_{1*}$ & Sections \ref{sec:GMM_global}, \ref{sec:vel_grad} \\
		$\Delta v_{2,\mathrm{B}}$ [km s$^{-1}$ arcmin$^{-1}$] & $-0.08^{+0.05}_{-0.05}$ & Heliocentric l.o.s.\ velocity gradient along $\phi_{2*}$ & Sections \ref{sec:GMM_global}, \ref{sec:vel_grad} \\
		$\sigma_{v,\mathrm{B}}$ [km s$^{-1}$] & $4.0^{+0.4}_{-0.3}$ & Heliocentric l.o.s.\ velocity dispersion & Sections \ref{sec:GMM_global}, \ref{sec:vel_disp_prof} \\
		$\mathrm{[Fe/H]}_{0,\mathrm{B}}$ [dex] & $-2.28^{+0.05}_{-0.05}$ & \BooI\ central metallicity & Section \ref{sec:GMM_global} \\
		$\Delta\mathrm{[Fe/H]}_{\mathrm{B}}$ [dex arcmin$^{-1}$] & $-0.010^{+0.003}_{-0.003}$ & Radial metallicity gradient & Sections \ref{sec:GMM_global}, \ref{sec:feh_grad} \\
		$\sigma_{\mathrm{[Fe/H],B}}$ [dex] & $0.28^{+0.03}_{-0.03}$ & Metallicity dispersion & Sections \ref{sec:GMM_global}, \ref{sec:feh_grad} \\
		$\langle\mu_{\alpha,*}\rangle_\mathrm{B}$ [mas yr$^{-1}$] & $-0.38^{+0.02}_{-0.02}$ & \BooI\ heliocentric proper motion (RA $\cos\delta$) & Section \ref{sec:GMM_global} \\
		$\langle\mu_\delta\rangle_\mathrm{B}$ [mas yr$^{-1}$] & $-1.08^{+0.01}_{-0.01}$ & \BooI\ heliocentric proper motion (Dec) & Section \ref{sec:GMM_global} \\
		$C_{\langle\mu_{\alpha,*}\rangle,{\langle\mu_{\delta}}\rangle}$ & $-0.08$ & Correlation between inferred $\langle\mu_{\alpha,*}\rangle_\mathrm{B}$ and $\langle\mu_\delta\rangle_\mathrm{B}$ & Section \ref{sec:GMM_global} \\
        \hline
        \multicolumn{4}{c}{Additional Derived Properties} \\
        \hline
		$N_{\text{mem},v}$ & 115 & Number of member stars with good velocities & Section \ref{sec:GMM_members} \\
		$N_\text{mem,[Fe/H]}$ & 92 & Number of member stars with good [Fe/H] & Section \ref{sec:GMM_members} \\
		$\langle v_\mathrm{B}\rangle$ [km s$^{-1}$] & $103.0^{+0.4}_{-0.4}$ & \BooI\ mean heliocentric l.o.s.\ velocity & Section \ref{sec:GMM_global} \\
		$\langle \text{[Fe/H]}_\mathrm{B}\rangle$ [dex] & $-2.43^{+0.04}_{-0.04}$ & \BooI\ mean metallicity & Section \ref{sec:GMM_global} \\
		$\Delta v_{\mathrm{B}}$ [km s$^{-1}$ arcmin$^{-1}$]  & $0.09^{+0.05}_{-0.05}$ & Magnitude of observed velocity gradient  & Section \ref{sec:vel_grad}  \\
		$\Delta v_{1,\mathrm{B}}'$ [km s$^{-1}$ arcmin$^{-1}$]  & $-0.09^{+0.03}_{-0.03}$ & Intrinsic l.o.s.\  velocity gradient along $\phi_{1*}$ & Section \ref{sec:vel_grad} \\
		$\Delta v_{2,\mathrm{B}}'$ [km s$^{-1}$ arcmin$^{-1}$]  & $-0.08^{+0.05}_{-0.05}$ & Intrinsic l.o.s.\  velocity gradient along $\phi_{2*}$  & Section \ref{sec:vel_grad} \\
		$\Delta v_{\mathrm{B}}'$ [km s$^{-1}$ arcmin$^{-1}$]  & $0.12^{+0.04}_{-0.03}$ & Magnitude of intrinsic velocity gradient  & Section \ref{sec:vel_grad} \\
        \hline
        \multicolumn{4}{c}{Dark Matter Properties} \\
        \hline
		$Q$ & $1.18^{+0.51}_{-0.59}$ & Axial ratio of dark matter halo & Section \ref{sec:jeans_modeling} \\
		$\log_{10}b_\mathrm{halo}$ [pc] & $3.07^{+0.87}_{-0.45}$ & Scale radius of dark matter halo & Section \ref{sec:jeans_modeling} \\
		$\log_{10}\rho_{0}$ [$M_\odot$ pc$^{-3}$] & $-1.88^{+0.90}_{-1.51}$ & Central density of dark matter halo & Section \ref{sec:jeans_modeling} \\
		$-\log_{10}(1-\beta_z)$ & $0.25^{+0.21}_{-0.21}$ & Velocity anisotropy parameter & Section \ref{sec:jeans_modeling} \\
		$\alpha$ & $1.76^{+0.82}_{-0.81}$ & Sharpness of dark matter slope transition & Section \ref{sec:jeans_modeling} \\
		$\beta$ & $6.39^{+2.20}_{-2.12}$ & Outer slope of dark matter halo & Section \ref{sec:jeans_modeling} \\
		$\gamma$ & $1.00^{+0.52}_{-0.60}$ & Inner slope of dark matter halo & Section \ref{sec:jeans_modeling} \\
		$i$ [deg] & $70.47^{+12.83}_{-13.23}$ & Inclination angle & Section \ref{sec:jeans_modeling} \\
            %$(M_\text{dyn}/L)_{r_\text{half}}$ [$M_\odot/L_\odot$] & & Dynamical mass-to-light ratio & Section \ref{sec:jeans_modeling} \\
            %$\log\rho_\text{DM}(150~\text{pc})$ [$M_\odot~\text{kpc}^{-3}$]& & Dark matter density at 150 pc & Section \ref{sec:jeans_modeling} \\
        \hline
        \multicolumn{4}{c}{GCE Properties} \\
        \hline
		$\tau_\text{SFH}$ [Gyr] & $0.2^{+0.1}_{-0.1}$ & Star formation history timescale & Section \ref{sec:GCE} \\
		$\tau_\text{SFE}$ [Gyr] & $13.9^{+7.6}_{-5.3}$ & Star formation efficiency timescale  & Section \ref{sec:GCE} \\
		$t_\text{trunc}$ [Gyr] & $>0.5$ & Time of star formation truncation & Section \ref{sec:GCE} \\
		$\eta$ & $203^{+27}_{-36}$ & Mass-loading factor & Section \ref{sec:GCE} \\
	\enddata
	\tablecomments{
	Selected properties of \BooI\ from the literature and inferred from our analysis of the combined dataset. Analogous results from the individual datasets are presented in Table \ref{tab:BooI_par_extra}. The ``intrinsic" velocity gradients are the velocity gradients corrected for perspective rotation effects (see Section \ref{sec:vel_grad}). For a discussion of the difference between central and mean systemic velocities and metallicities, see Section \ref{sec:GMM_global}.
	}
\end{deluxetable*}

In addition to requiring all stars to have 1700D spectra with median $\text{S/N} > 3$ pixel$^{-1}$ and valid \gaia\ proper motions, we also require all stars to have $|v|<600$~\kms. We also require that the \gaia\ renormalized unit weight error (RUWE) and the \gaia\ astrometric excess noise satisfy the conditions $\texttt{ruwe} < 1.4$ and $\texttt{astrometric\_excess\_noise} < 2$ respectively.

As previously implemented in \Sfive\ studies \citep[e.g.,][]{ji:2021}, we require $\sigma_v<10$~\kms\ and $\sigma_\FeH<0.5$~dex as larger uncertainties are frequently indicative of spectral fits impacted by large sky subtraction residuals. Instead of removing these stars entirely, we inflate both their metallicity and velocity uncertainties by an arbitrarily large amount in order to effectively mask out these measurements while still considering the star's proper motion.

Lastly, we account for binary, horizontal branch (HB), and RRL stars in our sample. Binary candidates identified in Section \ref{sec:binary} as well as any stars identified as RRL in \citet{clementini:2023} have their velocity uncertainties increased by an arbitrarily large amount so that they do not artificially inflate the inferred velocity dispersion of \BooI. Similarly, we inflate the metallicity uncertainties of all RRL and HB stars in our sample because the metallicities measured by \texttt{RVSpecFit} are less reliable for hotter and variable stars. As a result of this treatment, HB stars contribute only to the proper motion and line-of-sight velocity determination of \BooI, and RRL only contribute to the proper motion determination.

\subsection{Gaussian Mixture Model}
\label{sec:GMM_like}
We determine stellar membership and systemic properties of \BooI\ using a two-component GMM with one component representing \BooI's stellar population and one component representing MW foreground contaminants. GMMs have been applied extensively over the last $\sim$15 years to numerous stellar streams, dwarf galaxies, and other resolved stellar systems \citep[e.g.,][]{koposov:2011, martinez:2011, walker:2011, pace:2014, walker:2016, caldwell:2017, pace:2020, awad:2025, yang:2025}. We direct the interested readers to \citet{pace:2020} for a well-written and detailed description of the methodology.

In brief, the likelihood of a star in our sample can be written as mixture of its likelihood of belonging in the \BooI\ ($\mathcal{L}_\text{B}$) or MW ($\mathcal{L}_\text{MW}$) component of the GMM:
\begin{align}
    \mathcal{L}(\mathbf{D}_i | \theta) &= f_\text{B}\mathcal{L}_\text{B}(\mathbf{D}_i | \theta_\text{B}) \nonumber\\
    &+ (1 - f_\text{B})\mathcal{L}_\text{MW}(\mathbf{D}_i | \theta_\text{MW}), \label{eq:GMM1}
\end{align}
where $f_\text{B}$ is the fraction of stars in the dataset that are members of \BooI, $\theta_\text{B}$ and $\theta_\text{MW}$ are the model parameters describing the metallicity, velocity, and proper motion distributions of \BooI\ and the MW foreground components respectively. Here, $\mathbf{D}_i$ represents the measured quantities of the star in our dataset as described in Section \ref{sec:GMM_data}. In turn, the probability that a star in the sample belongs to \BooI\ (as opposed to the MW foreground) is
\begin{equation} \label{eq:MemP}
    p_\text{BooI,i} = \frac{f_\text{B}\mathcal{L}_\text{B}}{f_\text{B}\mathcal{L}_\text{B} + (1 - f_\text{B})\mathcal{L}_\text{MW}}, \\
\end{equation}
where the dependence on $\mathbf{D}_i$ and $\theta$ have been dropped for brevity.

Under the reasonable assumption that each measurement in our dataset is independent, the total likelihood of our stellar sample given the model is the product of all individual star likelihoods,
\begin{equation}
    \mathcal{L}(\mathbf{D} | \theta) = \prod_{i=1}^{N_*}\mathcal{L}(\mathbf{D}_i | \theta).
\end{equation}
Similarly, if the metallicity, velocity, and proper motion probability distributions are all independent of one another, then the component likelihoods in Equation \ref{eq:GMM1} can be expressed as a product of the three distributions,
\begin{equation}
    \label{eq:BooI_like}
    \mathcal{L}_\text{B}(\mathbf{D}_i | \theta_\text{B}) = \mathcal{P}_\text{B}^{\text{vel}}\mathcal{P}_\text{B}^{\text{[Fe/H]}}\mathcal{P}_\text{B}^{\text{PM}}
\end{equation}
and
\begin{equation}
    \mathcal{L}_\text{MW}(\mathbf{D}_i | \theta_\text{MW}) = \mathcal{P}_\text{MW}^{\text{vel}}\mathcal{P}_\text{MW}^{\text{[Fe/H]}}\mathcal{P}_\text{MW}^{\text{PM}}.
\end{equation}

We model the probability distribution of stellar heliocentric l.o.s. velocities in \BooI\ as a 2D linear function of position along and perpendicular to its on-sky projected direction of motion, $\phi_{1*}$ and $\phi_{2*}$, with Gaussian dispersion:
\begin{equation}
\mathcal{P}_\text{B}^{\text{vel}} = \mathcal{N}\left(v_{i} | v_\text{B}(\phi_{1*,i},~\phi_{2*,i}), ~\sqrt{\sigma^2_{v,i} + \sigma^2_{v, \text{B}}}\right),
\end{equation}
where $\mathcal{N}(x|\mu, \sigma)$ represents a normal distribution with mean, $\mu$, and standard deviation, $\sigma$, evaluated at $x$. Here the systemic velocity as a function of location is
\begin{equation}
    \text{v}_\text{B}(\phi_{1*}, \phi_{2*}) = \Delta v_\text{1,BooI}\phi_{1*} + \Delta v_\text{2,BooI}\phi_{2*} + v_{0, \text{B}}. \label{eq:BooI_vel}
\end{equation}
In this parameterization, $v_{0, \text{B}}$ is the central velocity of \BooI, $\Delta v_\text{1,BooI}$ and $\Delta v_\text{2,BooI}$ are the components of the velocity gradient along and perpendicular to \BooI's proper motion direction respectively, and $\sigma_{v, \text{B}}$ is the intrinsic velocity dispersion. This parameterization differs slightly from that used in previous studies of velocity gradients where the velocity distribution is modeled as a 1D linear function along a free angle \citep[e.g.,][]{ji:2021}. We find that Equation \ref{eq:BooI_vel} produces similar results to the previous parameterization but yields better-behaved (i.e., Gaussian) posteriors. The choice to set the velocity components along $\phi_{1*}$ and $\phi_{2*}$ rather than the major and minor axes of \BooI\ was driven by the expectation that perspective rotation would dominate the observed velocity gradient (see Section \ref{sec:vel_grad}) but does not meaningfully impact the results.

We model the probability distribution of stellar metallicities in \BooI\ as a linear function of elliptical radius, $r$, with Gaussian dispersion:
\begin{align}
    \mathcal{P}_\text{B}^{\text{[Fe/H]}} = \mathcal{N}\bigg(&\text{[Fe/H]}_{i} | \text{[Fe/H]}_\text{B}(r_i), \nonumber\\
        &\sqrt{\sigma^2_{\text{[Fe/H]},i} + \sigma^2_{\text{[Fe/H]}, \text{B}}}\bigg),
\end{align}
where the systemic metallicity as a function of radius is
\begin{equation}
\text{[Fe/H]}_\text{B}(r) = \Delta\text{[Fe/H]}_\text{B}r + \text{[Fe/H]}_{0, \text{B}}.
\end{equation}
Here the probability distribution is parameterized by the central metallicity of \BooI, $\text{[Fe/H]}_{0, \text{B}}$, its radial metallicity gradient, $\Delta\text{[Fe/H]}_\text{B}$, and its intrinsic dispersion, $\sigma_\text{[Fe/H],B}$. We experimented with non-Gaussian shapes for the metallicity distribution function and found no substantial change to the recovered members or systemic metallicity properties.

Lastly for \BooI, we adopt a bivariate Gaussian proper motion distribution:
\begin{equation}
    \mathcal{P}_\text{B}^{\text{PM}} = \mathcal{N}_2(\mu_{\alpha*, i},~\mu_{\delta, i} | \langle \mu_{\alpha*}\rangle_\text{B},~ \langle \mu_{\delta}\rangle_\text{B},~Cov_{i}),
\end{equation}
where $\mathcal{N}_2(\mu_1, \mu_2, Cov)$ represents a bivariate normal distribution with means, $\mu_1$ and $\mu_2$, and covariance matrix, $Cov$. In this case, the covariance matrix is a combination of measured proper motion uncertainties and correlated errors, $\sigma_{\mu_{\alpha*}}$, $\sigma_{\mu_{\delta}}$, and $C_{\mu_{\alpha*},\mu_{\delta}}$, and intrinsic proper motion dispersions, $\sigma_{\mu_{\alpha*},\text{B}}$ and $\sigma_{\mu_{\delta}, \text{B}}$:
\begin{equation}
    Cov_i = 
    \begin{pmatrix}
        \sigma_{\mu_{\alpha*},i}^2 + \sigma_{\mu_{\alpha*},\text{B}}^2 & C_{\mu_{\alpha*},\mu_{\delta},i}\sigma_{\mu_{\alpha*},i}\sigma_{\mu_{\delta},i} \\
        C_{\mu_{\alpha*},\mu_{\delta},i}\sigma_{\mu_{\alpha*},i}\sigma_{\mu_{\delta},i} & \sigma_{\mu_{\delta},i}^2 + \sigma_{\mu_{\delta},\text{B}}^2
    \end{pmatrix}.
\end{equation}
\BooI's proper motion probability distribution is thus parameterized by its mean systemic proper motions, $\langle \mu_{\alpha*}\rangle_\text{B}$ and $\langle \mu_{\delta}\rangle_\text{B}$, and its intrinsic dispersions, $\sigma_{\mu_{\alpha*},\text{B}}$ and $\sigma_{\mu_{\delta}, \text{B}}$. 

The probability distributions for the MW foreground component are simpler in comparison. We assume 1D Gaussian distributions for both the stellar velocity and metallicity distributions of the MW foreground:
\begin{equation}
    \mathcal{P}_\text{MW}^{\text{vel}} = \mathcal{N}(v_{i} - \langle v\rangle_\text{MW},~\sqrt{\sigma^2_{\text{vel},i} + \sigma^2_{\text{vel}, \text{MW}}})
\end{equation}
and 
\begin{align}
    \mathcal{P}_\text{MW}^{\text{[Fe/H]}} = \mathcal{N}\bigg(&\text{[Fe/H]}_{i} - \langle\text{[Fe/H]}\rangle_\text{MW}, \nonumber \\
    &\sqrt{\sigma^2_{\text{[Fe/H]},i} + \sigma^2_{\text{[Fe/H]}, \text{MW}}}\bigg)
\end{align}
where $\langle v\rangle_\text{MW}$ and $\langle\text{[Fe/H]}\rangle_\text{MW}$ are the mean velocity and metallicity of the foreground, and $\sigma_{v, \text{MW}}$ and $\sigma_\text{[Fe/H],MW}$ are the velocity and metallicity dispersions of the foreground. Finally, we assume the foreground has a uniform proper motion distribution.

In total, the fiducial GMM has 16 parameters:  11 for the \BooI\ component, $\theta_\text{B}=\{v_{0,\text{B}},~\Delta v_{1, \text{B}}, ~\Delta v_{2, \text{B}},~\sigma_{v, \text{B}},~\text{[Fe/H]}_{0,\text{B}},$ $\Delta\text{[Fe/H]}_\text{B},~\sigma_\text{[Fe/H],BooI},~\langle \mu_{\alpha*}\rangle_\text{B},~\langle \mu_{\delta}\rangle_\text{B},~\sigma_{\mu_{\alpha*},\text{B}},$ $~\sigma_{\mu_{\delta},\text{B}}\}$, 4 for the MW foreground component, $\theta_\text{MW}=\{\langle v\rangle_\text{MW},~\sigma_{v, \text{MW}},~\langle\text{[Fe/H]}\rangle_\text{MW},~\sigma_\text{[Fe/H],MW}\}$, and 1 describing the membership fraction: $f_\text{B}$. We aim to infer all of these parameter models with the exception of \BooI's intrinsic proper motion dispersions, which is unresolved given current \gaia\ DR3 measurement uncertainties. Instead, we assume that radial and tangential velocity dispersions of \BooI\ are approximately equal, and set the proper motion dispersions in terms of the heliocentric l.o.s.\ velocity dispersion as
\begin{equation}
    \sigma_{\mu_{\alpha*},\text{B}}~[\text{mas yr}^{-1}]= \sigma_{\mu_{\delta},\text{B}} = \frac{\sigma_{v, \text{B}}}{4.74d}, \label{eq:PM_disp}
\end{equation}
where $d$ is the heliocentric distance to \BooI\ in kpc (66.3~kpc; \citealt{dallora:2006}). For all other model parameters, we adopt weakly informative or uninformative priors, which are summarized in Table \ref{tab:priors}. Together, the product of these 14 priors with the likelihood function presented in Equation \ref{eq:GMM1} provides the posterior distribution of model parameters.

\begin{deluxetable}{cc}
	\centerwidetable
	\caption{Free Model Parameters and Priors}
	\label{tab:priors}
	\tablehead{
	    \colhead{Parameter} &  
            \colhead{Prior}
	}
	\startdata
            \hline
            \multicolumn{2}{c}{Gaussian Mixture Model} \\
            \hline
            $f_\text{mem}$ & $\mathcal{U}(0, ~1)$ \\
            $v_{0,\text{B}}$ & $\mathcal{N}(100~\text{km s}^{-1}, ~10~\text{km s}^{-1})$ \\
            $\Delta v_{1,\text{B}}$ & $\mathcal{U}(-20 ~\text{km s}^{-1}~\text{deg}^{-1}, ~20~\text{km s}^{-1}~\text{deg}^{-1})$ \\
            $\Delta v_{2,\text{B}}$ & $\mathcal{U}(-20 ~\text{km s}^{-1}~\text{deg}^{-1}, ~20~\text{km s}^{-1}~\text{deg}^{-1})$ \\
            $\log_{10}(\sigma_{v,\text{B}})$ & $\mathcal{U}(0.0, ~1.3)$ \\
            $\text{[Fe/H]}_{0,\text{B}}$ & $\mathcal{N}(-2.4, ~1.0)$ \\
            $\Delta\text{[Fe/H]}_\text{B}$ & $\mathcal{U}(-1 ~\text{dex deg}^{-1}, 1 ~\text{dex deg}^{-1})$ \\
            $\log_{10}(\sigma_\text{[Fe/H],BooI})$ & $\mathcal{U}(-1.0, ~0.3)$ \\
            $\langle\mu_{\alpha*}\rangle_\text{B}$ & $\mathcal{N}(-0.4 ~\text{mas yr}^{-1}, ~0.2~\text{mas yr}^{-1})$ \\
            $\langle\mu_{\delta}\rangle_\text{B}$ & $\mathcal{N}(-1.1 ~\text{mas yr}^{-1}, ~0.2~\text{mas yr}^{-1})$ \\
            $\langle v\rangle_\text{MW}$ & $\mathcal{N}(0 ~\text{km s}^{-1}, ~50~\text{km s}^{-1})$ \\
            $\log_{10}(\sigma_{v,\text{MW}})$ & $\mathcal{U}(-1.0, ~4.0)$ \\
            $\langle \text{[Fe/H]}\rangle_\text{MW}$ & $\mathcal{U}(-5.0, ~1.0)$ \\
            $\log_{10}(\sigma_\text{[Fe/H],MW})$ & $\mathcal{U}(-1.0, ~0.5)$ \\
            \hline
            \multicolumn{2}{c}{Axisymmetric Jeans Model} \\
            \hline
            $Q$ [Gyr] & $\mathcal{U}(0.1, ~2.20)$ \\
            $\log(b_\text{halo})$ [pc] & $\mathcal{U}(0, ~5)$ \\
            $\log(\rho_0)$ [$M_\odot~\text{pc}^{-3}$] & $\mathcal{U}(-5, ~-5)$ \\
            $-\log(1-\beta_z)$ & $\mathcal{U}(-1, ~1)$ \\
            $\alpha$ & $\mathcal{U}(0.5, ~3)$ \\
            $\beta$ & $\mathcal{U}(3, ~10)$ \\
            $\gamma$ & $\mathcal{U}(0, ~2)$ \\
            $i$ & $\mathcal{U}(\cos^{-1}q', ~90)$ \\
            \hline
            \multicolumn{2}{c}{Galactic Chemical Evolution Model} \\
            \hline
            $\tau_\text{SFH}$ [Gyr] & $\mathcal{U}(0.08, ~0.35)$ \\
            $\log(\tau_\text{SFE})$ [Gyr] & $\mathcal{U}(0, ~4)$ \\
            $t_\text{trunc}$ [Gyr] & $\mathcal{U}(0, ~2)$ \\
            $\eta$ & $\mathcal{U}(0, ~10^{4})$ 
	\enddata
	\tablecomments{
	Priors adopted for each of the free parameters in the Gaussian mixture model used in Section \ref{sec:GMM}, the axisymmetric Jeans model used in Section \ref{sec:jeans_modeling} and the galactic chemical evolution model used in Section \ref{sec:GCE}. $\mathcal{U}(a,~b)$ denotes uniform priors bounded by $a$ and $b$, while $\mathcal{N}(\mu,~\sigma)$ denotes a Gaussian prior centered at $\mu$ with standard deviation $\sigma$
	}
\end{deluxetable}

To sample this posterior distribution, we employ  the \textit{Preconditioned Monte Carlo} (PMC) method for Bayesian inference implemented in the publicly available Python package \texttt{pocoMC}\footnote{\url{https://github.com/minaskar/pocomc}} \citep{karamanis:2022a, karamanis:2022b}, which uses a combination of a normalizing flow with a sequential Monte Carlo sampling scheme to decorrelate and efficiently sample high-dimensional distributions with non-trivial geometry. We adopt default hyperparameters for \texttt{pocoMC} and run the sampler until it has converged (i.e., when the ``inverse temperature'' $\beta=1$) and reached an effective sample size of 5,000 samples.

\subsubsection{Non-mixture Model for the Combined and $N$-body Datasets}
As described previously in Section \ref{sec:data_comb}, we also consider in our analysis a combined dataset that includes member stars from \Sfive\ (presented in Section \ref{sec:GMM_members}), member stars from the archival AAT and MMT datasets (presented in Appendix \ref{app:archival}), and member stars from the archival VLT dataset (presented in \citetJenkins). The decision to combine the datasets after membership determination was driven by large differences in the targeting strategy of each observing program, which complicates modeling the MW foreground across all datasets.% (see Figure \ref{fig:GMM_Corner_Bkg}). 

Assuming all stars in the combined dataset are members, we can model the systemic properties of \BooI\ using just the \BooI\ likelihood given in Equation \ref{eq:BooI_like} without the need for a GMM.
All functional forms for \BooI's velocity, metallicity, and proper motion distributions as well as the priors on all relevant model parameters are the same as employed in the GMM applied to the individual datasets.

We also fit a model to the simulated dataset from an $N$-body simulation of \BooI. As in the combined dataset, the stars in the simulated dataset are all assumed to be member stars, so we can once again use the likelihood in Equation \ref{eq:BooI_like}. However, for the $N$-body simulation, we only consider the heliocentric l.o.s.\ velocity information of the sample.

\subsection{GMM Results: Membership}
\label{sec:GMM_members}
\begin{figure}[ht!] 
	\includegraphics[width=0.45\textwidth]{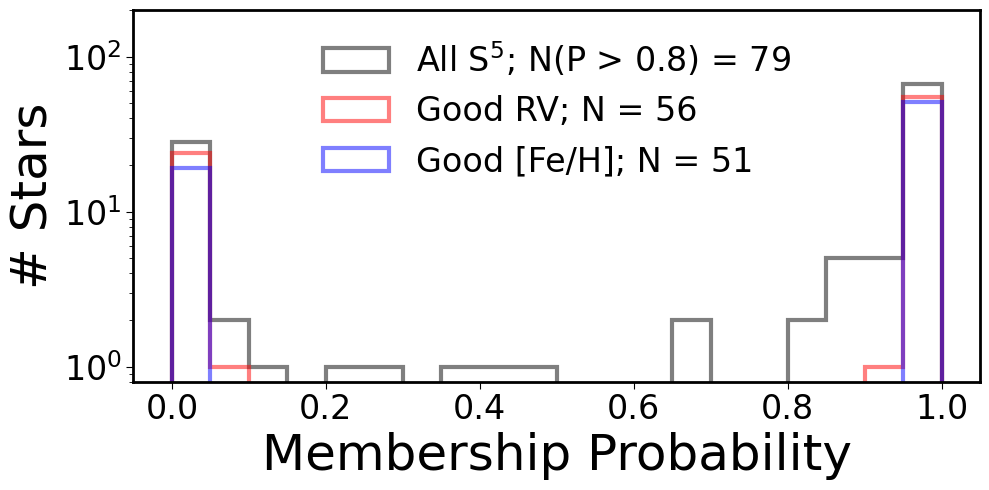}
    \caption{
        Distribution of median membership probabilities of our \Sfive\ sample. The subset of stars in our sample with good radial velocity measurements (i.e., $\sigma_v < 10$~\kms, $\sigma_\text{[Fe/H]} < 0.5$~dex, and neither a binary candidate or RRL star) and good [Fe/H] measurements (i.e., $\sigma_v < 10$~\kms, $\sigma_\text{[Fe/H]} < 0.5$~dex, and neither a HB or RRL star) are represented by the red and blue histograms respectively. A membership threshold of $p_\text{mem} > 0.8$ is adopted in this study to identify high-probability members of \BooI.
    } 
    \label{fig:mem_p}
\end{figure}

The membership probability, $p_\text{mem}$, of each star in the \Sfive\ dataset is calculated with Equation \ref{eq:MemP} using draws from the posterior distribution. The median and 1$\sigma$ uncertainties on $p_\text{mem}$ for each star are included in Table \ref{tab:stars}. Figure \ref{fig:mem_p} shows the distribution of median membership probabilities of this sample. Analogous figures for the archival AAT and MMT data are presented in Appendix \ref{app:archival}.
Adopting a median membership probability threshold of $p_\text{mem} > 0.8$, we identify 79 likely member stars. Of these, 56 have good radial velocity measurements (i.e., $\sigma_v < 10$~\kms, $\sigma_\text{[Fe/H]} < 0.5$~dex, and neither a binary candidate or RRL star) and 51 have good [Fe/H] measurements (i.e., $\sigma_v < 10$~\kms, $\sigma_\text{[Fe/H]} < 0.5$~dex and neither a HB or RRL star). 

\begin{figure*}[ht!] 
    \includegraphics[width=1.0\textwidth]{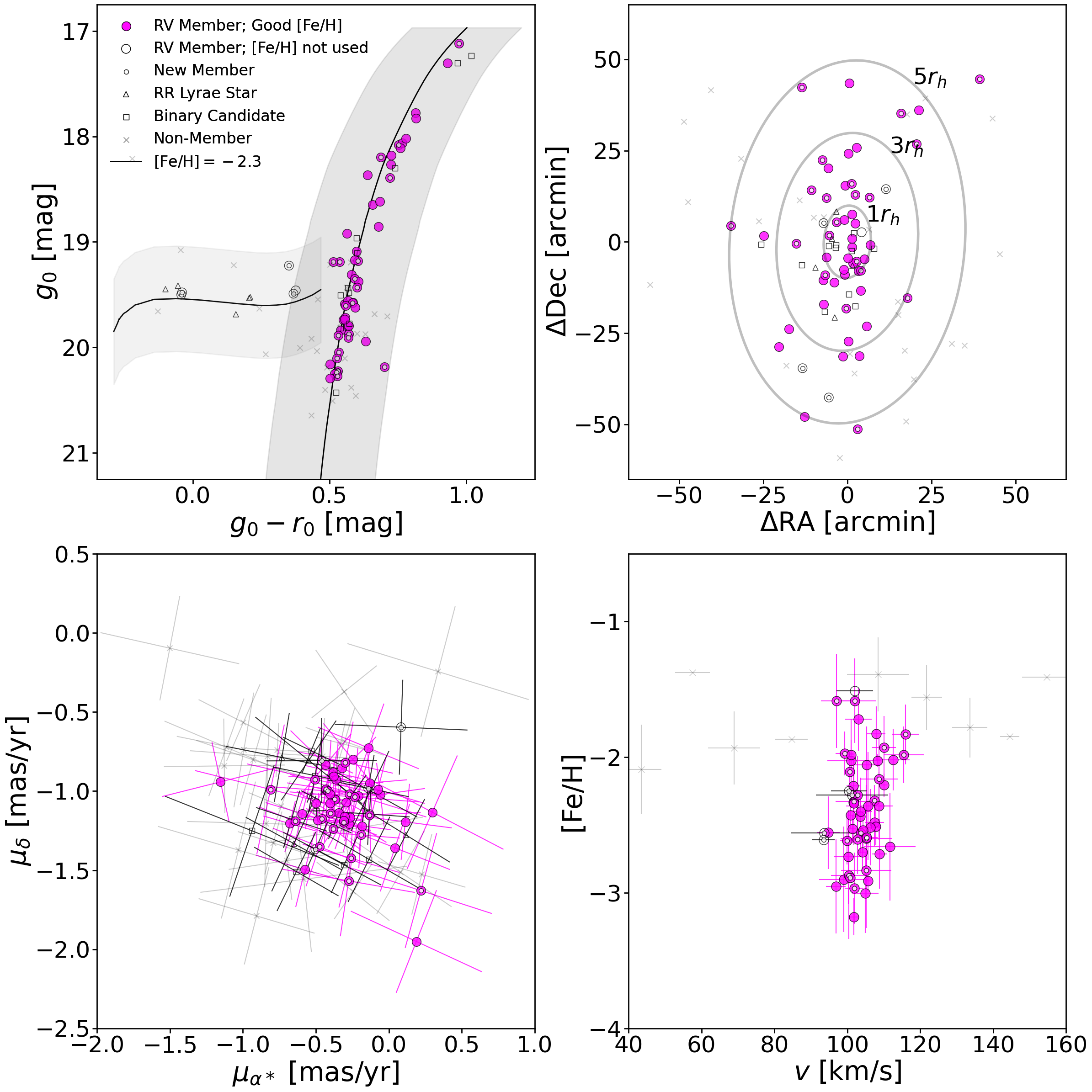}
    \caption{
        \textbf{Top Left.} Extinction-corrected  CMD of the \Sfive\ sample. High-probability \BooI\ members with good radial velocity (and [Fe/H]) measurements are represented by (filled) circles. Open triangle and square markers represent RRL stars and binary candidates respectively, while gray x's represent non-members. Smaller inset white circles denote new likely member stars, which were not identified in previous studies. %The 2 ambiguous binar stars, 1230834410380273664 and 1230833585745987968 are shown in yellow and cyan respectively.
        \textbf{Top Right.} On-sky spatial distribution of the \Sfive\ sample. Gray ellipses indicate the 1, 3, and 5 half-light radii of \BooI\ from \citet{munoz:2018}.
        \textbf{Bottom Left.} Proper motion  distribution of the \Sfive\ sample from \gaia\ DR3 \citep{gaiacollaboration:2022}.
        \textbf{Bottom Right.} Distribution of [Fe/H] vs.\ radial velocity for the \Sfive\ sample. Stars without reliable [Fe/H] measurements (open circles) are included using the nominal [Fe/H] value reported by \texttt{RVSpecFit}. 
    } 
    \label{fig:S5_4panel}
\end{figure*}

Figure \ref{fig:S5_4panel} shows the distribution of these likely members in several data dimensions, including the color-magnitude diagram (CMD; top left), spatial (top right), proper motion (bottom left), and metallicity and velocity (bottom right). In all panels, large circles represent member stars with good velocity measurements, while triangles, squares, and x's represent RRL stars, binary candidates, and non-member stars respectively. Filled circles denote likely members with good [Fe/H] measurements. Superimposed small white circles denote new likely member stars which have not previously been identified as members. %We also highlight the ambiguous binary stars 230834410380273664 and 1230833585745987968, which were identified in Section \ref{sec:binary}, using yellow and cyan circles respectively.

\begin{figure*}[ht!] 
    \includegraphics[width=1.0\textwidth]{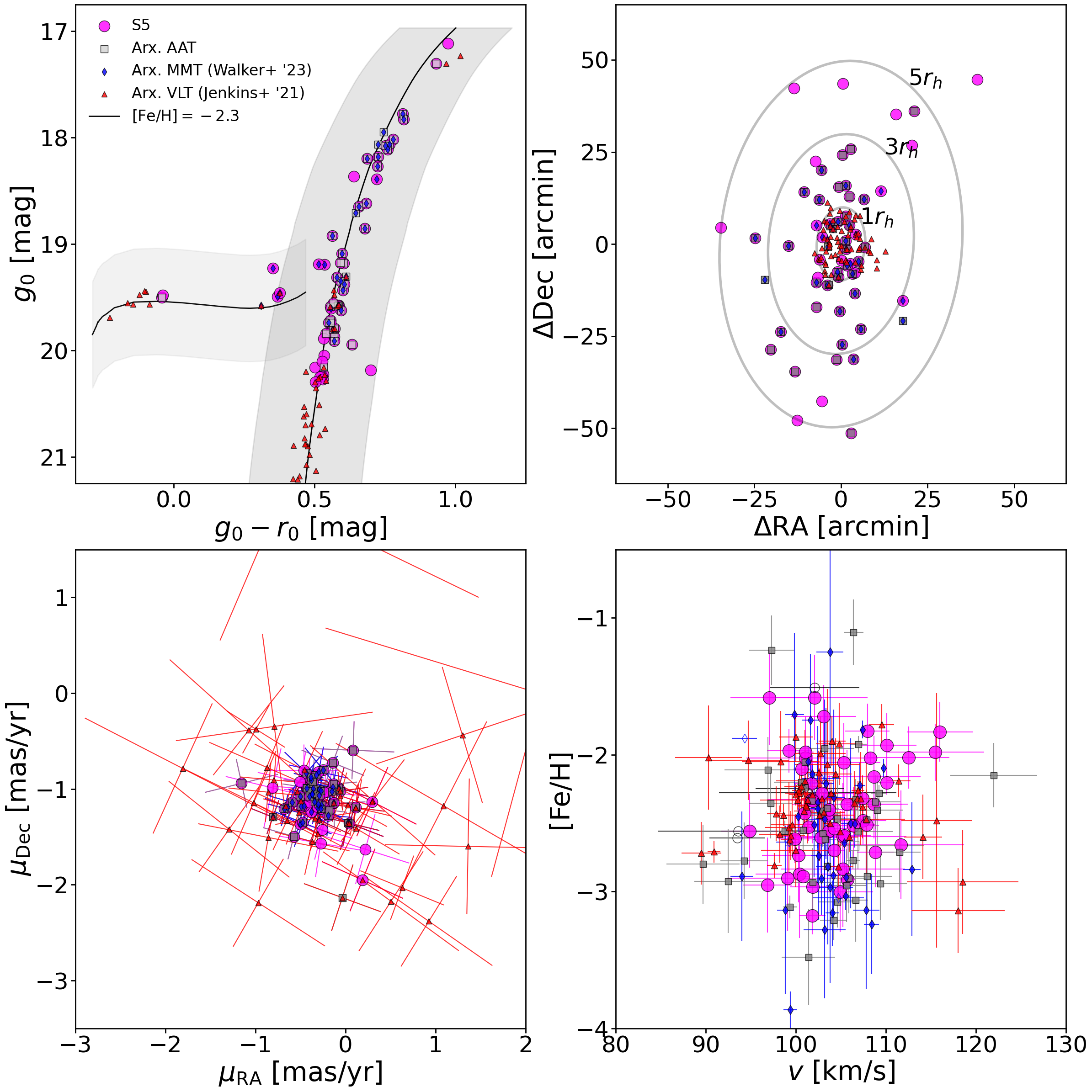}
    \caption{
        Same as Figure \ref{fig:S5_4panel} except showing only likely \BooI\ member stars with good radial velocity measurements from the respective \Sfive, archival AAT, archival MMT, and archival VLT datasets (magenta circles, gray squares, blue diamonds, and red triangles respectively respectively).
    } 
    \label{fig:Combined_4panel}
\end{figure*}

Of the 56 member stars with good velocities, 24 are newly identified by our \Sfive\ observations while 32 were previously identified by \citet{martin:2007}, \citet{norris:2010b}, \citetKoposov, \citetJenkins, and/or \citetLongeard. As can be seen in the top right panel of Figure \ref{fig:S5_4panel}, our sample includes likely member stars from the inner half-light radius of \BooI\ out to $\sim$7$r_h$ with the majority of new member stars found beyond one half-light radius.
%\update{Check to see if we recover any from \citet{pan:2025}.}

Merging this \Sfive\ sample of \BooI\ member stars with those that we (re-)identify from archival AAT and MMT observations and those identified from VLT observations by \citetJenkins, we create a combined dataset of 148 \BooI\ members, of which  115 have good non-variable radial velocities suitable for dynamical analyses and 92 have robust [Fe/H] measurements suitable for chemical evolution studies. This combined dataset includes 9 and 12 newly identified member stars with good radial velocities from the archival AAT and MMT datasets respectively, though all of these are also present in the \Sfive\ dataset. The distribution of member stars in this combined dataset is presented in Figure \ref{fig:Combined_4panel}.

\subsection{GMM Results: Systemic Properties}
\label{sec:GMM_global}
The systemic properties of \BooI\ inferred from our analysis of the combined data are summarized in Table \ref{tab:BooI_par}. %The posterior distribution of all of the model parameters is presented as a corner plot alongside analogous posteriors for GMM analysis of the archival AAT and MMT datasets in Figures \ref{fig:GMM_Corner_BooI} and \ref{fig:GMM_Corner_Bkg}. 
The inferred properties from the individual \Sfive\ and archival datasets are summarized in Table \ref{tab:BooI_par_extra}. A more thorough discussion of radially-dependent systemic properties (e.g. gradients) is presented below in Section \ref{sec:pop_gradients}.

\begin{deluxetable*}{lrrrrr}
	\centerwidetable
	\caption{Comparison of \BooI\ Properties Between Datasets}
	\label{tab:BooI_par_extra}
	\tablehead{
	    \colhead{Parameter} &  
            \colhead{\Sfive} & 
            \colhead{Arx.\ AAT} &
            \colhead{Arx.\ MMT} &
            \colhead{Arx.\ VLT} &
            \colhead{Combined}
	}
	\startdata
        \multicolumn{6}{c}{Inferred Properties} \\
        \hline
        $f_\mathrm{B}$ & $0.70^{+0.04}_{-0.05}$ & $0.81^{+0.05}_{-0.05}$ & $0.77^{+0.04}_{-0.05}$ & \nodata & \nodata \\
		$v_{0,\mathrm{B}}$ [km s$^{-1}$] & $103.7^{+0.6}_{-0.6}$ & $103.5^{+0.7}_{-0.7}$ & $103.5^{+0.7}_{-0.7}$ & $102.6^{+0.7}_{-0.8}$ & $103.0^{+0.4}_{-0.4}$ \\
		$\Delta v_{1,\mathrm{B}}$ [km s$^{-1}$ arcmin$^{-1}$] & $0.04^{+0.03}_{-0.03}$ & $0.03^{+0.04}_{-0.04}$ & $-0.03^{+0.06}_{-0.06}$ & $-0.10^{+0.14}_{-0.14}$ & $0.02^{+0.03}_{-0.03}$ \\
		$\Delta v_{2,\mathrm{B}}$ [km s$^{-1}$ arcmin$^{-1}$] & $-0.08^{+0.06}_{-0.06}$ & $-0.11^{+0.08}_{-0.08}$ & $-0.02^{+0.07}_{-0.07}$ & $-0.06^{+0.16}_{-0.16}$ & $-0.08^{+0.05}_{-0.05}$ \\
		$\sigma_{v,\mathrm{B}}$ [km s$^{-1}$] & $3.2^{+0.6}_{-0.5}$ & $3.9^{+0.7}_{-0.6}$ & $3.7^{+0.6}_{-0.4}$ & $5.10^{+0.70}_{-0.80}$ & $4.0^{+0.4}_{-0.3}$ \\
		$\mathrm{[Fe/H]}_{0,\mathrm{B}}$ [dex] & $-2.23^{+0.08}_{-0.08}$ & $-2.42^{+0.13}_{-0.13}$ & $-2.40^{+0.13}_{-0.13}$ & $-2.34^{+0.05}_{-0.05}$ & $-2.28^{+0.05}_{-0.05}$ \\
		$\Delta\mathrm{[Fe/H]}_{\mathrm{B}}$ [dex arcmin$^{-1}$] & $-0.008^{+0.003}_{-0.003}$ & $-0.006^{+0.006}_{-0.007}$ & $-0.016^{+0.008}_{-0.008}$ & \nodata & $-0.010^{+0.003}_{-0.003}$ \\
		$\sigma_{\mathrm{[Fe/H],B}}$ [dex] & $0.27^{+0.04}_{-0.04}$ & $0.40^{+0.06}_{-0.06}$ & $0.41^{+0.06}_{-0.05}$ & $-0.28^{+0.04}_{-0.03}$ & $0.28^{+0.03}_{-0.03}$ \\
		$\langle\mu_{\alpha,*}\rangle_\mathrm{B}$ [mas yr$^{-1}$] & $-0.39^{+0.02}_{-0.02}$ & $-0.39^{+0.02}_{-0.02}$ & $-0.39^{+0.02}_{-0.02}$ & $-0.45^{+0.04}_{-0.04}$ & $-0.38^{+0.02}_{-0.02}$ \\
		$\langle\mu_\delta\rangle_\mathrm{B}$ [mas yr$^{-1}$] & $-1.07^{+0.01}_{-0.01}$ & $-1.07^{+0.01}_{-0.01}$ & $-1.07^{+0.01}_{-0.01}$ & $-1.13^{+0.03}_{-0.03}$ & $-1.08^{+0.01}_{-0.01}$ \\
		$C_{\langle\mu_{\alpha,*}\rangle,{\langle\mu_{\delta}}\rangle}$ & $-0.10$ & $-0.10$ & $-0.14$ & \nodata & $-0.08$ \\
        \hline
        \multicolumn{6}{c}{Additional Derived Properties} \\
        \hline
		$N_\text{mem,vel}$ & 56 & 42 & 37 & 60 & 115 \\
		$N_\text{mem,[Fe/H]]}$ & 51 & 41 & 33 & 52 & 92 \\
		$\langle v_\mathrm{BooI}\rangle$ [km s$^{-1}$] & $103.8^{+0.6}_{-0.6}$ & $103.7^{+0.7}_{-0.7}$ & $103.4^{+0.6}_{-0.6}$ & $102.6^{+0.7}_{-0.8}$ & $103.0^{+0.4}_{-0.4}$ \\
		$\langle \text{[Fe/H]}_\mathrm{BooI}\rangle$ & $-2.38^{+0.05}_{-0.05}$ & $-2.51^{+0.07}_{-0.07}$ & $-2.60^{+0.10}_{-0.10}$ & $-2.34^{+0.05}_{-0.05}$ & $-2.43^{+0.04}_{-0.04}$ \\
		$\Delta v_{\mathrm{B}}$  & $0.09^{+0.06}_{-0.04}$ & $0.13^{+0.08}_{-0.06}$ & $0.08^{+0.05}_{-0.04}$ & $0.00^{+0.00}_{-0.00}$ & $0.09^{+0.05}_{-0.05}$ \\
		$\Delta v_{1,\mathrm{BooI}}'$ & $-0.07^{+0.03}_{-0.03}$ & $-0.07^{+0.04}_{-0.04}$ & $-0.13^{+0.06}_{-0.06}$ & $-0.20^{+0.14}_{-0.14}$ & $-0.09^{+0.03}_{-0.03}$ \\
		$\Delta v_{2,\mathrm{BooI}}'$ & $-0.08^{+0.06}_{-0.06}$ & $-0.11^{+0.08}_{-0.08}$ & $-0.02^{+0.07}_{-0.07}$ & $-0.06^{+0.16}_{-0.16}$ & $-0.08^{+0.05}_{-0.05}$ \\
		$\Delta v_{\mathrm{B}}'$  & $0.11^{+0.05}_{-0.04}$ & $0.15^{+0.06}_{-0.05}$ & $0.15^{+0.06}_{-0.05}$ & $0.11^{+0.00}_{-0.00}$ & $0.12^{+0.04}_{-0.03}$ \\
        \hline
        \multicolumn{6}{c}{Dark Matter Properties} \\
        \hline
            $Q$ & $1.14^{+0.54}_{-0.58}$ & \nodata & \nodata & $1.10^{+0.60}_{-0.60}$ & $1.18^{+0.51}_{-0.59}$ \\
		$\log_{10}b_\mathrm{halo}$ & $2.85^{+0.92}_{-0.60}$ & \nodata & \nodata & $3.70^{+0.90}_{-0.80}$ & $3.07^{+0.87}_{-0.45}$ \\
		$\log_{10}\rho_{0}$ & $-1.54^{+1.44}_{-1.81}$ & \nodata & \nodata & $-1.80^{+1.20}_{-0.90}$ & $-1.88^{+0.90}_{-1.51}$ \\
		$-\log_{10}(1-\beta_z)$ & $0.26^{+0.19}_{-0.20}$ & \nodata & \nodata & $0.10^{+0.50}_{-0.40}$ & $0.25^{+0.21}_{-0.21}$ \\
		$\alpha$ & $1.63^{+0.89}_{-0.77}$ & \nodata & \nodata & $1.80^{+0.80}_{-0.80}$ & $1.76^{+0.82}_{-0.81}$ \\
		$\beta$ & $6.22^{+2.30}_{-2.11}$ & \nodata & \nodata & $6.30^{+2.10}_{-2.10}$ & $6.39^{+2.20}_{-2.12}$ \\
		$\gamma$ & $1.27^{+0.48}_{-0.74}$ & \nodata & \nodata & $0.70^{+0.50}_{-0.60}$ & $1.00^{+0.52}_{-0.60}$ \\
		$i$ & $72.05^{+11.83}_{-13.77}$ & \nodata & \nodata & $74.20^{+11.20}_{-10.10}$ & $70.47^{+12.83}_{-13.23}$ \\
        \hline
        \multicolumn{6}{c}{GCE Properties} \\
        \hline
		$\tau_\text{SFH}$ & $0.2^{+0.1}_{-0.1}$ & \nodata & \nodata & \nodata & $0.2^{+0.1}_{-0.1}$ \\
		$\tau_\text{SFE}$ & $11.9^{+8.7}_{-5.1}$ & \nodata & \nodata & \nodata & $13.9^{+7.6}_{-5.3}$ \\
		$t_\text{trunc}$ & $>0.4$ & \nodata & \nodata & \nodata & $>0.5$ \\
		$\eta$ & $170^{+35}_{-46}$ & \nodata & \nodata & \nodata & $203^{+27}_{-36}$ \\
	\enddata
	\tablecomments{
	   Properties of \BooI\ inferred from the analysis of our different datasets. The units of each parameter are the same as in Table \ref{tab:BooI_par}. %The columns \Sfive\ (Alt.) and Combined (Alt.) provide the properties from the \Sfive\ and Combined datasets when stars 30834410380273664 and 123083358574598796 are classified as binaries. 
       Properties for the archival VLT dataset are taken from \citetJenkins\ except for the velocity gradients which are determined analogously to the combined dataset and the dark matter properties which are taken from \citet{hayashi:2023}.
	}
\end{deluxetable*}

The systemic proper motion that we infer from the combined dataset is $\langle\mu_{\alpha,*}\rangle_\mathrm{B}=-0.38\pm0.02$ mas yr$^{-1}$ in RA and 
$\langle\mu_\delta\rangle_\mathrm{B}=-1.08\pm0.01$ mas yr$^{-1}$ in declination. We find nearly identical results for the individual \Sfive\ and archival datasets with the sole exception of the VLT dataset, which lacks bright stars with precise \gaia\ proper motions (see \citetJenkins\ for additional discussion). These values are in good agreement with the recent proper motion measurements of \citet{mcconnachie:2020}, \citet{martinez-garcia:2021}, \citet{battaglia:2022a}, \citet{filion:2022}, and \citet{pace:2022}, but discrepant with the  measurements of \citet{li:2021}\footnote{Using \gaia\ EDR3, \citet{li:2021} find $\langle\mu_{\alpha,*}\rangle_\mathrm{B}=-0.307\pm0.052$ mas yr$^{-1}$ and $\langle\mu_\delta\rangle_\mathrm{B}=-1.157\pm0.043$ mas yr$^{-1}$.}.

% The systemic proper motion that we infer from the \Sfive\ dataset is $\langle\mu_{\alpha,*}\rangle_\mathrm{B}=-0.39\pm0.02$ mas yr$^{-1}$ in RA and 
% $\langle\mu_\delta\rangle_\mathrm{B}=-1.07\pm0.01$ mas yr$^{-1}$ in declination. We find nearly identical results when considering the combined dataset: $\langle\mu_{\alpha,*}\rangle_\mathrm{B}=-0.38\pm0.02$ mas yr$^{-1}$ and $\langle\mu_\delta\rangle_\mathrm{B}=-1.08\pm0.01$ mas yr$^{-1}$. Both sets of values are in good agreement with the recent proper motion measurements of \citet{mcconnachie:2020}, \citet{martinez-garcia:2021}, \citet{battaglia:2022a}, \citet{filion:2022}, and \citet{pace:2022}, but discrepant with the  measurements of \citet{li:2021}\footnote{Using \gaia\ EDR3, \citet{li:2021} find $\langle\mu_{\alpha,*}\rangle_\mathrm{B}=-0.307\pm0.052$ mas yr$^{-1}$ and $\langle\mu_\delta\rangle_\mathrm{B}=-1.157\pm0.043$ mas yr$^{-1}$.}. Our values are also in good agreement with the systemic proper motions found for the individual archival datasets in Appendix \ref{app:archival} with the sole exception of the VLT dataset, which lacks bright stars with precise \gaia\ proper motions (see \citetJenkins\ for additional discussion).

The mean heliocentric l.o.s.\ velocity of \BooI\ is not inferred directly in our GMM. Rather, we infer the systemic heliocentric l.o.s.\ velocity at its center: $v_{0,\text{B}}=103.0^{+0.4}_{-0.4}$ km s$^{-1}$ for the combined dataset. We expect the central and mean velocities to be similar if stars are distributed symmetrically about the galaxy's center, which is indeed the case ($\langle v_\mathrm{B}\rangle = 103.0^{+0.4}_{-0.4}$ km s$^{-1}$). This mean systemic velocity is in good agreement with the mean heliocentric l.o.s.\ velocity reported by \citetJenkins\ and \citetLongeard\ ($102.6^{+0.7}_{-0.8}$ km s$^{-1}$ and $103.0^{+0.6}_{-0.6}$ km s$^{-1}$ respectively) as well as the values found for the individual datasets in Appendix \ref{app:archival}.

% The mean heliocentric velocity of \BooI\ is not inferred directly in our GMM. Rather, we infer the systemic heliocentric velocity at its center, $v_{0,\mathrm{B}} = 102.9^{+0.8}_{-0.7}$ km s$^{-1}$ ($103.4^{+0.4}_{-0.4}$ km s$^{-1}$ for the combined dataset), though we expect these quantities to  be similar if stars are distributed symmetrically about the galaxy's center. Using only the sample of high probability members identified in Section \ref{sec:GMM_members}, we can separately infer \BooI's mean heliocentric velocity to be $\langle v_\mathrm{B}\rangle = 103.0^{+0.7}_{-0.7}$ km s$^{-1}$ ($103.1^{+0.4}_{-0.4}$ km s$^{-1}$ for the combined dataset), which is indeed consistent with the central velocity. Both the central and mean systemic velocities are in good agreement with the mean heliocentric velocity reported by \citetJenkins\ and \citetLongeard\ ($102.6^{+0.7}_{-0.8}$ km s$^{-1}$ and $103.0^{+0.6}_{-0.6}$ km s$^{-1}$ respectively) as well as the values found for the individual archival datasets in Appendix \ref{app:archival}.

From our combined dataset, we infer a marginally resolved heliocentric l.o.s.\ velocity gradient, which is oriented approximately along \BooI's semi-minor axis. The components of the velocity gradient parallel and perpendicular to \BooI's on-sky proper motion are $\Delta v_{1,\mathrm{B}} = 0.02\pm0.03$ km s$^{-1}$ arcmin$^{-1}$ and $\Delta v_{2,\mathrm{B}} = -0.08\pm0.05$ km s$^{-1}$ arcmin$^{-1}$ respectively---or equivalently $\Delta v_{1,\mathrm{B}} = 0.2\pm0.3$ km s$^{-1}$ $r_h^{-1}$ and $\Delta v_{2,\mathrm{B}} = -0.8\pm0.5$ km s$^{-1}$ $r_h^{-1}$. This velocity gradient is consistent with the gradients found for the individual datasets in Appendix \ref{app:archival}. However, it is substantially shallower than and approximately perpendicular to the gradient reported by \citetLongeard. \BooI's velocity gradient is discussed in more detail in Section \ref{sec:vel_grad}.

We infer the velocity dispersion of \BooI\ to be $\sigma_{v,\mathrm{B}} = 4.0^{+0.4}_{-0.3}$. This is consistent with, though slightly smaller than, the previous constraints from \citetJenkins\ and \citetLongeard\ of $5.1^{+0.7}_{-0.8}$ km s$^{-1}$ and $4.5^{+0.3}_{-0.3}$ km s$^{-1}$ respectively. The smaller velocity dispersion found in this work can likely be attributed to a more thorough exclusion of binary stars. For a discussion of a potential radial dependence of the velocity dispersion, see Section \ref{sec:vel_disp_prof}.

From our combined dataset, we infer the central metallicity of \BooI\ to be $\mathrm{[Fe/H]}_{0,\mathrm{B}}=-2.28\pm0.05$. The mean metallicity derived from the likely member stars is slightly lower, $\langle \text{[Fe/H]}_\mathrm{B}\rangle = -2.43\pm0.04$, as would be expected in the presence of a negative metallicity gradient. Indeed, we infer a weak but clearly resolved metallicity gradient of $\Delta\mathrm{[Fe/H]}_{\mathrm{B}} = -0.010\pm0.003$ dex arcmin$^{-1}$ or $-0.10\pm0.03$ dex $r_h^{-1}$ (see Section \ref{sec:feh_grad} for additional discussion). We infer a metallicity dispersion of $\sigma_{\mathrm{[Fe/H],B}}=0.28^\pm0.03$. The shape and physical interpretation of \BooI's MDF is presented in Section \ref{sec:GCE}.

Our inferred metallicity dispersion is in good agreement with the metallicity dispersions inferred by \citetJenkins\ and \citetLongeard\ of $\sigma_{\mathrm{[Fe/H],B}}=0.28^{+0.04}_{-0.03}$ and $0.26^{+0.04}_{-0.03}$ respectively. The metallicity gradient we infer is also consistent with the gradient reported by \citetLongeard\ ($\Delta\mathrm{[Fe/H]}_{\mathrm{B}} = -0.008^{+0.003}_{-0.003}$ dex arcmin$^{-1}$). However, the mean metallicity we infer is slightly lower than found by \citetJenkins\ ($-2.34\pm0.05$) and slightly higher than found by \citetLongeard\ ($-2.60\pm0.03$). The difference with \citetJenkins\ can easily be explained by the different spatial selection of the two datasets. The \citetJenkins\ sample is contained nearly entirely within \BooI's inner half-light radius, and is thus expected to be more metal-rich given \BooI's negative metallicity gradient. Differences in sample composition may also explain the difference with \citetLongeard, as we find that the archival AAT (and MMT) datasets have mean metallicities that are 0.1 (and 0.2) dex lower than the combined dataset. We note that a mean metallicity of $-2.43$ that we find in this analysis would place \BooI\ squarely on the Local Group dwarf galaxy mass- and luminosity-metallicity relationships of \citet{kirby:2013}, while a mean metallicity of $-2.60$ would place \BooI\ slightly below these relationships (though certainly still within the relationship's intrinsic scatter).

\section{Population Gradients}
\label{sec:pop_gradients}
In this section, we discuss in greater detail the population gradients (or lack thereof) that we infer from our datasets, including the velocity gradient (Section \ref{sec:vel_grad}), the velocity dispersion profile (Section \ref{sec:vel_disp_prof}), and the metallicity gradient and dispersion profile (Section \ref{sec:feh_grad}).

\subsection{Velocity Gradient}
\label{sec:vel_grad}
The presence, or lack-thereof, of a velocity gradient in a galaxy provides clues as to its internal dynamics and dynamical evolution. For example, a velocity gradient oriented along a galaxy's semi-major axis might indicate oblate (i.e., disk-like) rotation, while a velocity gradient oriented along a galaxy's semi-minor axis as observed in the M31 dwarf spheroidal (dSph) satellite, Andromeda II \citep{ho:2012}, the isolated dwarf irregular, Phoenix \citep{kacharov:2017}, and the MW dSph satellite, Ursa Minor \citep{pace:2020} would be indicative of prolate rotation about its minor axis, perhaps as the result of a merger \citep[e.g.,][]{ebrova:2017}. On the other hand, a velocity gradient oriented along a galaxy's orbit is potentially a sign of tidal disruption as reported in MW UFDs Hercules \citep{aden:2009, martin:2010, kupper:2017, ou:2024}, Ursa Major II \citep{simon:2007, smith:2013}, and Leo V \citep{collins:2017}, as well as diffuse MW satellites Antlia 2 and Crater 2 \citep{ji:2021}. However, due to small sample sizes and uncertain measurements, the detection of velocity gradients in most UFDs remains quite tenuous.

\begin{figure*}[ht!] 
	\includegraphics[width=1.0\textwidth]{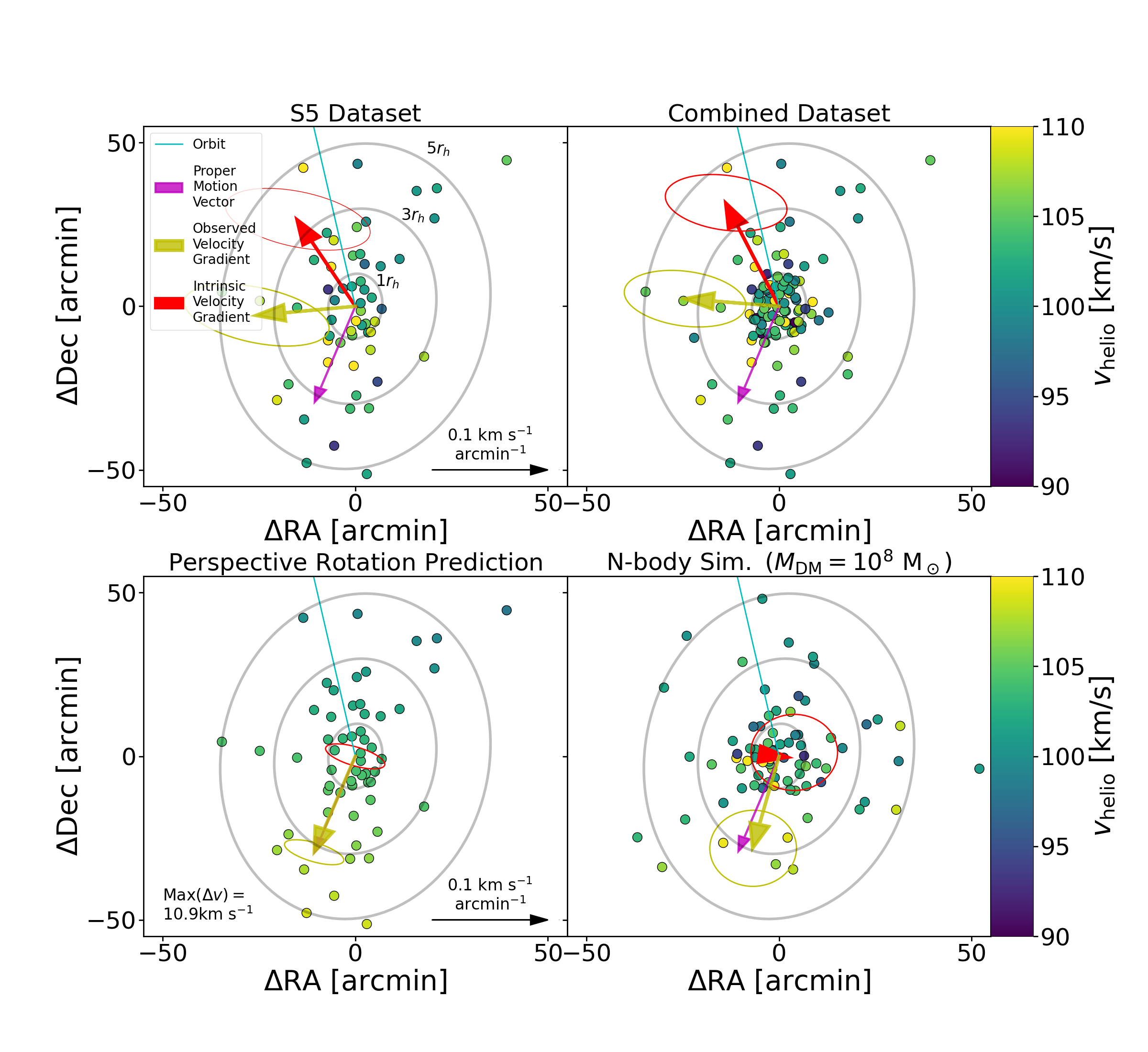}
    \caption{
        Each panel shows the on-sky spatial distribution of \BooI\ members colored by their heliocentric l.o.s.\ velocity. The orbit and proper motion vector of \BooI\ are represented in each panel by the cyan line and magenta arrow respectively, while the observed and intrinsic velocity gradient are represented by the yellow and red arrows respectively. Yellow and red ellipses show the 1$\sigma$ uncertainties on the corresponding velocity gradient vectors. The top two panels show the member stars and inferred velocity gradients for the \Sfive\ (left) and combined datasets (right). The bottom left panel illustrates the velocity gradients predicted from solid body perspective rotation in the absence of intrinsic velocity gradients. The bottom right panel illustrates the velocity gradients predicted by mock \Sfive-like observations of an N-body simulation of \BooI.
    } 
    \label{fig:vel_grad}
\end{figure*}

In the top two panels of Figure \ref{fig:vel_grad}, we illustrate the observed velocity gradients inferred from the \Sfive\ (top left) and combined (top right) \BooI\ datasets. The median and 1$\sigma$ uncertainties on the observed heliocentric l.o.s.\ velocity gradient are represented by the yellow arrows and ellipses respectively. High-probability member stars are included as circles color-coded by their heliocentric l.o.s.\ velocity. The on-sky proper motion vector is included as a magenta arrow, while the galaxy's orbit as calculated with \texttt{Galpy}\footnote{\url{http://github.com/jobovy/galpy}} \citep{bovy:2015} assuming the \texttt{MWPotential2014} is denoted by the cyan line. We stress that the observed proper motion vector is not aligned with \BooI's orbit because it has not been corrected for solar reflex motion (i.e., the Sun's motion with respect to \BooI).

As discussed briefly in Section \ref{sec:GMM_global}, we observe a shallow, marginally resolved heliocentric l.o.s.\ velocity gradient aligned roughly along \BooI's semi-minor axis in both the combined and \Sfive\ datasets. In the combined dataset, this gradient has a magnitude of $\Delta v_{\mathrm{B}} = 0.09\pm0.05$~km~s$^{-1}$ arcmin$^{-1}$ ($0.9\pm0.5$~km~s$^{-1}$ $r_h^{-1}$) and a position angle of $\theta_{\Delta v}=86^{+28}_{-19}$ degrees. We find similar results in the individual archival datasets, though with considerably larger uncertainties (see Appendix \ref{app:archival}).

It is important to note that the reported velocity gradients above are the observed velocity gradient and not the intrinsic or internal velocity gradient of \BooI\ as we have not yet accounted for projection effects related to the galaxy's relative motion. As discussed previously in the literature \citep[see][]{feast:1961, vandermarel:2002, kaplinghat:2008, walker:2008, pace:2020}, a satellite galaxy modeled as an extended solid body moving relative to the observer will exhibit an observed velocity gradient along its on-sky proper motion direction even in the absence of tidal disruption or intrinsic rotation simply because stars on one side of the galaxy will be moving towards or away from the observer faster than stars on the other. 

This ``perspective rotation" is more pronounced for systems that are nearby, have higher proper motions, and are more extended on the sky. As a result, it has primarily been considered in the analysis of classical dSph satellites due to their larger sizes (e.g., \citealt{kaplinghat:2008, walker:2008, pace:2020}). Recently, \citet{ji:2021} demonstrated the importance of accounting for perspective rotation when investigating the tidal disruption of the ultra diffuse galaxies Antlia~2 and Crater~2. At a distance of only $\sim$66~kpc, the extent of our \BooI\ sample ($\sim7r_h$) spans roughly $2^{\circ}$ on the sky, which results in a non-negligible perspective rotation signature. Here we explore the impact of that effect on \BooI's intrinsic velocity gradient. To our knowledge, this is the first consideration of perspective rotation in any UFD.

In the bottom left panel of Figure \ref{fig:vel_grad}, we illustrate the observed velocity gradient we would expect to observe in \BooI\ from this perspective rotation alone. As expected, this observed velocity gradient is oriented along \BooI's proper motion vector. The magnitude of this effect in \BooI\ is $0.105\pm0.004$ km s$^{-1}$ arcmin$^{-1}$, which results in a non-negligible 10.9~\kms\ difference in the heliocentric l.o.s.\ velocity over the 2-degree FoV of our sample. Indeed, we find from additional tests on bootstrapped data that in the absence of an intrinsic velocity gradient and given \BooI's intrinsic velocity dispersion and \Sfive's sample and velocity uncertainties, we would recover a non-zero gradient due to perspective rotation 99\% of the time.

In each panel of Figure \ref{fig:vel_grad}, we depict the intrinsic velocity gradients (i.e., with perspective rotation removed) and their uncertainties as red arrows and ellipses respectively. For the perspective rotation prediction (bottom left), the intrinsic velocity gradient is zero by construction. For the combined dataset, the intrinsic velocity gradient has a slope of $\Delta v_{\mathrm{B}'} = 0.12^{+0.04}_{-0.03}$ km s$^{-1}$ arcmin$^{-1}$ ($1.2^{+0.4}_{-0.3}$~km~s$^{-1}$ $r_h^{-1}$) and a position angle of $\theta_{\Delta v'}=34^{+9}_{-39}$ degrees. 
Similar, albeit less certain, intrinsic velocity gradients are found for the individual archival datasets (see Appendix \ref{app:archival}).

\subsubsection{Comparison to Literature}
Our results for the observed velocity gradient are in stark contrast with the observed velocity gradient reported by \citetLongeard\ of $\Delta v_{\mathrm{B}} = 0.40\pm0.10$ km s$^{-1}$ arcmin$^{-1}$ along \BooI's semi-major axis, which is equivalent to $\Delta v_{1,\mathrm{B}} = -0.39\pm0.10$ km s$^{-1}$ arcmin$^{-1}$ and $\Delta v_{2,\mathrm{B}} = -0.10\pm0.02$ km s$^{-1}$ arcmin$^{-1}$ if decomposed into two components along $\phi_{1*}$ and $\phi_{2*}$ as in our model. Notably, their velocity gradient predicts a difference in the mean velocity from one end of our \BooI\ sample to the other of $\sim$40 \kms. Because  \citetLongeard\ do not correct for perspective rotation, which acts opposite to their observed gradient, the intrinsic velocity gradient implied by their result would be even larger, $\Delta v_{\mathrm{B}}' \sim 0.50$ km s$^{-1}$ arcmin$^{-1}$.

We are unable to explain this disagreement as we cannot reproduce their reported velocity gradient when applying our methods to their provided catalog of member stars. We also find no evidence of the large change in mean velocity across either dataset that would be predicted by their gradient, though it should be easily detectable given the typical velocity uncertainty of the data and intrinsic velocity dispersion of \BooI. %\citetLongeard\ report that the inferred velocity dispersion of \BooI\ decreases from 5.8 \kms\ to 4.5 \kms\ with the inclusion of the velocity gradient, but we are skeptical that such a large velocity gradient would not result in an even larger difference. For example, if a $0.40\pm0.10$ km s$^{-1}$ arcmin$^{-1}$ velocity gradient were unaccounted for in our combined dataset, we would expect to recover velocity dispersion of $\sim$8 \kms.

It is possible that the inclusion of additional archival data sources \citep[e.g., from][]{martin:2007, lai:2011, gilmore:2013, ishigaki:2014, frebel:2016} or the presence of binary systems in the data analyzed by \citetLongeard\ is the source of the very steep gradient they infer. It is also possible that the strength of their velocity gradient is biased by foreground stars in the outskirts of their sample. We note that the binned kinematic analysis of \citetLongeard\ appears consistent with a flat or shallow velocity gradient for $r/r_h\lesssim2$ and only appears steeper in their outer-most radial bins, which are characterized by small sample sizes, substantially larger velocity dispersions than the inner radial bins ($\sim$10 vs.\ $\sim$4 \kms), and large $\sim$5--10 \kms\ uncertainties on both the mean velocity and velocity dispersion.

%Another possibility is that the mean velocity profile of \BooI\ is better described by a non-linear function that steepens with increasing radius as is seen in the $N$-body simulation discussed in Section \ref{sec:vel_grad_sim} (albeit on a larger spatial scale than found by \citetLongeard). If this were the case for \BooI, we would expect to see our recovered velocity dispersion increase at large radii due to the poor fit of our linear model. However, as shown in Section \ref{sec:vel_disp_prof}, our dataset actually favors a velocity dispersion that decreases at large radius. For these reasons, we believe that the shallower velocity gradients that we infer from the \Sfive\ and combined datasets are the more accurate results. Larger samples of member stars at all radii, but especially in \BooI's outskirts, would enable a more robust study of the velocity gradient's radial dependence.

\subsubsection{Comparison to Simulations}
\label{sec:vel_grad_sim}
In order to compare our observational results to expectations from dynamical simulations, we apply similar modeling techniques to an $N$-body simulation of \BooI. As described in Section \ref{sec:data_sim}, we adopt median \Sfive\ observational velocity uncertainties ($\sim$4 \kms) and down-sample to match the spatial distribution of member stars in the \Sfive\ dataset. We repeat our analysis with 10 bootstrapped samples in order to quantify the uncertainty of our inference on the sample size.

As shown in the bottom right panel of Figure \ref{fig:vel_grad}, we infer an observed velocity gradient of $0.10\pm0.02$ \kms\ arcmin$^{-1}$ aligned with proper motion vector of \BooI. This matches our expectations for perspective rotation in the absence of an intrinsic velocity gradient (bottom left panel) but not what we infer for the observational data (top two panels).
The lack of a detectable intrinsic velocity gradient, however, does not imply that the simulated \BooI\ has not experienced tidal disruption, nor does it imply that the simulated \BooI\ is without a velocity gradient. Indeed, when inspecting the full $N$-body simulation prior to down sampling (see Figure \ref{fig:nbody}), there is evidence of tidal debris along the simulated orbit. These tidal features, however, are diffuse and located beyond the footprint of our dataset. This is consistent with our earlier finding that $r_t \gtrsim8~\text{kpc}$.

\begin{figure*}[ht!] 
    \centering
	\includegraphics[width=1.0\textwidth]{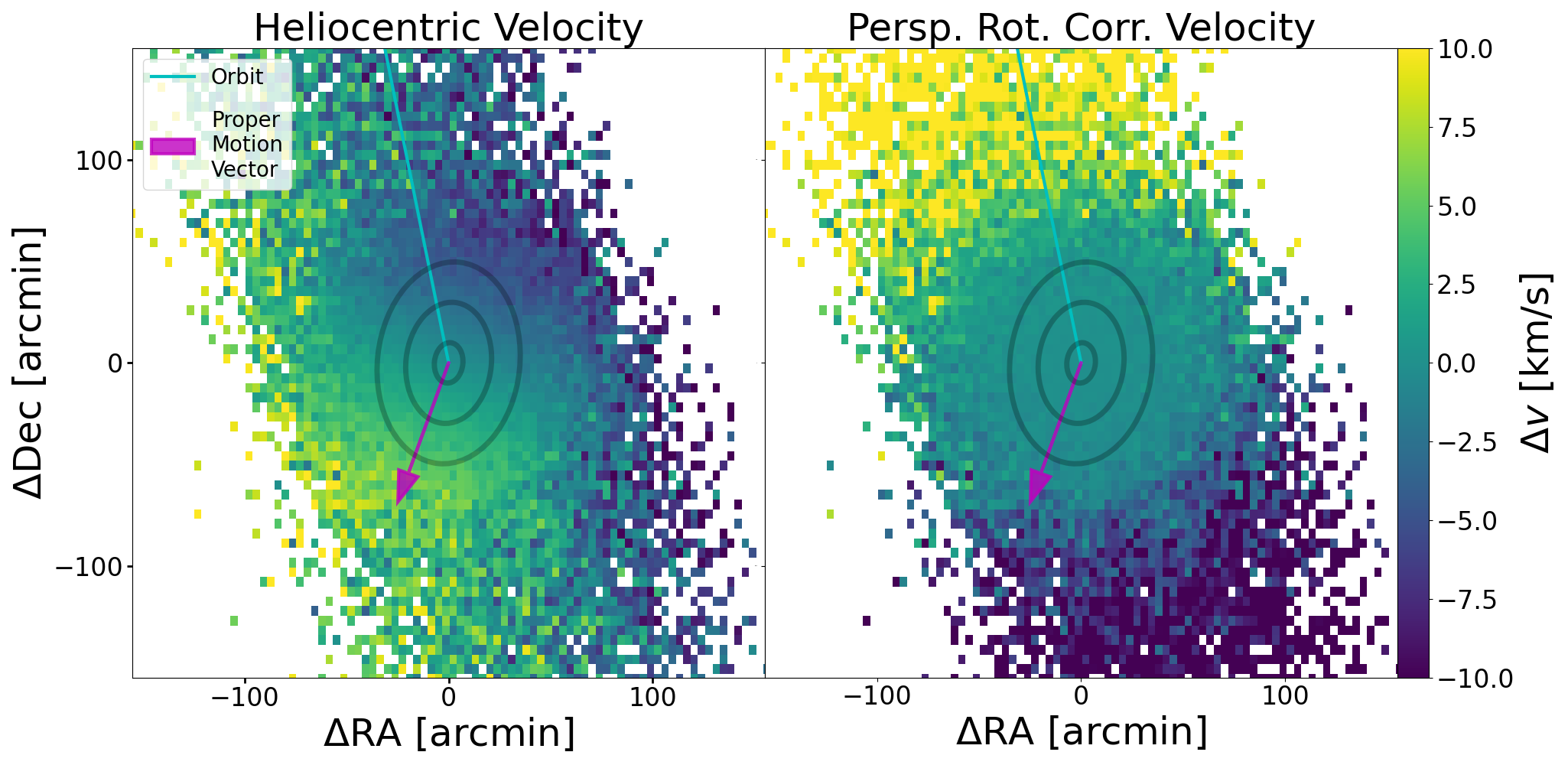}
    \caption{
        Velocity field of the \BooI\ $N$-body simulation without (left) and with (right) accounting for the effect of perspective rotation. In the left-hand panel, the velocity gradient is dominated by the perspective rotation, while in the right-hand panel the velocity gradient is dominated by tidal disruption at large radii and approximately flat within 7$r_h$.
    } 
    \label{fig:nbody_vel}
\end{figure*}

In Figure \ref{fig:nbody_vel}, we illustrate the line-of-sight velocity field of the $N$-body simulation. In the left panel, we show the median observed heliocentric l.o.s.\ velocity of $N$-body particles across the simulated \BooI. Here we can see that the velocity gradient within 5–10$r_h$ is dominated by perspective rotation and oriented along the simulated galaxy's proper motion vector. In the right panel, we show the median heliocentric l.o.s.\ velocity corrected for perspective rotation. With this correction applied, a clear velocity gradient aligned with the galaxy's tidal features and orbital motion is visible. While the magnitude of this gradient at large radii is similar to what we infer from our observations of \BooI, the velocity field is approximately constant within the inner $\sim$7$r_h$, which is why we do not detect it in our mock sample. We also do not find evidence of a tangential velocity gradient (i.e., in proper motions space) within the inner region of the simulated galaxy.

The difference between the results inferred from the observed dataset and the simulated dataset may suggest one or more of the following. It is possible that the intrinsic velocity gradient inferred from our observational data is not due to tidal disruption as initially expected but is instead evidence of rotation, either intrinsic or as the result of a dwarf-dwarf merger \citep[e.g.,][]{frebel:2016}. Alternatively, it is possible that the $N$-body simulation is less prone to disruption than \BooI, perhaps because it assumes a spherical rather than triaxial halo \citep[e.g.,][]{moore:2004}. To test these hypotheses, additional dynamical modeling is necessary, which we will undertake in a future study.

\subsubsection{Tides vs.\ Rotation}
\label{sec:vel_grad_discussion}
As introduced previously, a velocity gradient may be the result of intrinsic disk-like rotation, merger-induced rotation, and/or tidal disruption.
While the inferred intrinsic velocity gradient appears to be most closely aligned with \BooI's orbit, the uncertainties on the direction of the intrinsic velocity gradient vector and the near alignment of \BooI's semi-major axis and orbital motion make it difficult to confidently discriminate between a tidal or rotational origin of the gradient from the orientation alone. Of course, the alignment of \BooI's semi-major axis with its orbit is itself an indication of potential tidal disruption as noted by previous studies \citep[e.g.,][see also \citealt{munoz:2008} for additional discussion of elongation and tides]{munoz:2018, pace:2022, longeard:2022}. Moreover, oblate rotation in UFDs is exceedingly rare, with only Hydrus 1 showing tentative signs of disk-like rotation \citep{koposov:2018}. Given \BooI's early infall 7--10 Gyr ago \citep[e.g.,][]{rocha:2012, miyoshi:2020, barmentloo:2023}, it seems unlikely that it would have maintained any intrinsic disk-like morphology or rotation due to the effects of tidal stirring \citep[e.g.,][]{mayer:2001, kazantzidis:2011}. 

On the other hand, the lack of a velocity gradient in the inner $\sim$7$r_h$ of the simulated dataset complicates the conclusion that \BooI's velocity gradient is of tidal origin. However, as discussed in Section \ref{sec:vel_grad_sim}, it is possible that the $N$-body simulation experiences less tidal disruption than \BooI\ as a result of modeling choices (e.g., a spherical NFW dark matter density profile). To further investigate the potential tidal origins of the velocity gradient, we calculate \BooI's tidal radius at pericenter following the approach of \citet{pace:2025}. Assuming a flat rotation curve for the MW and a circular orbit for \BooI, the tidal radius at pericentric passage can be estimated as $r_t=r_\text{peri}(M_\text{BooI}/2M_\text{MW})^{1/3}$, where $r_\text{peri}$, $M_\text{BooI}$, and $M_\text{MW}$ are the pericentric distance, \BooI\ mass, and enclosed MW mass at pericenter respectively \citep{king:1962}. We adopt $r_\text{peri}=35~\text{kpc}$ from \citet{battaglia:2022a} and estimate the enclosed MW mass using the \texttt{MWPotential2014} potential from \citet{bovy:2015}. For the mass of \BooI, we integrate the dark matter density profile inferred in Section \ref{sec:jeans_modeling} out to the scale radius of the dark matter halo ($\sim$1~kpc). This yields a tidal radius of $r_t\sim2.4$~kpc, roughly 13 times \BooI's half-light radius. Accounting for 1$\sigma$ uncertainties in the dark matter density profile parameters, the tidal radius could be as small as $r_t=1.5$~kpc or $7.9r_h$. A tidal radius of $\sim$8--13$r_h$ is approximately consistent with the radius at which the velocity gradient becomes apparent in the simulated dataset (Figure \ref{fig:nbody_vel}). It thus seems unlikely that any of the \BooI\ members identified in this work are unbound, being actively stripped away, or are otherwise strongly affected by tidal forces. If this is true, then tides alone cannot explain \BooI's velocity gradient. To more definitively test the potential tidal origins of \BooI's velocity gradient, a dedicated search should be made for stripped extra-tidal member stars.
    
Another possibility is that \BooI's elongation is the result of a past merger as argued by \citet{frebel:2016}, which could have imparted a rotational signature in the galaxy's outskirts.  %However, we find no evidence of multiple kinematic or chemical populations in our datasets as one might expect from such a merger. In addition, 
However, this scenario would require the chance alignment of the merger event with \BooI's orbital motion. Nevertheless, it is difficult to rule out this scenario without either a larger sample of stars or additional measurements of stellar chemistry (e.g., $\alpha$ and/or neutron-capture elements) from high-resolution spectroscopy of stars in \BooI's outskirts. Such observations would need to be sufficiently precise to distinguish between the presence of multiple stellar populations and the existence of smooth radial gradients in \BooI's chemodynamic properties.
    
Unfortunately, even with the larger combined dataset presented in this work, \BooI's velocity gradient remains enigmatic and of uncertain origin. To confidently discrimination between rotational (either intrinsic or merger-induced) and tidal origins additional observations of \BooI\ at large radii are required.

\subsection{Velocity Dispersion Profile}
\label{sec:vel_disp_prof}
The radial velocity dispersion profile of a dwarf galaxy provides valuable insight into its dynamical evolution, including the presence of multiple dynamic populations \citep[e.g.,][]{pace:2020}, the impacts of tidal disruption \citep[e.g.,][]{read:2006, penarrubia:2008}, and the shape of its underlying dark matter halo \citep[e.g.,][]{lokas:2002, walker:2008, hayashi:2020}. In this section, we perform a simple binned analysis to investigate \BooI's empirical velocity dispersion profile. An un-binned analysis providing updated constraints on \BooI's dark matter profile is presented in Section \ref{sec:jeans_modeling}.

For each of our datasets, we separate our sample of likely member stars with good velocities into three radial bins: $r/r_h < 1$, $1\leq r/r_h < 3$, and $r/r_h \geq 3$. In the \Sfive\ dataset the number of stars in each bin is approximately equal with $\sim$20 stars in each bin, whereas for the combined dataset, the number of stars decreases from $\sim$60 in the inner bin to $\sim$20 in the outer bin, owing to the large contribution of member stars within one half-light radius from archival VLT observations. From the sample of likely member stars, we subtract out the systemic velocity inferred from our GMM (including the observed velocity gradient), and fit for the velocity dispersion in each bin. %We perform this analysis both with and without the ambiguous binary sources 230834410380273664 and 123083358574598796.

\begin{figure*}[ht!] 
	\includegraphics[width=1.0\textwidth]{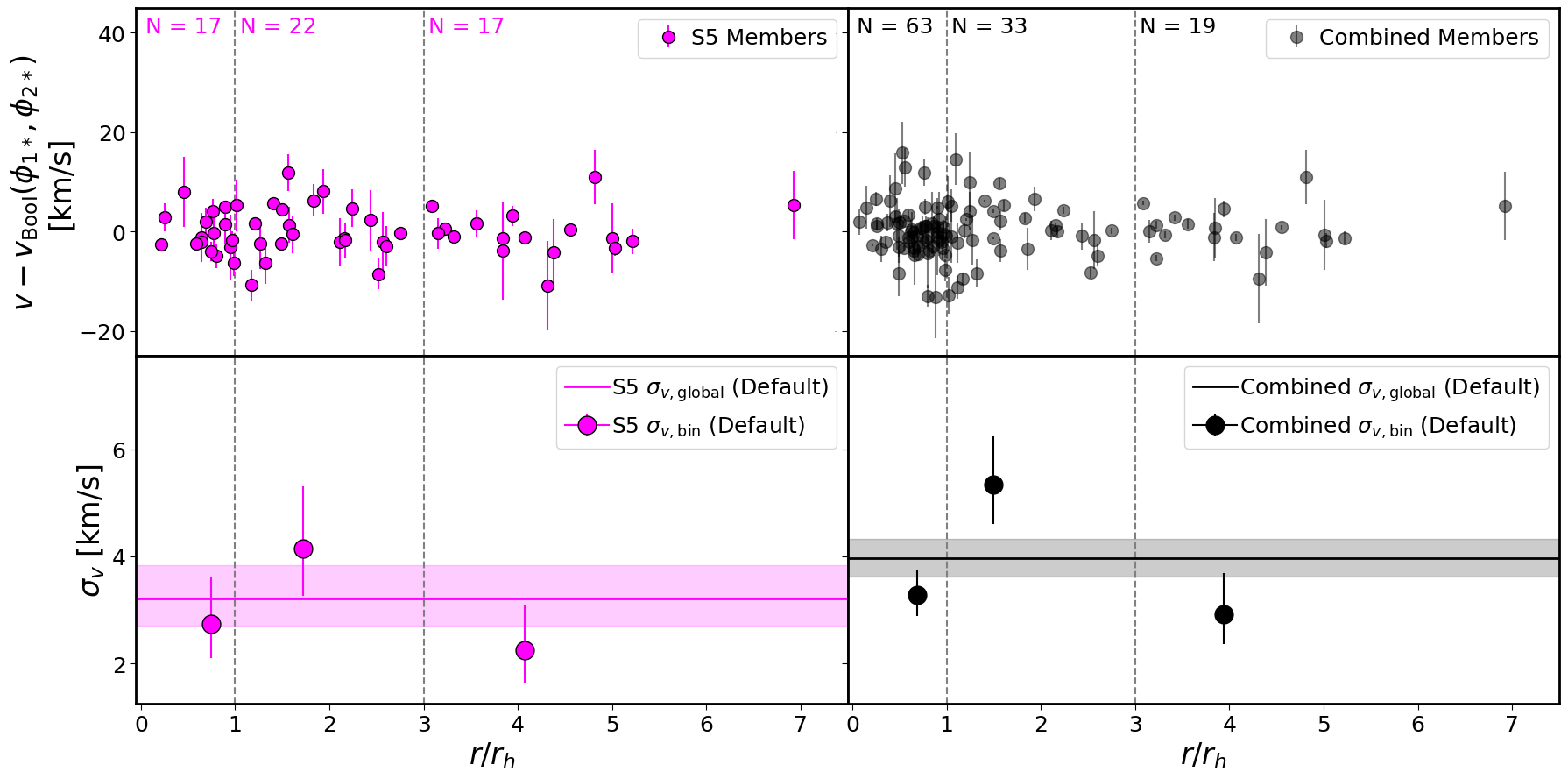}
    \caption{
        \textbf{Top.} Dispersion of heliocentric l.o.s.\ velocities around the inferred systemic velocity gradient as a function of radius for high probability \BooI\ members in the \Sfive\ (left; magenta circles) and combined (right; black circles) datasets. %As in Figure \ref{fig:S5_4panel}, we highlight the presence of ambiguous binary sources 1230834410380273664 and 1230833585745987968 (yellow and cyan circles respectively).
        \textbf{Bottom.} Velocity dispersion of \BooI\ members in radial bins of $r/r_h < 1$, $1\leq r/r_h < 3$, and $r/r_h \geq 3$ for the \Sfive\ (left; magenta circles) and combined (right; black circles) datasets. %Open circles denote the velocity dispersion of the inner bin when sources 1230834410380273664 and 1230833585745987968 are excluded from the analysis. 
        Markers are plotted at the median radius of the stars in each bin. For comparison, we include the global velocity dispersion inferred from the data (solid lines and shaded regions).
    } 
    \label{fig:vdisp_grad}
\end{figure*}

Figure \ref{fig:vdisp_grad} displays the results of this binned velocity dispersion analysis. The top two panels display the residual heliocentric l.o.s.\ velocities of the \Sfive\ (left) and combined (right) samples. In the bottom 2 panels, we plot the inferred velocity dispersion of each bin. % with (filled circles) and without (open circles) sources 230834410380273664 and 123083358574598796. 
The global velocity dispersions inferred from our GMM analysis (see Section \ref{sec:GMM_global}) are included as horizontal lines for reference. Analogous results for the individual archival datasets are presented in appendix Figure \ref{fig:comb_disp_grad} and discussed in Appendix \ref{app:archival}.

While a slightly lower global velocity dispersion is inferred from the \Sfive\ dataset compared to the combined dataset ($3.2^{+0.6}_{-0.5}$~km~s$^{-1}$ vs. $4.0^{+0.4}_{-0.3}$~km~s$^{-1}$), both datasets exhibit a similarly shaped velocity dispersion profile. For the combined (\Sfive) dataset, the velocity dispersion within one half-light radius is $3.2^{+0.5}_{-0.4}$~km~s$^{-1}$ ($2.7^{+0.8}_{-0.6}$~km~s$^{-1}$). This increases to $5.3^{+0.9}_{-0.7}$~km~s$^{-1}$ ($4.2^{+1.2}_{-0.8}$~km~s$^{-1}$) between one and three half-light radii and decreases to $2.9^{+0.8}_{-0.6}$~km~s$^{-1}$ ($2.2^{+0.8}_{-0.6}$~km~s$^{-1}$) beyond three half-light radii.

% The binned velocity dispersion of the default \Sfive\ dataset indicates a decreasing velocity dispersion with increasing radius such that the velocity dispersion in the outermost bin is well below the global velocity dispersion inferred by the GMM analysis. When sources 230834410380273664 and 123083358574598796 are removed from the binned analysis, the velocity dispersion of the inner bin decreases to be consistent with that of the outer bin, indicating a potential peak in the velocity dispersion at $\sim$2 $r_h$, though a flat dispersion profile is also consistent with the data. Meanwhile the binned velocity dispersion profile of the combined dataset also suggests a potential peak in the dispersion profile at $\sim$2 $r_h$ both with and without the inclusion of sources 230834410380273664 and 123083358574598796, though it is more prominent in the latter case. 

The peak in the velocity dispersion profile between one and three $r_h$ is qualitatively consistent with the radially increasing velocity dispersion reported by \citetLongeard. \citetLongeard, however, infers a much larger velocity dispersion ($\sim$ $10\pm5$ \kms) at $\sim$3 $r_h$, perhaps due to the smaller number of members and increased foreground contamination %in their sample at this radius 
or perhaps related to their treatment of the velocity gradient.
% L22 has enormous error bars at large radii owing to small number statistics. May be related to a degenercy with the large velocity gradients they find, as they fit a GMM to each radius bin.

%\elaborate{Provide brief interpretation of the velocity dispersion profile. Note that a larger sample and more precise velocity measurements would better discriminate between a dispersion profile that is flat and one that peaks and turns over. Acquiring additional radial velocity measurements of stars 1230834410380273664 and 123083358574598796 would }

\subsection{Metallicity Gradient and Dispersion Profile}
\label{sec:feh_grad}

The metallicity gradients of dwarf galaxies have long been a subject of interest as they are thought to encode a variety of physical processes, including secular inside-out evolution, stellar feedback-induced radial migration, mergers, and tidal stripping \citep[e.g.,][]{schroyen:2013, pontzen:2014, benitez-llambay:2016, revaz:2018, mercado:2021, tarumi:2021}. While metallicity gradients are commonplace and varied in massive MW and M31 satellites, they have only been detected in four UFDs. \citet{chiti:2021} detected a gradient of $\Delta\text{[Fe/H]}\sim-0.1~\text{dex}~r_h^{-1}$ along with an extended stellar halo in Tucana II, which is thought to be the product of a major merger \citep[e.g.,][]{tarumi:2021, chiti:2023}. \citet{fu:2024} detected gradients of $\Delta\text{[Fe/H]}\sim-0.2~\text{dex}~r_h^{-1}$ and $-0.5~\text{dex}~r_h^{-1}$ in Andromeda XVI and XXVIII, which are consistent with radial migration driven by their extended star formation histories. Lastly, \citetLongeard\ detected a shallow gradient of $\Delta\text{[Fe/H]}\sim-0.08~\text{dex}~r_h^{-1}$ in \BooI.

\begin{figure*}[ht!] 
	\includegraphics[width=1.0\textwidth]{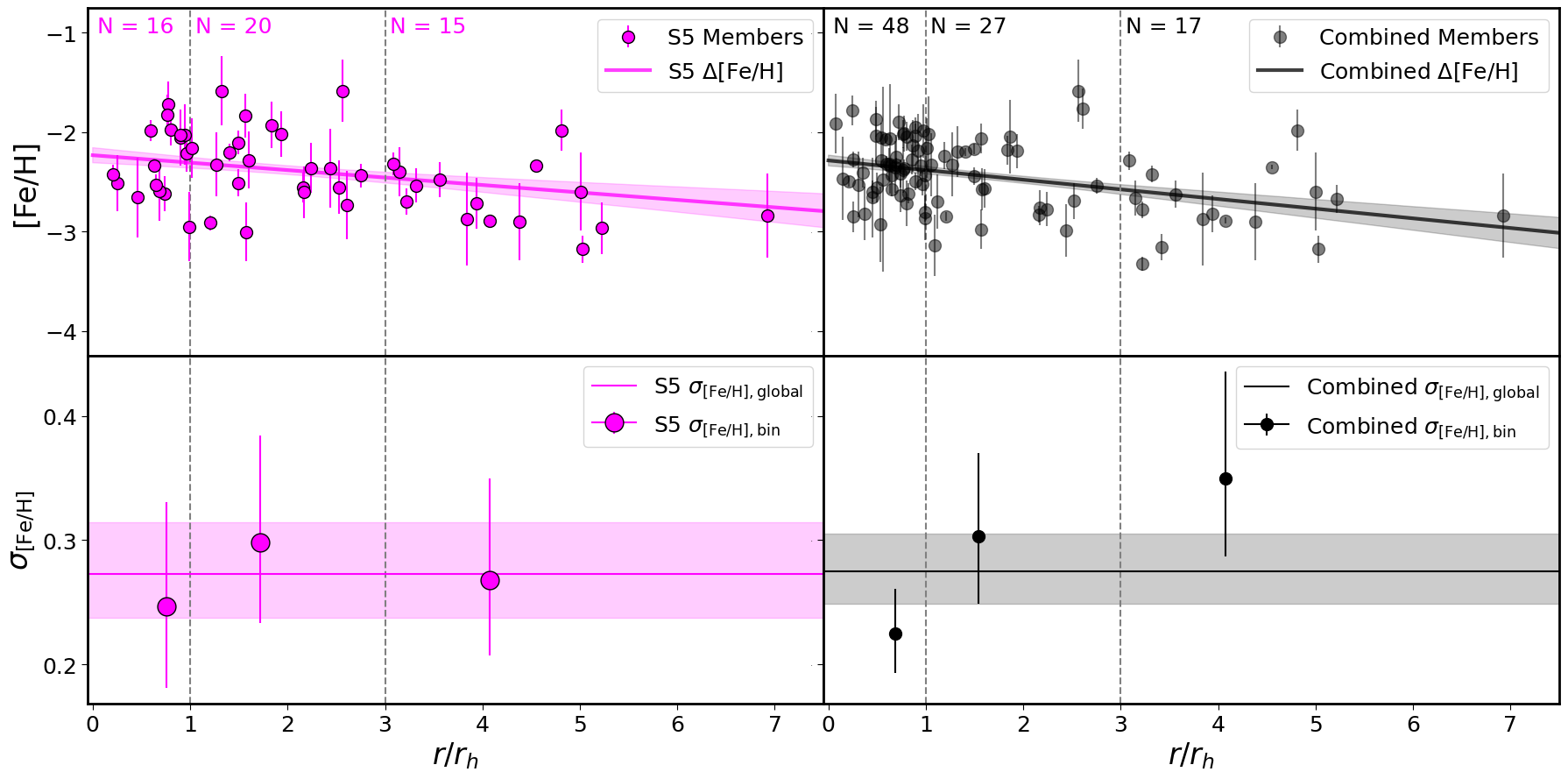}
    \caption{
        \textbf{Top.} 
        Distribution of stellar metallicity as a function of radius for high probability \BooI\ members in the \Sfive\ (left; magenta circles) and combined (right; black circles) datasets. The inferred metallicity gradients and their 1$\sigma$ uncertainty are represented by the solid line and shaded region.
        \textbf{Bottom.} Metallicity dispersion of \BooI\ members in radial bins of $r/r_h < 1$, $1\leq r/r_h < 3$, and $r/r_h \geq 3$ for the \Sfive\ (left; magenta circles) and combined (right; black circles) datasets. For comparison, we include the global metallicity dispersions inferred from these datasets (solid lines and shaded regions).
    } 
    \label{fig:fehdisp_grad}
\end{figure*}

In the top two panels of Figure \ref{fig:fehdisp_grad}, we illustrate the metallicity gradients inferred from the \Sfive\ (top left) and combined (top right) \BooI\ datasets. As discussed briefly in Section \ref{sec:GMM_global}, we recover a weak, but resolved, negative metallicity gradient of $\Delta\mathrm{[Fe/H]}_{\mathrm{B}} = -0.008\pm0.003$ dex arcmin$^{-1}$ for the \Sfive\ dataset and $\Delta\mathrm{[Fe/H]}_{\mathrm{B}} = -0.010\pm0.003$ dex arcmin$^{-1}$ for the combined dataset. Measured in terms of \BooI's half-light radii, these are equivalent to $\sim-0.08~\text{dex}~r_h^{-1}$ and $\sim-0.10~\text{dex}~r_h^{-1}$ respectively, which is in good agreement with the gradient previously reported by \citetLongeard. We find similar results for the individual archival datasets as well (see Appendix \ref{app:archival}).

In the bottom two panels of Figure \ref{fig:fehdisp_grad}, we display the binned metallicity dispersion profile of \BooI, following the same approach as done for the velocity dispersion profile in the previous section. We find that both the \Sfive\ and combined datasets are consistent with a flat metallicity dispersion profile. The combined dataset hints at an increasing metallicity dispersion with radius, but the uncertainties are large enough to be 1$\sigma$ consistent with the inferred global metallicity dispersion. 

The similarity between \BooI\ and Tucana II in both metallicity gradient and extended stellar population, suggests that \BooI, like Tucana II, may have experienced a past major merger event as proposed by \citealt{frebel:2016} and \citealt{tarumi:2021}. However, it is worth noting that when compared with more massive classical dwarfs of a similar age, \BooI's metallicity gradient is similar to those without evidence of a major merger (e.g., Cetus, Draco, Tucana, Pegasus) and shallower than those that do (e.g., Phoenix, Sextans; \citealt{taibi:2022}). Of course, any number of confounding factors, including tidal stripping, could have played a subsequent role in flattening a merger-created gradient.

Whether the product of a merger or secular evolution, it is quite likely that the dearth of resolved metallicity gradients in UFDs is at least in part driven by the limited samples of metallicity measurements at large radii. Determining the origin of these gradients will require additional high-precision metallicity and individual elemental abundance measurements beyond $\sim3r_h$.

\section{Dark Matter Density Profile}
\label{sec:jeans_modeling}
As a nearby dark matter-dominated galaxy, \BooI---like other UFDs---has been the subject of multiple dynamic studies concerned with the nature and distribution of dark matter \citep[e.g.,][]{pace:2019, hayashi:2021a, hayashi:2021b, hayashi:2023, horigome:2023}. However, each of these past analyses have been limited to the archival VLT datasets of \citetKoposov\ or \citetJenkins, which only trace the stellar dynamics within one half-light radius and feature higher contamination from binary systems. In this section, we revisit the dark matter content of \BooI, using our larger and more spatially extended \Sfive\ and combined datasets. 

Specifically, we apply the axisymmetric Jeans mass modeling methods developed in \citet{hayashi:2012}, \citet{hayashi:2015}, and \citet{hayashi:2020}. These methods were previously applied to MW UFDs, including \BooI, in \citet{hayashi:2023}. A detailed description and discussion of this methodology is provided in the aforementioned papers, and an abridged summary is provided in Appendix \ref{app:jeans}. In short, this mass modeling assumes that \BooI\ resides in a dark matter halo with a generalized axisymmetric Hernquist profile 
\citep[see Equations \ref{eq:dm_density1} and \ref{eq:dm_density2};][]{hernquist:1990, zhao:1996}
that is in dynamical equilibrium such that the motions of its stars can be described by the axisymmetric Jeans equations with a constant velocity anisotropy 
\citep[see Equations \ref{eq:jeans1}--\ref{eq:jeans3}; e.g.,][]{cappellari:2008}. The stellar distribution is assumed to follow an axisymmetric Plummer profile 
\citep[see Equations \ref{eq:plummer1} and \ref{eq:plummer2}]{plummer:1911} that is aligned with the underlying dark matter halo. This mass model has 8 free parameters:
\begin{itemize}
    \item $Q$, the axial ratio of the axisymmetric dark matter halo.
    \item $b_\text{halo}$, the scale radius of the dark matter halo.
    \item $\rho_0$, the central density of the dark matter halo.
    \item $\beta_z$, the velocity anisotropy parameter.
    \item $\beta$, the outer slope of the dark matter halo.
    \item $\gamma$, the inner slope of the dark matter halo.
    \item $\alpha$, the sharpness of the dark matter slope transition.
    \item $i$, the inclination angle.
\end{itemize}
We fit this model to the un-binned velocity dispersion profile of our data assuming a Gaussian line-of-sight velocity dispersion. % according to the likelihood function given in Equation \ref{eq:jeans_like}. 
We adopt broad uniform priors (see Table \ref{tab:priors}) 
and estimate the posterior distribution using Markov chain Monte Carlo (MCMC) sampling.
In our analysis, we consider only the 56 (115) \BooI\ member stars in our \Sfive\ (combined) dataset that have good velocity measurements and are not flagged as RRL or binary stars. %As shown in Section \ref{sec:vel_disp_prof}, \BooI's velocity dispersion profile is sensitive to the inclusion of the ambiguous binary star candidates 230834410380273664 and 1230833585745987968. To test their impact on the recovered dark matter density profile, we repeat our analysis both with and without these sources.

The parameters inferred from our dynamical analysis are summarized in Tables \ref{tab:BooI_par} and \ref{tab:BooI_par_extra}. The posterior distributions are presented in Figure \ref{fig:jeans_corner}. Despite the expanded dataset, we find that $Q$, $\alpha$, $\beta$, and $i$ remain largely unconstrained and recover the well-known $b_\text{halo}$-$\rho_0$ and $Q$-$\beta_z$ degeneracies \citep[e.g.,][]{binney:2008, cappellari:2008, geringer-sameth:2015, hayashi:2015, hayashi:2020, hayashi:2023}. Mild degeneracies are also found for $\gamma$ with $\rho_0$. Nevertheless, the combined dataset is able to reasonably constrain $b_\text{halo}$ ($\log_{10}b_\mathrm{halo}=3.07^{+0.87}_{-0.45}$), $\rho_0$ ($\log_{10}\rho_{0}=-1.88^{+0.90}_{-1.51}$), and $\beta_z$ ($-\log_{10}(1-\beta_z)=0.25^{+0.21}_{-0.21}$). $Q$, $\alpha$, $\beta$, and $i$, however, remain largely unconstrained.

We find that the inner dark matter slope is weakly constrained by the combined (\Sfive) dataset to be $\gamma=1.00^{+0.52}_{ -0.60}$ ($1.27^{+0.48}_{ -0.74}$). Though this result favors a cuspy dark matter profile ($\gamma\gtrsim1$) consistent consistent with expectations from cold dark matter cosmology and our current understanding of galaxy evolution \citep[e.g.,][and references therein]{bullock:2017}, we are unable to completely rule out the presence of a core ($\gamma\sim0$).

\begin{figure*}[ht!] 
	\includegraphics[width=1.0\textwidth]{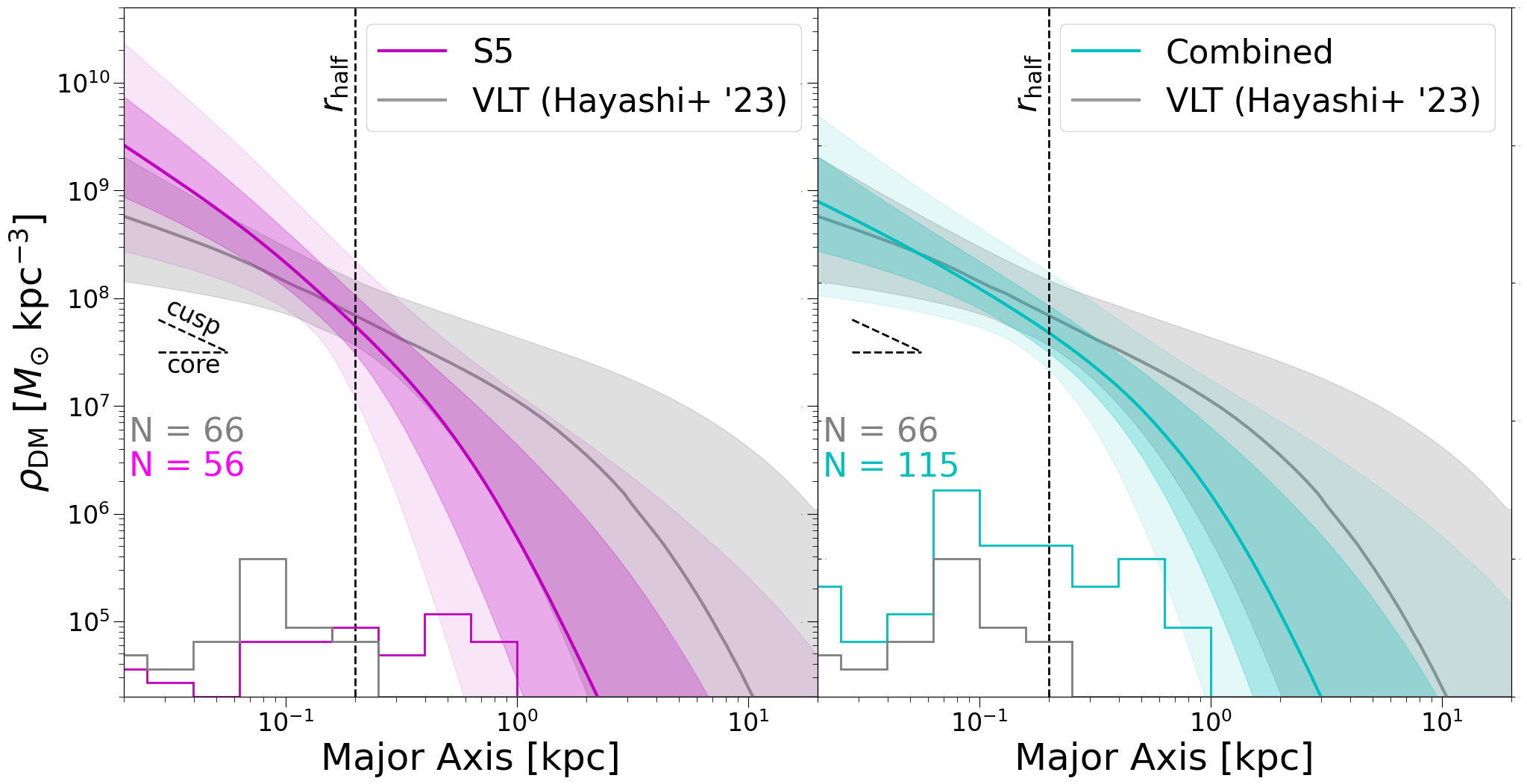}
    \caption{
        Inferred dark matter density profile along the major axis for the default \Sfive\ (left) and combined (right) datasets. %The top right and bottom right panels show the same for when the ambiguous binary sources 1230834410380273664 and 123083358574598796 are removed from the \Sfive\ and combined datasets respectively. 
        In each panel, the solid lines show the median value of posterior samples, while the dark and light shaded regions show the 68\% and 95\% confidence intervals respectively. The gray solid lines and shaded regions in each panel provide the median and 68\% confidence intervals from  \citet{hayashi:2023}, which only uses the archival VLT data. The vertical dashed lines indicate the \BooI's half-light radius from \citet{munoz:2018}. The inset histograms illustrate the spatial distribution of our samples (colored lines) along the semi-major axis compared to the previous VLT-only sample (gray line).
    }
    \label{fig:jeans_dm_profile}
\end{figure*}

In Figure \ref{fig:jeans_dm_profile}, we compare the dark matter profile inferred from our analysis of the \Sfive\ (left) and combined datasets (right) to the dark matter profile previously inferred by \citet{hayashi:2023} using VLT data alone (gray line and shaded region). The size and spatial distribution of each sample is illustrated by the inset histograms in each panel. Though our inferred parameters are consistent within $1\sigma$ uncertainties of those reported by \citet{hayashi:2023}, our analysis favors a cuspier and less spatial extended halo. This is in large part due to improved constraints on $b_\text{halo}$ and $\beta_z$ (see Table \ref{tab:BooI_par_extra}) enabled by our larger and more radially extended samples.

%As can be seen in Figure \ref{fig:jeans_dm_profile}, %even in the less constraining cases, 
%our results (colored lines) favor cuspier and less spatially extended halos than the previous results found by \citet{hayashi:2023} using VLT data alone ($\gamma=0.7^{+0.5}_{ -0.6}$; gray lines). This is in large part due to improved constraints on $b_\text{halo}$ and $\beta_z$ enabled by our more radially extended samples, which is illustrated by the inset histograms of each panel. %This is most pronounced for the default \Sfive\ dataset, which is to be expected given the centrally-peaked velocity dispersion profile found in Section \ref{sec:vel_disp_prof}. However, the consistency between the other three datasets in both the velocity dispersion profile and dynamical modeling suggest that these constraints are likely more robust. 

% Potential Extension

This analysis highlights that even with a doubling of the l.o.s.\ velocity measurements and one of the largest kinematics samples in any UFD, the shape of \BooI's dark matter halo remains poorly constrained. This result is inline with several recent theoretical studies have concluded that much larger samples (i.e., $\mathcal{O}(10^{3})-\mathcal{O}(10^4)$ stars) are required to discriminating between cusped and cored dark matter halos with Jeans modeling \citep[e.g.,][]{chang:2021, guerra:2023}.

\subsection{Jeans Modeling Limitations}
\label{sec:jeans_caveats}
% The presence of an intrinsic velocity gradient behooves some caution when interpretting these results as the axisymmetric model assumes dynamical equilibrium and doesn't model rotational motion.

There are two primary factors that complicate the interpretation of the dynamical modeling presented above, both of which are related to presence of a velocity gradient discussed in Section \ref{sec:vel_grad}. First, the model of \citet{hayashi:2023} used in this work adopts the steady-state Jeans equations, which assumes dynamical equilibrium. However, if the velocity gradient found in \BooI\ is indicative of tidal disruption, then this assumption is invalid. While the presence of tidal disruption does not necessarily bias or invalidate the results of dynamical modeling of dwarf galaxies, particularly if contamination from unbound stars is minimal as is likely the case for our dataset \citep[e.g.,][]{read:2018, genina:2020, deleo:2024, nguyen:2025}, we have not tested our model on simulations of galaxies impacted by tides. Thus we have not quantified what effects tides might have on the inferred dynamical parameters. Second, the model of \citet{hayashi:2023} used in this work assumes that the streaming motion of stars in UFDs is negligible and thus does not account for the possibility that \BooI\ may be rotating about its minor axis. In principle, this can be addressed by relaxing the assumption of zero streaming motion in the axisymmetric Jeans equations.

The results of our Jeans analysis are also limited by other modeling assumptions outlined in \citet{hayashi:2020} and \citet{hayashi:2023}, including the choice of priors, shape of the stellar distribution, and Gaussian line-of-sight velocity distributions. As discussed in these works and in the literature \citep[e.g.,][]{read:2021}, these assumptions may bias our results to favor cuspier dark matter profiles. Lastly, the presence of binaries in our data can inflate the velocity dispersion of \BooI. Though we have done our best to remove binaries, it is inevitable that there remain undetected binaries in our dataset.

In light of these limitations, we recommend that additional modeling of the \BooI\ mass profile be undertaken that more appropriately takes into account the potential presence of tidally induced disequilibrium and/or rotational motion. Additional radial velocity monitoring should also be undertaken to identify and confirm the binarity of stars in \BooI. Such analysis is beyond the scope of this paper.

\section{Chemical Evolution Modeling}
\label{sec:GCE}
The stellar metallicity distribution function (MDF) of a galaxy encodes valuable insight into the physical processes governing the galaxy's formation and chemical evolution (e.g., star formation, nucleosynthesis, stellar feedback, and gas in/outflows). As the most luminous nearby UFD, \BooI's well-sampled MDF has been the subject of numerous qualitative and quantitative analyses \citep[e.g.,][]{lai:2011, gilmore:2013, vincenzo:2014, romano:2015, webster:2015, romano:2019, lacchin:2020, rossi:2024}. The additional member stars identified in this work yield a substantial increase to the number of stars in \BooI\ with well-measured [Fe/H] abundances. With 51 [Fe/H] measurements, the \Sfive\ dataset is one of the largest homogeneous samples of [Fe/H] in \BooI\, while the combined dataset of 92 [Fe/H] measurements represent a $\sim$30\% increase over existing samples and the largest sample of spectroscopic metallicities in any UFD by a factor of $\sim$2. As such, a re-analysis of \BooI's chemical evolution is warranted. Here, we only consider measurements of [Fe/H] from our \Sfive\ and combined datasets, saving a joint fit with other individual elemental abundances (e.g., [C/Fe], [Mg/Fe]) measured from high-resolution observations \citep[e.g.,][]{norris:2010a, gilmore:2013, ishigaki:2014, waller:2023} for a future analysis.%

To constrain the chemical enrichment history of \BooI, we fit the MDF of \BooI\ with the analytic galactic chemical evolution (GCE) model of \citet{weinberg:2017} (hereafter \citetWAF), which was previously applied to the UFD Eridanus II by \citet{sandford:2024}. We provide a brief summary of the model and fitting methods in Appendix \ref{app:GCE}, and direct the reader to \citet{weinberg:2017} and \citet{sandford:2024} for a more detailed description and discussion of the model.
In short, this GCE model tracks the time evolution of elements in a fully-mixed, one-zone system undergoing gas accretion, star formation, prompt enrichment from core-collapse supernovae (CC SNe), delayed enrichment from Type Ia supernovae (SN Ia), and stellar feedback-driven galactic outflows. It is parameterized in terms of 4 parameters:
\begin{itemize}
    \item $\tau_\text{SFH}$, the star formation history (SFH) timescale assuming a linear-exponential form for the SFH ($\dot{M}_* \propto e^{-t/\tau_\text{SFH}}$).
    \item $t_\text{trunc}$, the time at which the SFH truncates (e.g., due to abrupt quenching from ram-pressure stripping or reionization).
    \item $\tau_\text{SFE}$, the star formation efficiency (SFE) timescale assuming a linear star formation law ($\tau_\text{SFE}=\text{SFE}^{-1}=M_g/\dot{M}_*$).
    \item $\eta$, the mass-loading factor describing the ejection of gas from the interstellar medium, assuming a linear scaling with the SFR ($\eta=\dot{M}_\text{outflow}/\dot{M}_*$).
\end{itemize}
To infer the values of these parameters, we adopt the likelihood function given by Equation \ref{eq:gce_like} and adopt weakly informative priors motivated by the CMD-based SFH from \citet[][see Table \ref{tab:priors}]{durbin:2025}. Sampling is performed using \texttt{pocoMC} in the same manner as described in Section \ref{sec:GMM_like}.

The GCE parameters of \BooI\ inferred from our analysis are summarized in Tables \ref{tab:BooI_par} and \ref{tab:BooI_par_extra}. Their posterior distributions are presented in Appendix \ref{app:GCE}. In Figure \ref{fig:MDF}, we present the results of our fit to the \Sfive\ (top left; magenta) and combined (top right; black) MDFs. The red dashed line shows the ``best-fit" model MDF predicted by the \textit{maximum a posteriori} GCE parameters from the sampling. To better compare the continuous and noiseless model MDF to the observed MDF, we make mock MDFs by drawing models repeatedly from the posterior distribution and sampling stars from their predicted MDF and the uncertainty distribution of the observed data. The median and 68\%/95\% confidence intervals of this posterior predictive check (PPC) are represented by the solid blue line and shaded regions. The bottom panels show the same data and models, except using the cumulative MDF (cMDF) to provide an alternative visualization. In both cases, we find excellent agreement between draws from our posteriors and the observed (c)MDFs.

\begin{figure*}[ht!] 
	\includegraphics[width=1.0\textwidth]{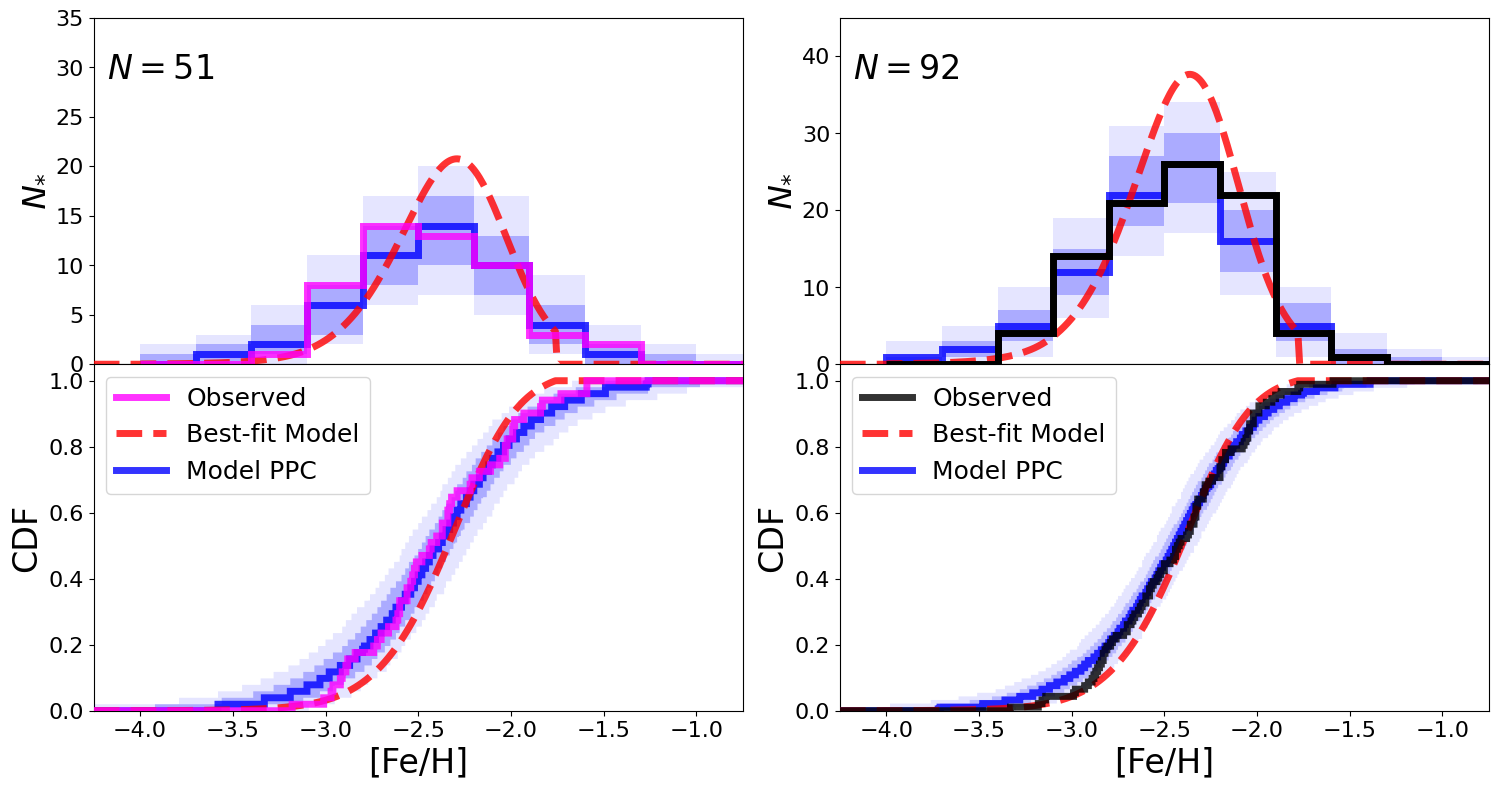}
    \caption{
        \textbf{Left.} MDF (top) and cMDF (bottom) of the \Sfive\ dataset (magenta lines). The dashed red lines show the MDF and cMDF corresponding to the \textit{maximum a posteriori} values from the GCE modeling, while the solid blue lines and shaded regions show the median and 68/95\% confidence intervals of the posterior sample when adopting characteristic observational uncertainties.
        \textbf{Right.} Same for the combined dataset with the observed MDF and cMDF represented by black lines. 
    } 
    \label{fig:MDF}
\end{figure*}

From both the \Sfive\ and combined MDFs, we infer the SFH timescale to be $\tau_\text{SFH}=0.2\pm0.1$ Gyr. While the data is sufficient to disfavor the shortest SFH timescales ($<150$~Myr), the inference of $\tau_\text{SFH}$ is dominated by the tight priors set by CMD-based SFH (80--350~Myr). We are also unable to recover a clear truncation of the SFH before 2 Gyr from either the \Sfive\ or combined dataset, though we are able to place 95\% confidence lower limits of $t_\text{trunc} > 0.4$ Gyr and $>0.5$ Gyr from the respective datasets. As previously discussed in the context of Eridanus II by \citet{sandford:2024}, the inability to detect or constrain a sharp truncation in the SFH is expected for systems with short SFH timescales as the abrupt cessation of star formation imprints only a subtle signature on the MDF after a few times the $\tau_\text{SFH}$.

For the SFE timescale, we infer $\tau_\text{SFE}=11.9^{+8.7}_{-5.1}$ Gyr and $13.9^{+7.6}_{-5.3}$ Gyr from the \Sfive\ and combined datasets respectively. This is equivalent to constraints on the SFE ($\tau_\text{SFE}^{-1}$) of $\text{SFE}=0.08^{+0.06}_{-0.04}$ Gyr$^{-1}$ and $0.07^{+0.04}_{-0.03}$ Gyr$^{-1}$. As can be seen in Figure \ref{fig:ChemEv_Corner}, the $\tau_\text{SFE}$ is mildly degenerate with $\tau_\text{SFH}$ such that a more extended period of star formation would require more inefficient star formation to match the data and vice versa.

Lastly, we infer a mass-loading factor of $\eta=170^{+35}_{-46}$ from the \Sfive\ MDF and $203^{+27}_{-36}$ from the combined MDF. In other words, for every 1 $M_\odot$ of star formation, $\sim$200 $M_\odot$ of gas is ejected from the ISM. We note that the \Sfive\ MDF allows for a lower mass-loading factor if $t_\text{trunc}\lesssim0.5$ Gyr, but this regime is strongly disfavored by the combined dataset.

\subsection{Comparison to Previous GCE Studies of \BooI}

\citet{lai:2011} and \citetJenkins\ previously analyzed MDFs of \BooI\ using three toy models: a ``pristine" or ``leaky box" model, a ``pre-enriched" model, and an ``extra gas" model (see \citealt{kirby:2011a} for details). Both studies found a slight preference for the extra gas model, indicating the importance of including the accretion of pristine gas over the star forming lifetime of \BooI. Due to the difference in parametrization of these models compared to the \citetWAF\ model, it is difficult to draw direct quantitative comparisons between these analyses and our own. However, we note that the extra gas model is the most qualitatively similar to the \citetWAF\ GCE model. Additionally, both \citet{lai:2011} and \citetJenkins\ infer very low values for their models' effective yield parameter, which is qualitatively consistent with the large mass-loading factor inferred in this work. A large degree of mass loss was also argued for by \citet{gilmore:2013}. % For the extra gas models, \citet{lai:2011} and \citetJenkins\ that $M_\text{inf}\sim4-6$ times the final stellar mass, which is much lower (I think) than we recover. This is probably because the effective yield parameter can offset the need for the extra dilution.

More complex, semi-analytic and numerical models have also been used to study the chemical enrichment of \BooI\ \citep[e.g.,][]{vincenzo:2014, romano:2015, lacchin:2020}. However, unlike in \citet{lai:2011}, \citetJenkins, and our own analysis, these previous studies did not perform a statistical fit to the data, but rather investigate a coarse grid of model parameters to see which matched the observed data most closely. \citet{vincenzo:2014}, \citet{romano:2015}, and \citet{lacchin:2020} all fix the mass-loading factor to $\eta\sim10$, which is much smaller than value we infer. %\footnote{\citet{romano:2015} assumes that the winds are metal-enriched, which could explain a difference in mass-loading factors by a factor of a few, but not a factor of 10 or more.}. 
For this fixed mass-loading factor, \citet{vincenzo:2014} and \citet{lacchin:2020} found that the stellar chemistry of \BooI\ was best reproduced with an SFE of 0.01 Gyr$^{-1}$, while \citet{romano:2015} found that slightly large SFE values between 0.013 and 0.053 Gyr$^{-1}$ produced the best results. Our inferred SFE is higher, though still consistent with \citet{romano:2015}, which we attribute to the higher mass-loading factor found in our analysis.
%All three studies found that $\sim$1--2$\times10^7$~M$_\odot$ of accreted gas was required to match the observed properties of \BooI.  %, which is _____ compared to what we infer (XXX).

Our GCE model, like that of \citet{romano:2015}, is unable to explain the absence of gas in \BooI\ at present day, requiring that an external process like re-ionization or ram pressure stripping remove the gas. This is in contrast to the findings of \citet{vincenzo:2014}, which found that stellar feedback is sufficient to expel \BooI's remaining gas.

In an investigation of potential Pop III enrichment scenarios in \BooI, \citet{rossi:2024} reported a mass-loading factor of $\eta=85\pm12$. The factor of $\sim$2 difference can easily be explained by differences in the adopted nucleosynthetic yields or the treatment of direct ejection of supernova products \citep[see][]{romano:2019, weinberg:2023}.
Lastly, our constraints on $\tau_\text{SFH}$ and $t_\text{trunc}$ are broadly consistent with previous analyses requiring the duration of star formation in \BooI\ to be at least 100 Myr \cite[e.g.,][]{gilmore:2013, webster:2015}.

\subsection{Comparison to Other Galaxies and Simulations}
Here we compare our inferences of \BooI's SFE and mass-loading factor to values measured or inferred for other galaxies across a range in stellar mass. Though the methods underlying each of these values vary substantially, a cohesive picture of how these galaxy properties correlate with stellar mass nevertheless emerges.

\begin{figure*}[ht!] 
	\includegraphics[width=1.0\textwidth]{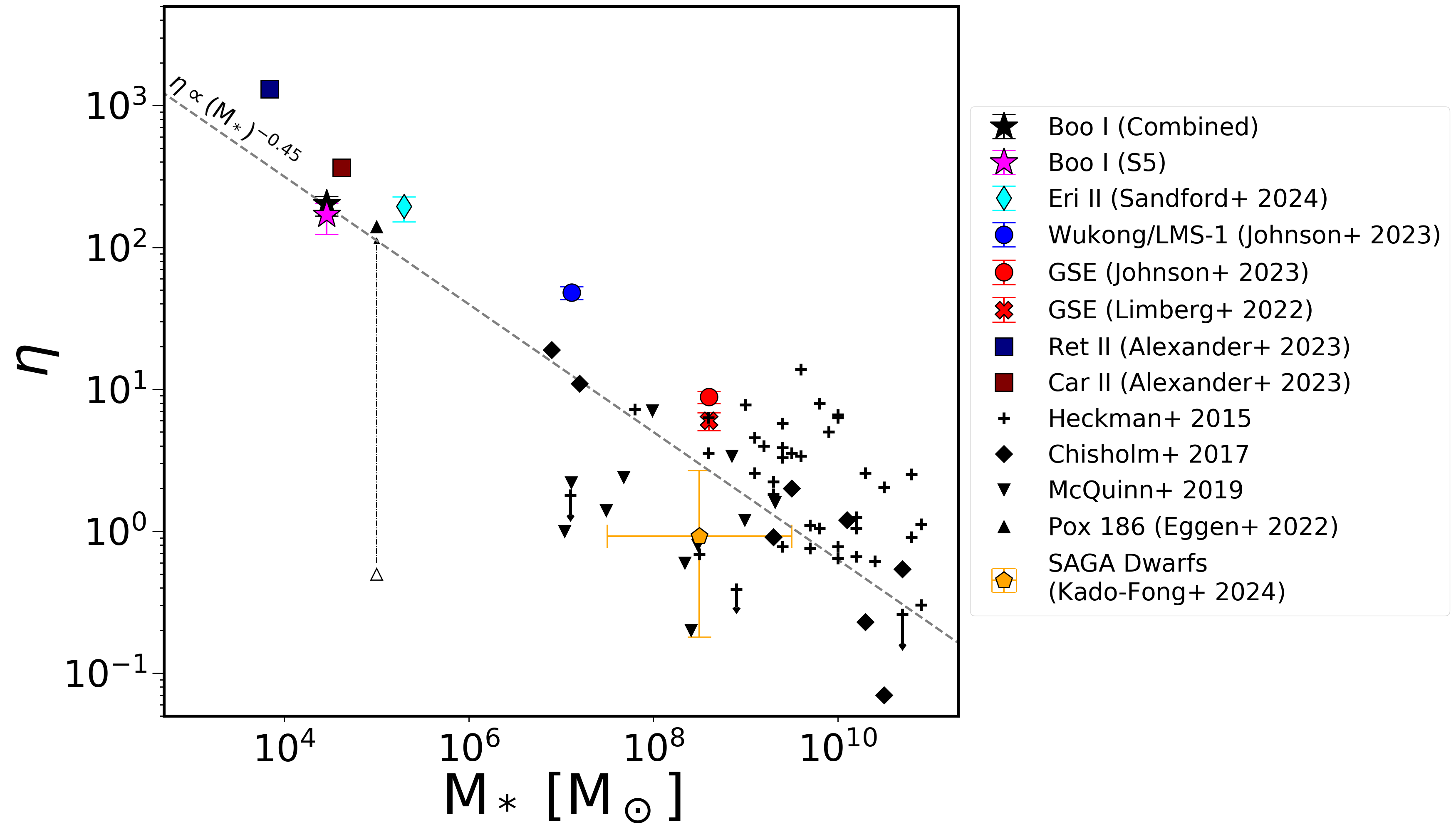}
    \caption{
        The inferred mass-loading factor of \BooI\ from the \Sfive\ and combined datasets (magenta and black stars respectively) compared to the mass-loading factors inferred by the chemical evolution studies of \citet{johnson:2023} for Wukong/LMS-1 and GSE (blue and red circles respectively), \citet{limberg:2022} in GSE (red ``x"), \citet{alexander:2023} for Carina II and Reticulum II (navy and maroon squares respectively), and \citet{sandford:2024} for Eridanus II (cyan diamond) as a function of stellar mass. The mass-loading factors inferred by the evolution of the mass-metallicity relation in the SAGA background dwarf galaxy sample by \citet{kado-fong:2024} is included as an orange pentagon. Mass-loading factors for galaxies observed by \citet{heckman:2015}, \citet{chisholm:2017}, and \citet{mcquinn:2019} are included as black pluses, diamonds, and triangles respectively. The current observed mass-loading factor of Pox 186 and its previous estimated mass-loading factor from \citet{eggen:2022} are represented by the open and filled black triangle respectively. The scaling found by \citet{pandya:2021} in FIRE-2 simulations indicative of energy-driven winds ($\eta\propto M_*^{-0.45}$) is included for reference as the dashed gray line.
    } 
    \label{fig:Eta_Mstar}
\end{figure*}

In Figure \ref{fig:Eta_Mstar}, we compare the mass-loading factor we infer for \BooI\ from the \Sfive\ and combined dataset (magenta and black stars respectively) to the mass-loading factor inferred by other dwarf galaxy chemical evolution studies \citep[e.g.,][]{limberg:2022, johnson:2023, kado-fong:2024, sandford:2024}, as well the mass-loading factors inferred from direct observations of galactic outflows in nearby star forming galaxies \citep[e.g.,][]{heckman:2015, chisholm:2017, mcquinn:2019, eggen:2022}. %Previous GCE studies included in this figure include \citet{johnson:2023} who use the \texttt{VICE} one-zone chemical evolution model to analyze the disrupted dwarf galaxies \textit{Gaia}-Sausage Enceladus (GSE) and Wukong/LMS-1 (red and blue circles respectively), \citet{alexander:2023} who use the \texttt{i-getool} inhomogenous chemical evolution model to study the Carina II and Reticulum II UFDs (maroon and navy squares respectively), \citet{sandford:2024} who uses the \citetWAF\ analytic chemical evolution model to study the Eridanus UFD (cyan diamond), and \citet{kado-fong:2024} who models the evolution of the mass-metallicity relation in the SAGA background dwarf galaxy sample (orange pentagon). 
%Mass-loading factors for direct observational indicators include values reported by \citet{heckman:2015}, \citet{chisholm:2017}, and \citet{mcquinn:2019} (black pluses, diamonds, and triangles respectively). 
%We also include the current observed and previous estimated mass-loading factor of the dwarf starburst galaxy Pox 186 from \citet{eggen:2022} (open and filled black triangles respectively). 
In addition, we include the scaling relation between stellar mass and mass-loading factor $\eta=0.6\times(M_*/10^{10}~\text{M}_\odot)^{-0.45}$ measured by \citet{pandya:2021} in FIRE-2 cosmological simulations \citep{hopkins:2018}. %The slope of this relationship is consistent with the physical scenario of energy-driven winds from SNe \citep[e.g.,][]{chevalier:1985}, which is thought to dominate over momentum-driven winds in low-mass galaxies \citep[e.g.,][]{murray:2005, murray:2010, hopkins:2012, dave:2013}.

As discussed in \citet{sandford:2024}, drawing direct comparisons between mass-loading factors measured through direct observational indicators, those measured in hydrodynamic simulations, and those inferred from GCE models is challenging. This is in part because the parameterization of outflows in GCE models frequently does not map neatly to observable quantities tracing galactic outflows or to the quantities tracked in hydrodynamic simulations. For example, in the \citetWAF\ model, $\eta$ parameterizes gas that is permanently removed from the reservoir of gas available for star formation, but in both reality and hydrodynamic simulation, a portion of this gas may be retained by a galaxy in a hot-gas phase that is similarly unavailable for star formation even if it is not ejected.

Additionally, the strength of galactic outflows experienced by galaxies observed at $z=0$ may not be representative of star formation rate-averaged outflows experienced by the same galaxy at high redshift. Furthermore, GCE-based mass-loading factors are strongly degenerate with the adopted stellar yields, which are uncertain to a factor of $\sim$2 \citep[e.g.,][]{griffith:2021, weinberg:2023}, and the degree to which metals are directly ejected from the galaxy by SN explosions. Propagating these uncertainties leads to an additional factor of $\sim$2--3 uncertainty in the inferred mass-loading factor\footnote{This may explain why the CGE results are systematically higher than the predictions of \citet{pandya:2021}.}. Nevertheless, our result for \BooI\ is in good agreement with both the theoretical predictions from hydrodynamic simulations and the emerging trend of increasing mass-loading factor with decreasing stellar mass. 

\begin{figure*}[ht!] 
	\includegraphics[width=1.0\textwidth]{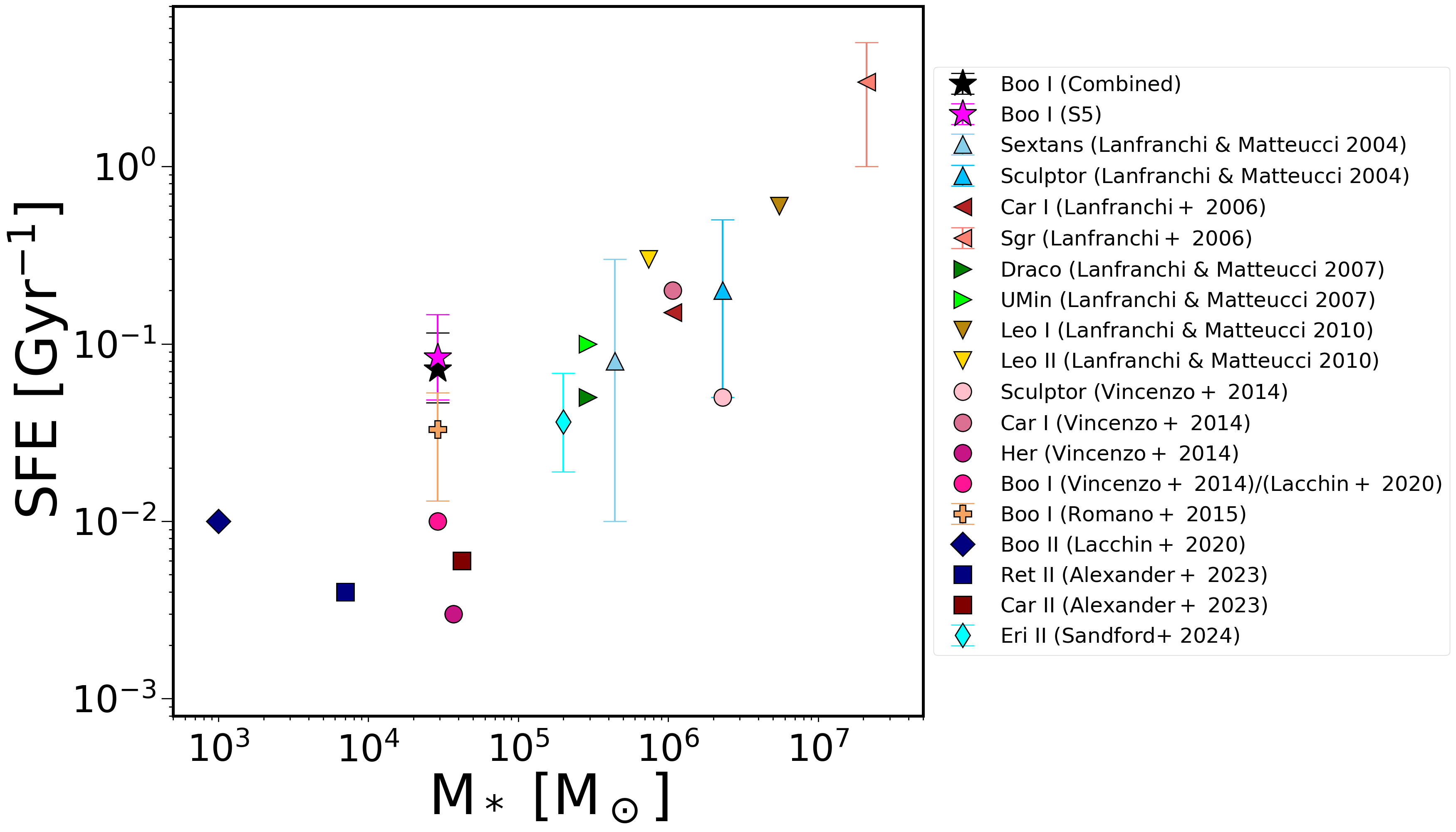}
    \caption{
        The inferred SFE of \BooI\ from the \Sfive\ and combined datasets (magenta and black stars respectively) compared to the SFEs reported by previous chemical evolution studies of LG dwarf galaxies \citep{lanfranchi:2004, lanfranchi:2006, lanfranchi:2007, lanfranchi:2010, vincenzo:2014, romano:2015, lacchin:2020, alexander:2023, sandford:2024}.
    } 
    \label{fig:SFE_Mstar}
\end{figure*}

In Figure \ref{fig:SFE_Mstar}, we compare the SFE we infer for \BooI\ to the SFE reported by GCE analyses of 16 Local Group dwarf galaxies spanning a wide range in stellar masses \citep{lanfranchi:2004, lanfranchi:2006, lanfranchi:2007, lanfranchi:2010, vincenzo:2014, romano:2015, lacchin:2020, alexander:2023, sandford:2024}. Though the assumptions made by the included GCE models vary greatly, there is a clear trend between SFE and galaxy mass---albeit with appreciable scatter. While slightly high given \BooI's mass, the SFE we infer is in reasonable agreement with expectations that less massive galaxies are less efficient at converting gas to stars.

%\subsection{Caveats}

% Constant time-independent SFE and eta

% One zone, instantaneous recycling/mixing

% Spatial Gradients

% Stochastic events and inhomogenous mixing.

% Debate over direct ejection of metals
% \citep{romano:2019, lanfranchi:2024}

\section{Summary}
\label{sec:summary}
In this paper, we present a comprehensive analysis of new and archival spectroscopic observations in the ancient, dark matter-dominated UFD \BooI. Our key results when using the combined dataset are as follows:
\begin{itemize}
    \item We create a combined dataset of 148 member stars extending our to $\sim7r_h$, including 24 newly confirmed members, 18 binary candidates, and 15 RRL stars. Of these member stars, 115 have good non-variable radial velocities suitable for dynamical analyses, and 92 have good [Fe/H] measurements suitable for chemical evolution analyses.
    \item We leverage the 16-year time baseline of our dataset to identify 15 new binary candidates, which we remove from our dynamical analysis. %and consider the impact that the misclassification of binaries has on our dynamical analysis.
    \item We infer a mean systemic velocity of $\langle v\rangle=103.0\pm0.4~\text{km}~\text{s}^{-1}$ and a velocity dispersion of $\sigma_{v}=4.0^{+0.4}_{-0.3}~\text{km}~\text{s}^{-1}$. % (or $3.9^{+0.4}_{-0.3}~\text{km}~\text{s}^{-1}$ depending on the treatment of ambiguous binary candidates). 
    Accounting for the effects of perspective rotation, we infer an intrinsic l.o.s.\ velocity gradient of $\Delta v = 0.12^{+0.04}_{-0.03}~\text{km}~\text{s}^{-1}~\text{arcmin}^{-1}$ ($1.2^{+0.4}_{-0.3}~\text{km}~\text{s}^{-1}~r_h^{-1}$) aligned with \BooI's orbit, which is significantly shallower than previously reported by \citetLongeard. 
    \item We infer a mean metallicity of $\langle\text{[Fe/H]}\rangle=-2.43\pm0.4$, a metallicity dispersion of $\sigma_\text{[Fe/H]}=0.28\pm0.03$, and a weak but resolved radial metallicity gradient of $\Delta\text{[Fe/H]}=-0.010\pm0.003~\text{dex}~\text{arcmin}^{-1}$ ($-0.10\pm0.03~\text{dex}~r_h^{-1}$).
    \item We provide updated constraints on \BooI's dark matter density profile using axisymmetric Jeans modeling. Our larger and more spatially extended dataset favors a cuspy dark matter halo ($\gamma=1.0^{+0.52}_{-0.60}$) but still cannot rule out a cored profile.
    \item We find that \BooI's MDF can be reproduced with a one-zone GCE featuring rapid ($\tau_\text{SFH}=0.2\pm0.1~\text{Gyr}$) and inefficient ($\text{SFE}=0.07^{+0.04}_{-0.03}~\text{Gyr}^{-1}$) star formation and large stellar feedback-driven outflows ($\eta=203^{+27}_{-36}$) in agreement with previous GCE modeling of other UFDs.
\end{itemize}

In addition to the above results, we discuss the implications of \BooI's intrinsic velocity gradient, but are unable to conclusively rule out either rotational or tidal origins. The alignment of \BooI's elongation, orbit, and velocity gradient, %along with its recent pericentric passage ($r_\text{peri}\sim35~\text{kpc}$) 300 Myr ago, 
support previous arguments that \BooI\ is undergoing tidal disruption---or has at least experienced tidal deformation as a result of its repeated pericentric passages \citep[e.g.,][]{roderick:2016, munoz:2018, longeard:2022}. 
%Additionally, we do not find evidence for multiple distinct chemodynamic populations as would be expected from the proposed major-merger scenario, but rather a smooth gradient in both metallicity and velocity, as well as their dispersions. 
We also find that despite \BooI's rapid star formation, its metallicity dispersion can be explained by self-enrichment without requiring either mergers or inhomogeneous mixing. These results suggest that the extended stellar population is more likely the result of tidal effects than a dry merger.
However, it is difficult with the current data to definitively confirm the presence of tidal disruption and rule out the alternative or additional scenario of a past merger. For example, our observations, though wide field, still do not extend beyond \BooI's estimated tidal radius, precluding the detection of extra-tidal stars. Moreover, dynamical simulations of a \BooI-like galaxy on a \BooI-like orbit are unable to reproduce the observed velocity gradient, predicting instead that a tidally-induced velocity gradient should only be observed at larger radii than our observations cover.

More conclusive statements will require a larger homogeneous dataset in which the selection function is well-characterized as will be the case from upcoming spectroscopic surveys (DESI; \citealt{desicollaboration:2016}, PFS; \citealt{takada:2014}). In particular, datasets that will provide robust and homogeneous [$\alpha$/Fe] (or neutron capture elements) both within one half-light radius and out beyond 5$r_h$ will be needed to test whether the extended features are in fact chemically distinct. Detection of extra-tidal stars that have been stripped from \BooI\ will also provide valuable insight into \BooI's dynamic history.
%In either case, the results of the DM profile analysis of \BooI\ should be treated with caution as \BooI\ is either undergoing tidal disruption (and hence not in dynamical equilibrium), partially rotationally supported, and/or composed of multiple dynamic components. 
In conjunction with additional spectroscopic data, more rigorous dynamical modeling will also aid in the interpretation of \BooI's chemodynamic evolution. Nevertheless, the work presented in this paper firmly establishes \BooI\ as a premier astrophysical laboratory for studying dark matter and galaxy evolution in low-mass satellite galaxies.

%\begin{acknowledgments}
\section*{Acknowledgments}  % for arXiv
We thank Marla Geha for valuable feedback and discussion regarding this work. NRS acknowledges support from an Arts \& Science Postdoctoral Fellowship at the University of Toronto. NRS, GEM and TSL acknowledge financial support from Natural Sciences and Engineering Research Council of Canada (NSERC) through grant RGPIN-2022-04794. NRS and JB acknowledge financial support from NSERC (funding reference number RGPIN-2020-04712). SEK acknowledges support from the Science \& Technology Facilities Council (STFC) grant ST/Y001001/1. KH acknowledges supports from Grant-in-Aid for Scientific Research from the Ministry of Education, Culture, Sports, Science, and Technology (MEXT), Japan, grant numbers JP24K00669 and JP25H01553.

This paper includes data obtained with the Anglo-Australian Telescope in Australia and hosted on AAO Data Central (\url{datacentral.org.au}). We acknowledge the traditional owners of the land on which the AAT stands, the Gamilaraay people, and pay our respects to elders past and present.

This work has made use of data from the European Space Agency (ESA) mission \gaia\ (\url{https://www.cosmos.esa.int/gaia}), processed by the \gaia\ Data Processing and Analysis Consortium (DPAC, \url{https://www.cosmos.esa.int/web/gaia/dpac/consortium}). Funding for the DPAC has been provided by national institutions, in particular the institutions participating in the \gaia\ Multilateral Agreement.

This work has also made use of data collected at the European Organisation for Astronomical Research in the Southern Hemisphere under ESO programs 82.B-0372(A), 185.B0946(A) and 185.B-0946(B).

Lastly, this work has made use of data in the Local Volume Database \citep{pace:2024}.

For the purpose of open access, the authors have applied a Creative Commons Attribution (CC BY) license to any Author Accepted Manuscript version arising from this submission.
%\end{acknowledgments}

\vspace{5mm}
\facilities{
    AAT (AAOmega+2dF),
    MMT (Hectochelle),
    VLT (GIRAFFE), 
    \gaia
}

\software{
    agama \citep{vasiliev:2019},
    astropy \citep{astropycollaboration:2013, astropycollaboration:2018, astropycollaboration:2022},
    astroquery \citep{ginsburg:2019},
    corner \citep{foreman-mackey:2016},
    emcee \citep{foreman-mackey:2013},
    GADGET-3 \citep{springel:2005},
    galpy \citep{bovy:2015},
    matplotlib \citep{Hunter:2007},
    numpy \citep{harris:2020},
    pandas \citep{mckinney:2010, thepandasdevelopmentteam:2024},
    pocoMC \citep{karamanis:2022a, karamanis:2022b},
    RVSpecFit \citep{koposov:2011, koposov:2019},
    scipy \citep{virtanen:2020}
}

\appendix
\restartappendixnumbering

\section{Catalog of Stars}
\label{app:catalog}
In Table \ref{tab:stars}, we provide a description of the columns for the the combined stellar catalog analyzed in this work. The dataset is made available in machine-readable format.

\begin{deluxetable}{rcl}
\label{tab:stars}
\tablecaption{Combined Catalog of Stars}
\tablehead{
   \colhead{Column} & \colhead{Units} & \colhead{Description}
}
\startdata
\texttt{gaia\_id} & \nodata  & \gaia Source ID \\
\texttt{alt\_id} & \nodata & Source ID from \citetKoposov\ and \citetJenkins \\
\texttt{RA} & deg & Right ascension \\
\texttt{Dec} & deg & Declination \\
\texttt{pmra} & mas yr$^{-1}$ & Proper motion (RA) \\
\texttt{pmra\_err} & mas yr$^{-1}$ & Proper motion (RA) \\
\texttt{pmdec} & mas yr$^{-1}$ & Proper motion uncertainty (dec) \\
\texttt{pmdec\_err} & mas yr$^{-1}$ & Proper motion uncertainty (dec) \\
\texttt{pmra\_pmdec\_corr} & \nodata & Correlation coefficient between \texttt{pmra} and \texttt{pmdec} \\
\texttt{vel\_avg} & kms s$^{-1}$ & Heliocentric l.o.s.\ velocity averaged over datasets \\
\texttt{vel\_err\_avg} & kms s$^{-1}$ & Uncertainty on \texttt{vel\_avg} \\
\texttt{vel\_s5} & kms s$^{-1}$ & Heliocentric l.o.s.\ velocity from the \Sfive\ dataset \\
\texttt{vel\_err\_s5} & kms s$^{-1}$ & Uncertainty on \texttt{vel\_err\_s5} \\
\texttt{vel\_q\_s5} & \nodata & Flag for velocities used in the dynamical analysis \\
\texttt{vel\_aat} & kms s$^{-1}$ & Corrected heliocentric l.o.s.\ velocity from the archival AAT dataset \\
\texttt{vel\_err\_aat} & kms s$^{-1}$ & Uncertainty on \texttt{vel\_err\_aat} \\
\texttt{vel\_q\_aat} & \nodata & Flag for velocities used in the dynamical analysis \\
\texttt{vel\_mmt} & kms s$^{-1}$ & Corrected heliocentric l.o.s.\ velocity from the archival MMT dataset \\
\texttt{vel\_err\_mmt} & kms s$^{-1}$ & Uncertainty on \texttt{vel\_err\_mmt} \\
\texttt{vel\_q\_mmt} & \nodata & Flag for velocities used in the dynamical analysis \\
\texttt{vel\_vlt} & kms s$^{-1}$ & Corrected heliocentric l.o.s.\ velocity from the archival VLT dataset \\
\texttt{vel\_err\_vlt} & kms s$^{-1}$ & Uncertainty on \texttt{vel\_err\_vlt} \\
\texttt{vel\_q\_vlt} & \nodata & Flag for velocities used in the dynamical analysis \\
\texttt{feh\_avg} & dex & [Fe/H] averaged over datasets. \\
\texttt{feh\_err\_avg} & dex & Uncertainty on \texttt{feh\_avg} \\
\texttt{feh\_s5} & dex & [Fe/H] from the \Sfive\ dataset. \\
\texttt{feh\_err\_s5} & dex & Uncertainty on \texttt{feh\_err\_s5} \\
\texttt{feh\_q\_s5} & \nodata & Flag for [Fe/H] used in the chemical analysis \\
\texttt{feh\_aat} & dex & Corrected [Fe/H] from the archival AAT dataset \\
\texttt{feh\_err\_aat} & dex & Uncertainty on \texttt{feh\_err\_aat} \\
\texttt{feh\_q\_aat} & \nodata & Flag for [Fe/H] used in the chemical analysis \\
\texttt{feh\_mmt} & dex & Corrected [Fe/H] from the archival MMT dataset \\
\texttt{feh\_err\_mmt} & dex & Uncertainty on \texttt{feh\_err\_mmt} \\
\texttt{feh\_q\_mmt} & \nodata & Flag for [Fe/H] used in the chemical analysis \\
\texttt{feh\_vlt} & dex & Corrected [Fe/H] from the archival VLT dataset \\
\texttt{feh\_err\_vlt} & dex & Uncertainty on \texttt{feh\_err\_vlt} \\
\texttt{feh\_q\_vlt} & \nodata & Flag for [Fe/H] used in the chemical analysis \\
\texttt{pval} & \nodata & $p$-value from null-hypothesis testing of RV variability \\
\texttt{binary} & \nodata & Flag indicating binary candidate ($\texttt{pval} < 0.1$) \\
\texttt{mem\_p\_s5} & \nodata & Membership probability in the \Sfive\ GMM analysis\\
\texttt{mem\_p\_aat} & \nodata & Membership probability in the archival AAT GMM analysis\\
\texttt{mem\_p\_mmt} & \nodata & Membership probability in the archival MMT GMM analysis\\
\texttt{mem\_p\_vlt} & \nodata & Membership probability from \citetJenkins\ archival VLT analysis \\
\texttt{member} & \nodata & Flag indicating likely \BooI\ member \\
\enddata
\tablecomments{Table \ref{tab:stars} is published in its entirety in the electronic 
edition of the {\it Astrophysical Journal}.}
\end{deluxetable}

\section{Analysis of Archival Datasets}
\label{app:archival}
In this section, we present the independent analysis of the archival AAT and MMT datasets and compare the results to the analysis of the \Sfive-only and combined datasets. For each of the archival AAT and MMT datasets, we apply the same data quality cuts and GMM as described in Sections \ref{sec:GMM_data} and \ref{sec:GMM_like}. Prior to fitting with the GMM, the velocity- and [Fe/H]-offsets presented in Table \ref{tab:offsets} are applied.

%In Figure \ref{fig:arx_mem_p}, we present the distribution of median membership probabilities for the archival AAT (left) and MMT (right) datasets. 
Using the same membership threshold of $p_\text{mem} > 0.8$ to identify likely members, we find 42 likely members with good velocities in the archival AAT dataset, of which 41 have good metallicities. In the archival MMT dataset, we find 37 likely members with good velocities, of which 33 have good metallicities. We find good agreement between the likely members between the \Sfive, archival AAT, and archival MMT datasets---that is, all stars that are found to be members in one dataset are also found to be members in the other datasets in which they are present. The sample of likely \BooI\ members for each dataset is illustrated in Figure \ref{fig:Combined_4panel}.

% \begin{figure}[ht!] 
% 	\includegraphics[width=1.0\textwidth]{lit_GMM_MemP.png}
%     \caption{
%         Same as Figure \ref{fig:mem_p} except showing the median membership probabilities of the archival AAT (left) and MMT (right) datasets.
%     } 
%     \label{fig:arx_mem_p}
% \end{figure}

%In Figure \ref{fig:Combined_4panel}, we present the likely \BooI\ members of the \Sfive\ (magenta circles), archival AAT (gray squares), and archival MMT (blue diamonds) in several data dimensions, including the CMD (top left), spatial (top right), proper motion (bottom), and metallicity and velocity (bottom right). We include likely members in archival VLT from \citetJenkins\ as red triangles for reference.

The systemic properties of \BooI\ inferred from our GMM analysis of the archival datasets are summarized in Table \ref{tab:BooI_par_extra}. %The posterior distribution of all of the model parameters is presented as corner plot alongside analogous posteriors for analysis of the \Sfive\ and combined dataset in Figures \ref{fig:GMM_Corner_BooI} and \ref{fig:GMM_Corner_Bkg}. 
For nearly all \BooI\ model parameters, we find the inferred values to be in good agreement between the individual and combined datasets. The only exceptions are the mean metallicity, which we infer to be $\sim$0.1 and $\sim$0.2 dex lower for the archival AAT and MMT datasets compared to the \Sfive, archival VLT, and combined datasets, as well as the metallicity dispersion which we infer to be $\sim$0.1 dex larger for the archival AAT and MMT datasets compared to the \Sfive, archival VLT, and combined datasets. The MW foreground model parameters differ considerably between the datasets as a result of the different target selections employed by each observing campaign.

% \begin{figure*}[ht!] 
%     \includegraphics[width=1.0\textwidth]{figures/GMM_Corner_BooI.png}
%     \caption{
%         Pair-wise plot of the posterior distribution for the parameters describing the \BooI\ component of the GMM as fit to the \Sfive, archival AAT, and archival MMT datasets (magenta, gray, and blue respectively). Posterior distributions from \citetJenkins\ for archival VLT and from a non-mixture model applied to our combined dataset are included for comparison where applicable (red and black respectively). Results for all datasets are in reasonable agreement.
%     } 
%     \label{fig:GMM_Corner_BooI}
% \end{figure*}

% \begin{figure}[ht!] 
%     \includegraphics[width=0.5\textwidth]{figures/GMM_Corner_Bkg.png}
%     \caption{
%         Same as Figure \ref{fig:GMM_Corner_BooI} except for the MW foreground component. The differences in the posteriors for these model parameters is indicative of the different target selections employed by each observing campaign.
%     } 
%     \label{fig:GMM_Corner_Bkg}
% \end{figure}

\begin{figure*}[ht!] 
	\includegraphics[width=1.0\textwidth]{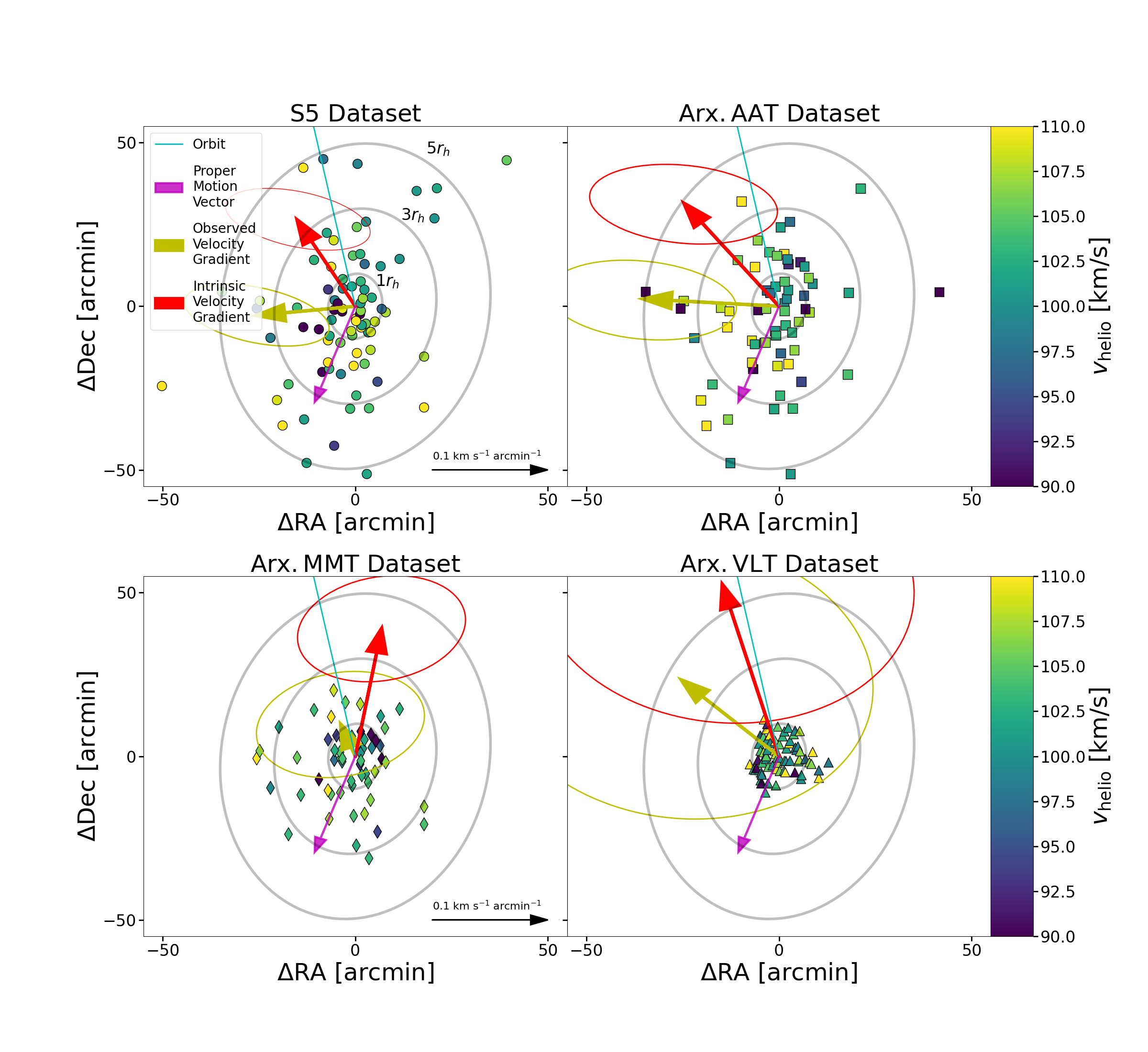}
    \caption{
        Same as Figure \ref{fig:vel_grad} for subsets of the combined dataset, including \Sfive\ (top left), archival AAT (top right), archival MMT (bottom left), and archival VLT (bottom right). Velocity gradients for the \Sfive, AAT, and MMT datasets are inferred from the GMM model described in Section \ref{sec:GMM} and Appendix \ref{app:archival}, while the velocity gradient for the VLT dataset is inferred directly from the high probability members found in \citetJenkins.
    } 
    \label{fig:allobs_vel_grad}
\end{figure*}

Figure \ref{fig:allobs_vel_grad} illustrates the velocity gradient inferred from the \Sfive\ (top left), archival AAT (top right), archival MMT (bottom left), and archival VLT (bottom right) datasets. We find the velocity gradients in all of the archival datasets to be consistent within 1$\sigma$ to the results from the \Sfive\ and combined datasets (see Section \ref{sec:vel_grad}). All datasets appear to independently support the existence of a velocity gradient aligned with \BooI's orbit and/or semi-major axis. The velocity gradient inferred from the archival VLT dataset is very uncertain due to the limited spatial extent of the sample. 

\begin{figure*}[ht!] 
\label{fig:comb_disp_grad}
\gridline{\fig{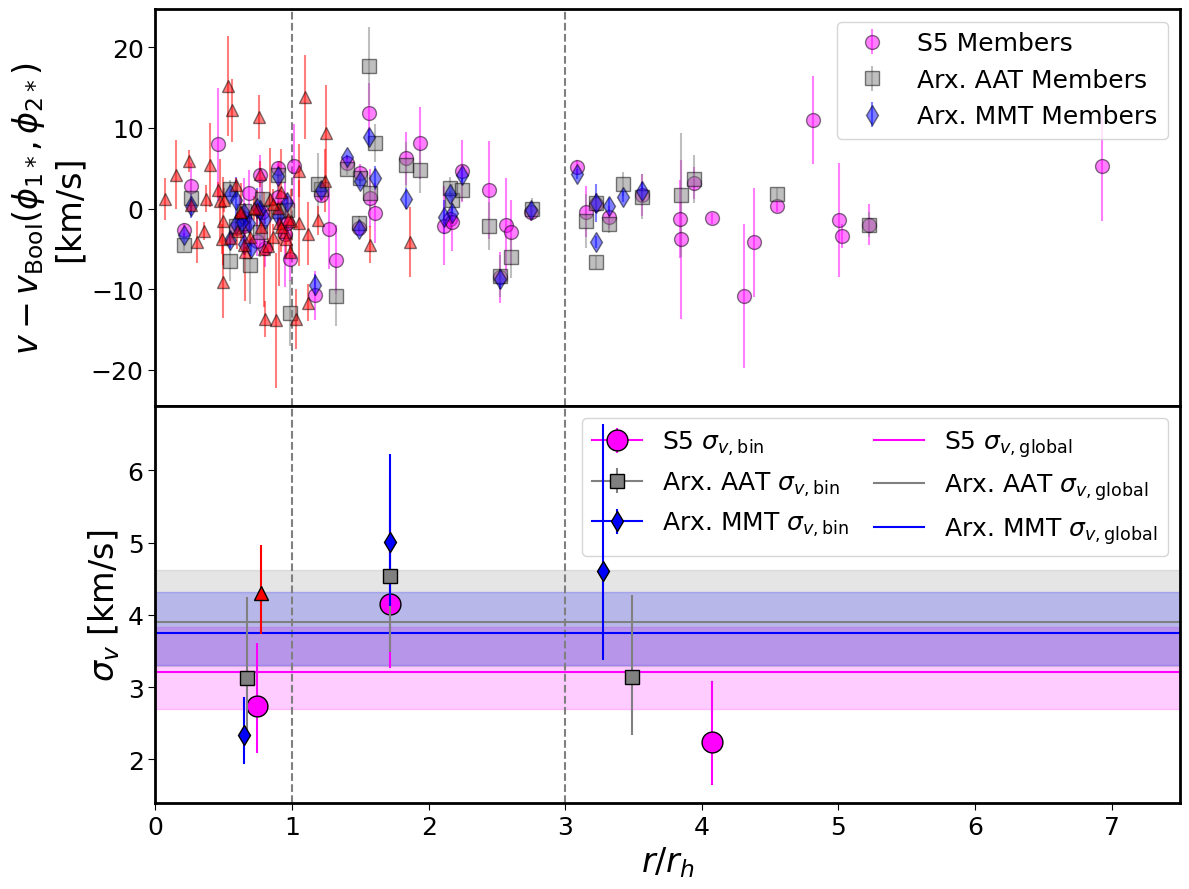}{0.5\textwidth}{}
          \fig{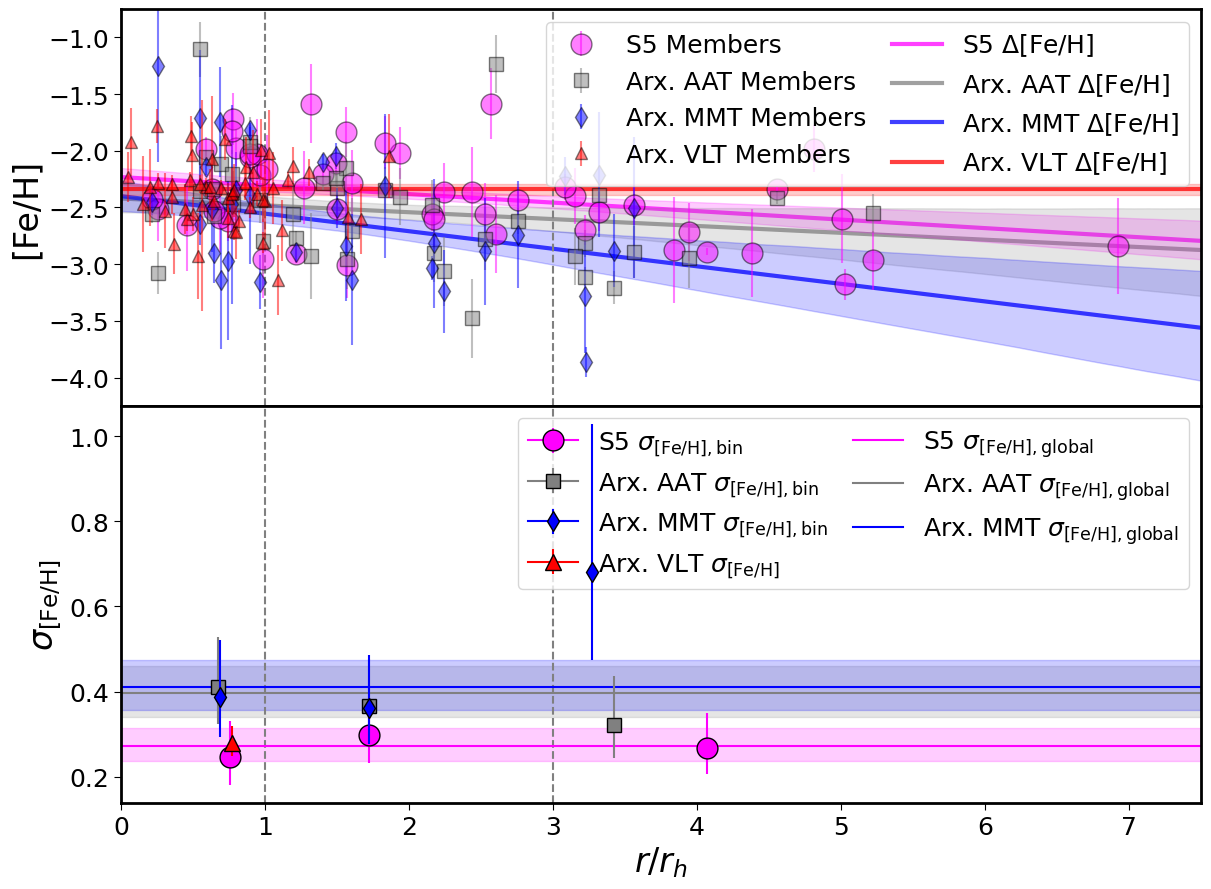}{0.5\textwidth}{}}
\caption{
    \textbf{Left.} Same as Figure \ref{fig:vdisp_grad} including \BooI\ members from \Sfive, archival AAT, \citetWalker, \citetJenkins\, and combined  datasets (magenta circles, gray squares, blue diamonds, red triangles, and black circles respectively respectively).
    \textbf{Right.} Same as Figure \ref{fig:fehdisp_grad} including \BooI\ members from \Sfive, archival AAT, \citetWalker, \citetJenkins\, and combined  datasets (magenta circles, gray squares, blue diamonds, red triangles, and black circles respectively respectively). 
}
\end{figure*}

As in Section \ref{sec:vel_disp_prof}, we perform a binned velocity dispersion analysis for the individual datasets, which we present in the left panels of Figure \ref{fig:comb_disp_grad}. We find general agreement between the velocity dispersion in each bin for each dataset. %, though the velocity dispersion of the inner bin for the archival AAT and MMT datasets are smaller than for the \Sfive\ dataset and in better agreement with the combined dataset as expected given the presence of the two ambiguous binary stars, 1230834410380273664 and 12308335857459879. 
Both of the datasets with $>$10 stars beyond 3$r_h$ (\Sfive\ and archival AAT) exhibit a smaller velocity dispersion in the outer bin, in contrast to the increasing velocity dispersion reported by \citetLongeard.
%This makes sense when considering that the two ambiguous binary stars, 1230834410380273664 and 12308335857459879, are the result of differences between the \Sfive\ and archival velocity measurements of these sources and have not been removed from the analysis for this figure. 

In the top right panel of Figure \ref{fig:comb_disp_grad}, we present results for the metallicity gradient of the individual datasets. The \Sfive, archival AAT, and archival MMT datasets all independently support the existence of a negative radial metallicity gradient with the archival MMT (AAT) dataset exhibiting the strongest (weakest) gradient. We do not attempt to fit a metallicity gradient to the archival VLT dataset due to its small spatial extent, instead we include the mean metallicity reported by \citetJenkins\ for reference. 
In the bottom right panel of Figure \ref{fig:comb_disp_grad}, we present the binned metallicity dispersions of each dataset. We find no evidence for a radially increasing or decreasing metallicity dispersion in any of the datasets. %From this binned analysis, it is apparent that the higher metallicity dispersion for the archival AAT and MMT datasets is being driven primarily by stars in the inner bin. This may also point to the source of the disagreement with our inferred metallicity dispersion and that reported by \citetLongeard.

\section{Axisymmetric Jeans Modeling}
\label{app:jeans}
Here, we provide a brief summary of the axisymmetric Jeans modeling performed in Section \ref{sec:jeans_modeling}. For a more complete description of the development of this methodology and its limitations, see \citet{hayashi:2012}, \citet{hayashi:2015}, \citet{hayashi:2020} and \citet{hayashi:2023}. In brief, we assume that \BooI\ is an axisymmetric dark matter-dominated system in dynamical equilibrium such that its stellar motions can be described by the axisymmetric Jeans equations:
\begin{align}
\label{eq:jeans1}
\overline{v_z^2}&=\frac{1}{\nu(R,Z)}\int_{z}^{\infty}\nu\frac{\partial\Phi_\text{DM}}{\partial z}dz~\text{and} \\
\label{eq:jeans2}
\overline{v_\phi^2}&=\frac{1}{1-\beta_z}\left[\overline{v_z^2} + \frac{R}{\nu}\frac{\partial(\nu \overline{v_z^2})}{\partial R}\right] + R\frac{\partial \Phi_\text{DM}}{\partial R},
\end{align}
where 
\begin{equation}
\label{eq:jeans3}
\beta_z = 1 - \overline{v_z^2}/\overline{v_R^2}
\end{equation}
is the radially independent velocity anisotropy parameter introduced by \citet{cappellari:2008}, $\nu$ is the three-dimensional stellar density, and $\Phi_\text{DM}$ is the dark matter halo's gravitational potential. 

We model the stellar density profile by with an axisymmetric Plummer model \citep{plummer:1911}:
\begin{equation}
\label{eq:plummer1}
\nu(R,z) = (3/4\pi b_*^3)(1+r_*^2/b_*^2)^{-5/2},
\end{equation}
where
\begin{equation}
\label{eq:plummer2}
r_*^{2} = R^{2} + z^{2}/q^{2}, 
\end{equation}
$b_*$ is the half-light radius, and $q$ is the axial ratio of the stellar distribution. This distribution is converted into a surface density profile using an Abel transform and the inclination angle, $i$, which we define to be the angle between the symmetry axis and the line of sight such that $i=0^{\circ}$ corresponds to looking at the galaxy face-on. As in \citet{hayashi:2023}, we leave the inclination angle as a free parameter and adopt the projected half-light radius and axial ratio for \BooI\ from \citet{munoz:2018}, where the intrinsic axial ratio, $q$, is related to the projected axial ratio, $q'$, as $q=\sqrt{q'^{2} - \cos^2i}/\sin i$ \citep{hubble:1926}. 

For the dark matter halo's gravitational potential, we adopt a generalized axisymmetric Hernquist profile \citep{hernquist:1990, zhao:1996}:
\begin{equation}
\label{eq:dm_density1}
\rho_\text{DM}(R, z) = \rho_0\left(\frac{r}{b_\text{halo}}\right)^{-\gamma}\left[1 + \left(\frac{r}{b_\text{halo}}\right)^{\alpha}\right]^{-\frac{\beta-\gamma}{\alpha}},
\end{equation}
where
\begin{equation}
\label{eq:dm_density2}
r^{2} = R^{2} + z^{2}/Q^{2}.
\end{equation}
Here, $\rho_0$ and $b_\text{halo}$ are the scale density and radius of the dark matter halo, respectively; $\alpha$ describes the sharpness of the transition from the inner dark matter slope, $\gamma$, and the outer dark matter slope, $\beta$; and $Q$ is the dark matter halo's axial ratio. We assume that the dark matter halo has the same orientation and symmetry axis as the stellar distribution. 
In summary, this model has 8 free parameters, $Q$, $\rho_0$, $b_\text{halo}$, $\beta_z$, $\alpha$, $\beta$, $\gamma$, and $i$. 

\begin{figure*}[ht!] 
\gridline{
    \fig{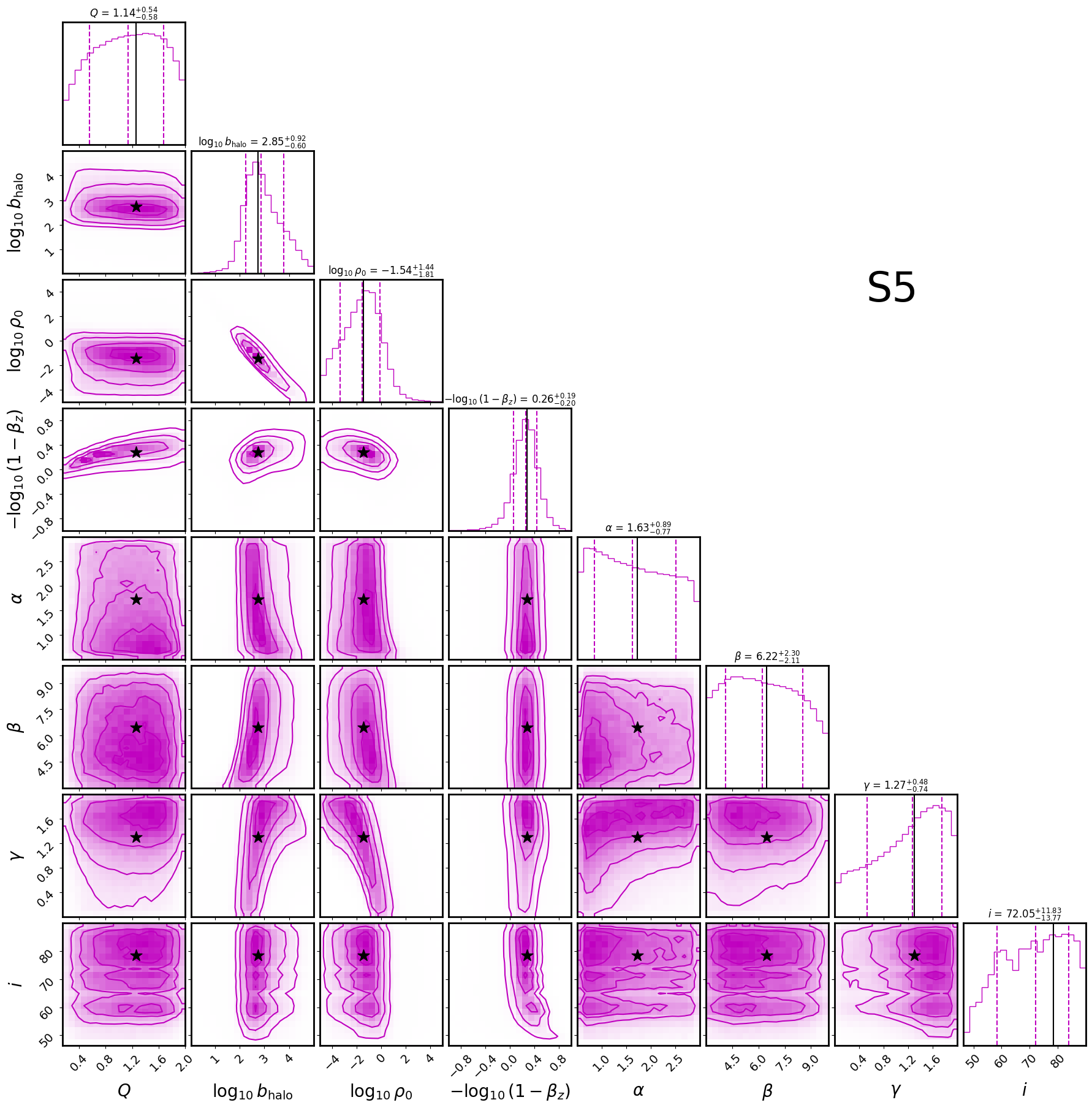}{0.5\textwidth}{}
    \fig{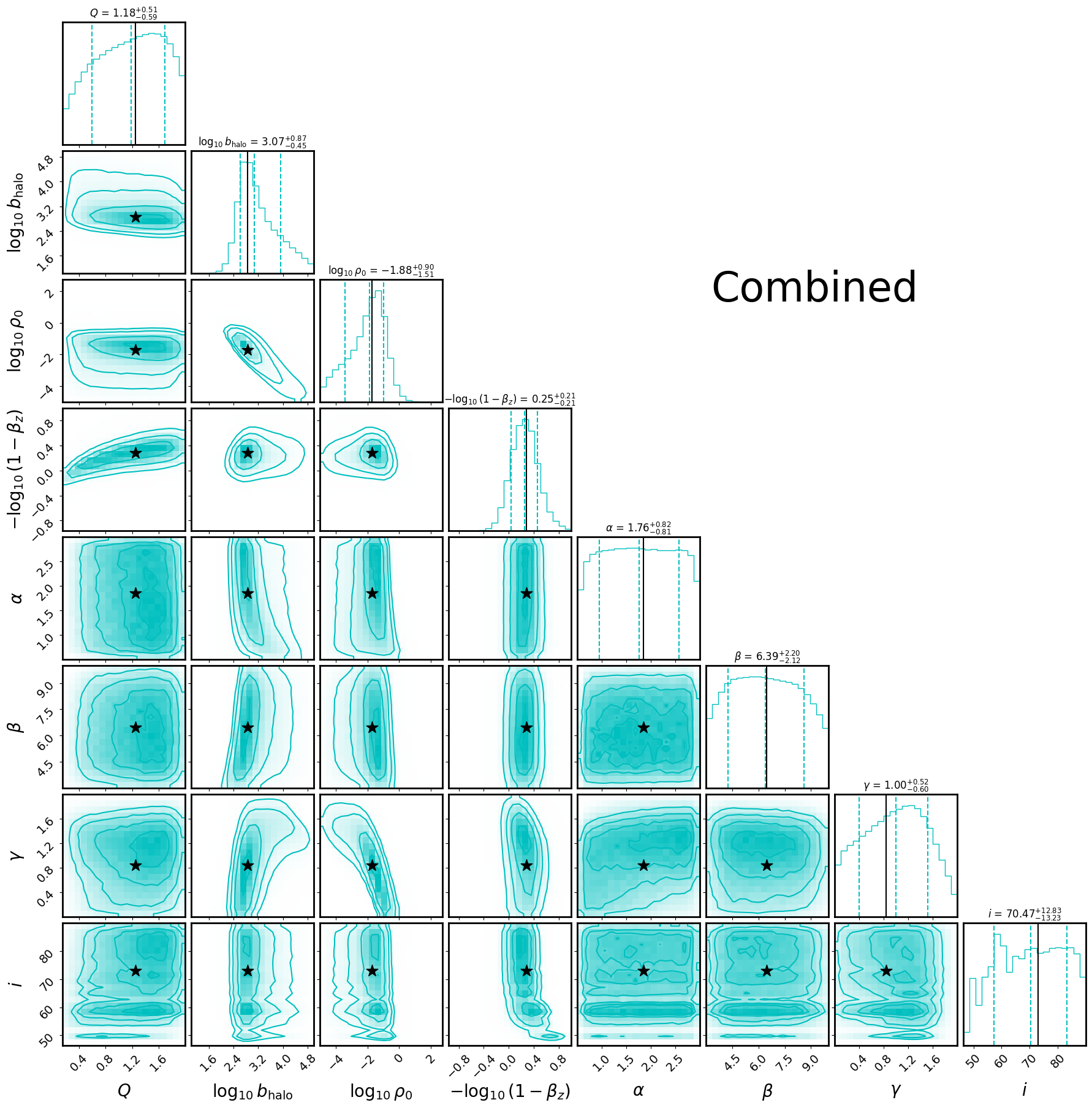}{0.5\textwidth}{}
}
% \gridline{\fig{jeans_corner_s5}{0.5\textwidth}{}
%           \fig{jeans_corner_s5_wo2}{0.5\textwidth}{}}
% \gridline{\fig{jeans_corner_comb}{0.5\textwidth}{}
%           \fig{jeans_corner_comb_wo2}{0.5\textwidth}{}}
\caption{
    Posterior probability distributions for the dark matter profile parameters for the default \Sfive\ (left) and combined (right) datasets. 
    % The top right and bottom right panels show the same for when the ambiguous binary sources 1230834410380273664 and 1230833585745987968 are removed from the \Sfive\ and combined datasets respectively. 
    Vertical dashed lines provide the median and 1$\sigma$ confidence intervals for each parameter. Black stars and vertical solid lines indicate the \textit{maximum a posteriori} values.
}
\label{fig:jeans_corner}
\end{figure*}

To fit this mass model to our datasets, we assume that that the line-of-sight velocity distribution at the location of each star is a Gaussian centered on the mean velocity of the galaxy such that the likelihood function is given by
\begin{equation}
    \label{eq:jeans_like}
    -2\ln(\mathcal{L}) = \sum_i\left[\frac{(v_i-\langle v_\text{B} \rangle)^2}{\sigma_i^2}+\ln(2\pi\sigma_i^2)\right],
\end{equation}
where $\sigma_i$ is the quadrature sum of the velocity uncertainty of the star and the line-of-sight velocity dispersion predicted by the model at the star's position in the plane of the sky. The mean velocity of \BooI\ $\langle v_\text{B} \rangle$ is a nuisance parameter.
We adopt the same flat or log-flat priors as \citet{hayashi:2023}, which we present in Table \ref{tab:priors}. Posterior distributions are estimated with MCMC sampling using the Metropolis–Hastings algorithm \citep{metropolis:1953, hastings:1970}. The resulting posterior distributions of the model parameters for the \Sfive\ (left) and combined (right) are illustrated in Figure \ref{fig:jeans_corner}.
 %illustrates the resulting posterior distributions of the model parameters for the \Sfive\ (top) and combined (bottom) datasets both with (left) and without (right) sources 1230834410380273664 and 1230833585745987968. 
 Vertical dashed lines provide the median and 1$\sigma$ confidence intervals for each parameter, while the black stars and vertical solid lines indicate the \textit{maximum a posteriori} values. 
These results are summarized in Table \ref{tab:BooI_par_extra}.

\section{Chemical Evolution Modeling}
\label{app:GCE}
In this appendix, we provide a brief summary of the galactic chemical evolution (GCE) model from \citet{weinberg:2017} (hereafter \citetWAF) that we use in Section \ref{sec:GCE}. For a more complete description of the model and its previous application to UFDs, see \citet{weinberg:2017} and \citet{sandford:2024} respectively. 

In brief, the \citetWAF\ model assumes that the SFR of \BooI\ can be approximated with a continuous analytic function. In this work, we adopt a linear-exponential functional form for the SFR:
\begin{equation}
    \label{eq:sfr}
    \dot{M}_*\propto
    \begin{cases}
        t\exp{(-t/\tau_\text{SFH})},& \text{if } t \leq t_\text{trunc} \\
        0,& \text{if } t > t_\text{trunc}
    \end{cases},
\end{equation}
where $\tau_\text{SFH}$ is the SFH timescale and $t_\text{trunc}$ is the time at which all star formation ceases (e.g., from ram pressure stripping or reionization). Furthermore, the model assumes that star formation is governed by a linear star formation law law parameterized by the star formation efficiency (SFE) timescale,
\begin{equation}
    \tau_\text{SFE}\equiv \text{SFE}^{-1}\equiv M_g/\dot{M}_*,
\end{equation}
where $M_g$ and $\dot{M}_*$ are the gas mass and star formation rate (SFR) respectively. 

At each timestep, a fraction, $r$, of the mass formed into stars is immediately returned to the ISM unprocessed without chemical enrichment (e.g., by core-collapse supernovae and asymptotic giant branch stars). As in \citet{sandford:2024}, we choose a mass recycling fraction of $r=0.37$, which \citetWAF\ showed to be an appropriate approximation for the adopted \citet{kroupa:2001} IMF. Additionally, at each timestep, gas is ejected from the ISM via stellar feedback according to a linear scaling with the SFR,
\begin{equation}
    \eta = \dot{M}_\text{outflow}/\dot{M}_*,
\end{equation}
where $\eta$ is the mass loading factor.
In the \citetWAF\ model, the rate of pristine gas accretion from the intergalactic medium is set implicitly such that the depletion of gas by star formation and galactic outflows is sufficiently balanced by gas recycling and galactic inflows:
\begin{equation} \label{eq:mdotinfall}
    \dot{M}_\text{inf}=(1+\eta-r)\dot{M}_* + \tau_\text{SFE} \ddot{M}_*
\end{equation}
(see Equation 9 in \citetWAF).

Enrichment from CC SNe is assumed to occur instantaneously following star formation, while the enrichment from SNe Ia is assumed to follow a $t^{-1.1}$ power-law delay time distribution with a minimum time delay of $t_D=0.05$ Gyr as determined empirically by \citet{maoz:2012}. We adopt the instantaneous mixing approximation, which has been shown to be a reasonable assumption for CC SNe and SNe Ia products in low-mass, ancient galaxies like \BooI\ \citep[e.g.,][]{escala:2018}. The \citetWAF\ model parametrizes chemical enrichment using dimensionless IMF-weighted metallicity-independent yield parameters, which represent the mass of elements produced per unit mass of star formation. We adopt the same empirically motivated values as \citet{sandford:2024}: a CC SN Fe yield of $y_\text{Fe}^\text{CC} = 6.0\times10^{-4}$ and a SN Ia Fe yield of $y_\text{Fe}^\text{Ia} = 1.2\times10^{-3}$.

In summary, this model has 4 free parameters, $\tau_\text{SFH}$, $t_\text{trunc}$, $\tau_\text{SFE}$, and $\eta$. The priors adopted for each of these parameters are summarized in Table \ref{tab:priors}. For $\tau_\text{SFH}$, we adopt a uniform prior from 0.08 to 0.35 Gyr, and for $t_\text{trunc}$, we adopt a uniform prior from 0 to 2 Gyr. These choices of priors are informed by \citet{durbin:2025} who derived a SFH for \BooI\ from deep \textit{HST}/ACS photometry and determined that 50\% (90\%) of its stars formed by $13.49^{+0.00}_{-0.04}$ ($13.37^{+0.00}_{-1.03}$) Gyr ago. For both $\log(\tau_\text{SFE})$ and $\eta$, we adopt broad, uniform priors of $0 < \log(\tau_\text{SFE}) < 4$ and $0 <\eta < 10^4$ respectively.

As in previous analyses of dwarf galaxy MDFs \citep[e.g.,][]{kirby:2011a}, we adopt the following likelihood function:
\begin{equation}
\label{eq:gce_like}
\ln \mathcal{L} = \sum_{i}\ln\int_{-\infty}^{\infty}\frac{dN}{d\text{[Fe/H]}}\frac{1}{\sqrt{2\pi}\sigma_{\text{[Fe/H]},i}}\exp\left(-\frac{(\text{[Fe/H]} - \text{[Fe/H]}_i)^2}{2\sigma_{\text{[Fe/H]},i}^2}\right)d\text{[Fe/H]},
\end{equation}
where $\frac{dN}{d\text{[Fe/H]}}$ is the normalized MDF predicted by the model. Here, $\text{[Fe/H]}_i$ and $\sigma_{\text{[Fe/H]},i}$ are once again the measured iron abundance and its uncertainty for each star. Simply put, this equation is the convolution of the model MDF with the observational Gaussian errors. For the \citetWAF\ model, 
\begin{equation}
\frac{dN}{d\text{[Fe/H]}} \propto \frac{\dot{M}_*}{d\text{[Fe/H]/dt}}.
\end{equation}
Sampling of the posterior distribution is performed \texttt{pocoMC} as described in Section \ref{sec:GMM_like}.

\begin{figure*}[ht!] 
    \centering
	\includegraphics[width=0.5\textwidth]{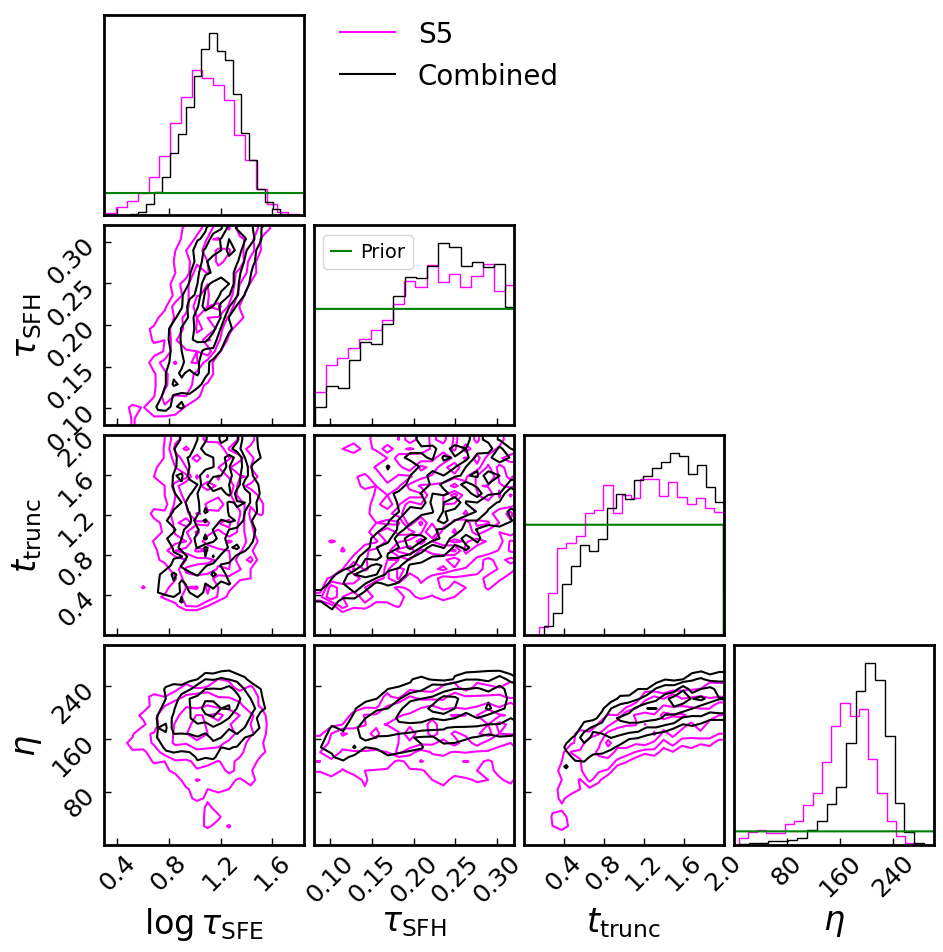}
    \caption{
        Posterior probability distribution for the galactic chemical evolution model parameters fit to the \Sfive\ (magenta) and combined (black) datasets. The adopted prior distributions are included for reference as solid green lines.
    } 
    \label{fig:ChemEv_Corner}
\end{figure*}

Figure \ref{fig:ChemEv_Corner} illustrates the resulting posterior distributions of the GCE model parameters for the \Sfive\ (magenta) and combined (black) datasets. The adopted prior distributions are included for reference as solid green lines.

\bibliography{BooI}{}
\bibliographystyle{aasjournal}



\end{document}